\documentclass[amsmath,twocolumn,amssymb,floatfix,showpacs,
superscriptaddress,nofootinbib,longbibliography]{revtex4-2}
\usepackage{mathtools}
\usepackage{braket}
\usepackage[dvipsnames]{xcolor}
\usepackage{float}
\usepackage{subfigure}
\usepackage{graphics}
\usepackage{bm}
\usepackage{MnSymbol}
\usepackage{braket}
\usepackage{float}
\usepackage{tikz}
\usepackage{blkarray}
\usepackage{pgffor}
\usepackage{float}
\usepackage{silence}
\usepackage[colorlinks=true,linktoc=page,linkcolor=blue,citecolor=blue,urlcolor=blue]{hyperref}

\mathchardef\mhyphen="2D % Define a "math hyphen"

\newcommand{\ie}{{\it i.e.,\,\,}}

\newcommand{\ua}{{\uparrow }}
\newcommand{\da}{{\downarrow }}

\newcommand\bea{\begin{eqnarray}}
\newcommand\eea{\end{eqnarray}}
\newcommand\beq{\begin{equation}}  
\newcommand\eeq{\end{equation}}

\newcommand{\non}{\nonumber}  
  
\usepackage[normalem]{ulem}
\definecolor{lime}{HTML}{A6CE39}
\usepackage{sidecap,tikz}
\usepackage{enumitem}
\DeclareRobustCommand{\orcidicon}{\hspace{-1.0mm}
	\begin{tikzpicture}
		\draw[lime, fill=lime] (0.0,0.0) 
		circle [radius=0.15] 
		node[white] {{\fontfamily{qag}\selectfont \tiny \,ID}};
		\draw[white, fill=white] (-0.0525,0.095) 
		circle [radius=0.007];
	\end{tikzpicture}
	\hspace{-3.0mm}
}
\foreach \x in {A, ..., Z}{\expandafter\xdef\csname orcid\x\endcsname{\noexpand\href{https://orcid.org/\csname orcidauthor\x\endcsname}
		{\noexpand\orcidicon}}
}

\newcommand{\R}{\mathcal{R}}
\newcommand{\I}{\mathcal{I}}
\newcommand{\kd}{k_{d}}

\providecommand{\K}{\mathcal{K}}
\providecommand{\A}{\mathcal{A}}

\begin{document}
%=============START of MAIN PAPER===============
\title{From Traveling Waves to Synthetic Gauge Fields: A Field-Free Route to Superconducting Diode Effect}
\author{Latu Kalita
$^\S$
}
\email{latu26@iisertvm.ac.in}
\affiliation{School of Physics, Indian Institute of Science Education and Research Thiruvananthapuram, Kerala 695551, India}
\author{Koustav Roy$^\S$}
\email{koustav.roy@iitg.ac.in}
\affiliation{Department of Physics, Indian Institute of Technology Guwahati-Guwahati, 781039 Assam, India}
\author{Bilal Tanatar}
\email[Corresponding author: ]{tanatar@fen.bilkent.edu.tr}
\affiliation{Department of Physics, Bilkent University, 06800 Bilkent, Ankara, Turkey}
\author{Saurabh Basu}
\affiliation{Department of Physics, Indian Institute of Technology Guwahati-Guwahati, 781039 Assam, India}

\begin{abstract}
The superconducting diode effect requires the forward and reverse bias currents to differ, a scenario achieved via inversion and time-reversal symmetries broken together. The latter, and the more crucial one, is inevitably achieved by a magnetic field. Here, we show that a spatio-temporally driven chemical potential of the form, $\mu_j(t)=\mu+V\cos(\omega t+\kd j)$, achieves both at once. Applied to a helical Shiba chain, the drive transmutes a
traveling Peierls phase via generating an odd-in-momentum term, equivalent to an imaginary next-nearest-neighbour Haldane-like hopping in real space. Remarkably, the protocol is \textit{spin-blind}, which renders
the time-reversal breaking via a synthetic \textit{gauge} flux rather than an effective Zeeman field.
The rectification is odd in the drive wavevector, with exact zeros at $\kd=n\pi$ and an analytical optimum at $\kd=2\pi/3$. It is however, even in the exchange coupling, $J$ and persists even at $J=0$. Furthermore, the rectification scales inversely with the pair-breaking gap, yielding approximately $50\%$ rectification in the topological phase and $10\%$ in the trivial phase. 
Symmetry breaking is thus drive-controlled, allowing it to be switched on, tuned, and reversed without an external magnetic field.
% The symmetry breaking is thus controlled by the driving protocol rather than the material itself, allowing it to be switched on, tuned, and reversed through the drive, without requiring an external magnetic field.
\end{abstract}

\maketitle
\def\thefootnote{$^\S$}\footnotetext{These authors contributed equally to this work}
%----------------------------------------------------------------------
\textit{Introduction.-} A superconducting diode transmits a net dissipationless current more easily in one direction than in the other, such that the two critical currents differ, $I_c^{+}\neq|I_c^{-}|$ \cite{Ando2020,Nadeem2023}. Since its observation in a Rashba superlattice \cite{Ando2020}, the superconducting diode effect (SDE) has been reported in Josephson junctions \cite{Baumgartner2022,Pal2022,Wu2022,Trahms2023}, van der Waals and twisted materials \cite{Bauriedl2022,Lin2022,Wu2022}, and tunnel and multilayer devices \cite{Strambini2022,Narita2022,Jeon2022}. Generally, an asymmetry between the positive and negative critical currents is permitted only when inversion ($\mathcal{P}$) and time-reversal ($\mathcal{T}$) symmetries are broken \emph{together} \cite{Rikken2001,Wakatsuki2017,Tokura2018}, a condition commonly realized through magnetochiral anisotropy that generates a finite Cooper-pair momentum \cite{YuanFu2022,Daido2022,DaidoYanase2022,Ilic2022,He2022,Davydova2022,Zhang2022,Misaki2021}. 
\par That innocuous ``together'' is the bottleneck: in practice the two symmetries are broken by two
separate agents, and the second ($\mathcal{T}$) almost always has a magnetic origin. Noncentrosymmetric crystals break
$\mathcal{P}$ by spin-orbit coupling and $\mathcal{T}$ by external fields
\cite{Smidman2017,Ilic2022,Legg2022}. Similar scenarios prevail in the topological platforms where study of
diode effect is most interesting. Recent studies have demonstrated a pronounced SDE in topological insulators and Rashba nanowires, with its magnitude closely correlated with proximity to the topological phase transition \cite{Legg2022}. Its relation to normal-state nonreciprocity was traced across the trivial and topological phases \cite{Angehrn2024}, and the role of topology was examined in quasi-one-dimensional wires and topological-insulator edges \cite{dePicoli2023,Huang2024,Lu2023,Cayao2024,LiuHuangWang2024}. Chains of magnetic adatoms on superconductors (Shiba chains) \cite{Yu1965,Shiba1968,Rusinov1969,Choy2011,NadjPerge2013,Klinovaja2013,Braunecker2013,Vazifeh2013,Pientka2013,NadjPerge2014,Ruby2017,Kim2018,Schneider2021} have taken this one step further. In a helical Shiba chain, a Zeeman field stabilizes the Fulde-Ferrell-Larkin-Ovchinnikov (FFLO) pairing \cite{Fulde1964,Larkin1965} that simultaneously hosts Majorana zero modes (MZMs) and a finite SDE \cite{Bhowmik2025}. Related proposals optimize the efficiency through SOC \cite{Bhowmik2025b}, replace the field by either altermagnetic or unconventional magnetic order \cite{Samanta2025,Pal2025}, extend the effect to multichannel wires \cite{Santra2026}, or combine a periodic drive with in-plane field \cite{Bhowmik2026}.
\par \par Several questions nevertheless remain open. First, every one of these mechanisms relies on a \textit{static} source of $\mathcal{T}$ breaking, either an external field or built-in magnetic order. A field competes with superconductivity, and in both cases, the diode polarity is locked to the orientation of the moments, so that reversed magnetic domains rectify in opposite directions. Second, $\mathcal{P}$ breaking rests on SOC or on a particular spin texture, which are material properties rather than in-situ knobs. Third, the connection between topology and rectification remains ambiguous, and no unifying principle has been identified. In this context, periodic driving offers a promising route to overcome these limitations \cite{Oka2009,Lindner2011,Kitagawa2010,Rudner2020,Eckardt2017,Bukov2015,Goldman2014,Roy2023creutz,Roy2025dissipative,Roy2025NHcreutz,Roy2025ckt,Roy2026hybrid} and has already been employed to generate Majorana modes \cite{Jiang2011,Thakurathi2013,Roy2024DKC,Roy2024rashba}. Yet, its application to SDE has so far been restricted to Josephson junctions \cite{RoyJJ2025,Soori2023,OrtegaTaberner2023} or systems assisted by static fields.
\par Our starting point originates from a recent observation \cite{RoyJJ2025}: two Kitaev leads driven with a relative phase $\zeta$ generate an artificial Zeeman field and an SOC-like term at the junction, replacing both static ingredients and yielding rectification up to $\sim70\%$. A phase mismatch \textit{between two} drives is, however, nothing but a discretized phase gradient. Associating such a phase to \textit{every} bond turns a pair of standing (simple harmonic) drives into a single \textit{traveling} wave. This suggests a tantalizing possibility: can a traveling wave replace effects of both the magnetic field and the SOC in a topological superconductor, and thereby produce a genuinely field-free bulk diode?
\par In this Letter, we explore this question with an affirmative outcome. Driving the chemical potential of a helical Shiba chain with a traveling wave produces a robust SDE for any wavevector associated with the drive ($k_d$) except for $\kd =0,\pi$, with rectification up to $\sim50\%$. In a rotating frame, the wave is a traveling Peierls phase (a synthetic gauge rather than a real Zeeman field) which makes the mechanism \textit{blind} to spin degrees of freedom and operative even without magnetic adatoms. We identify the spectral features in the \textit{tilted Floquet replica lattice} that imprints an odd, Haldane-like harmonic. We derive exact rectification rules and demonstrate that the drive frequency can dynamically shift the rectification window.
\begin{figure}[t]
\includegraphics[width=\columnwidth]{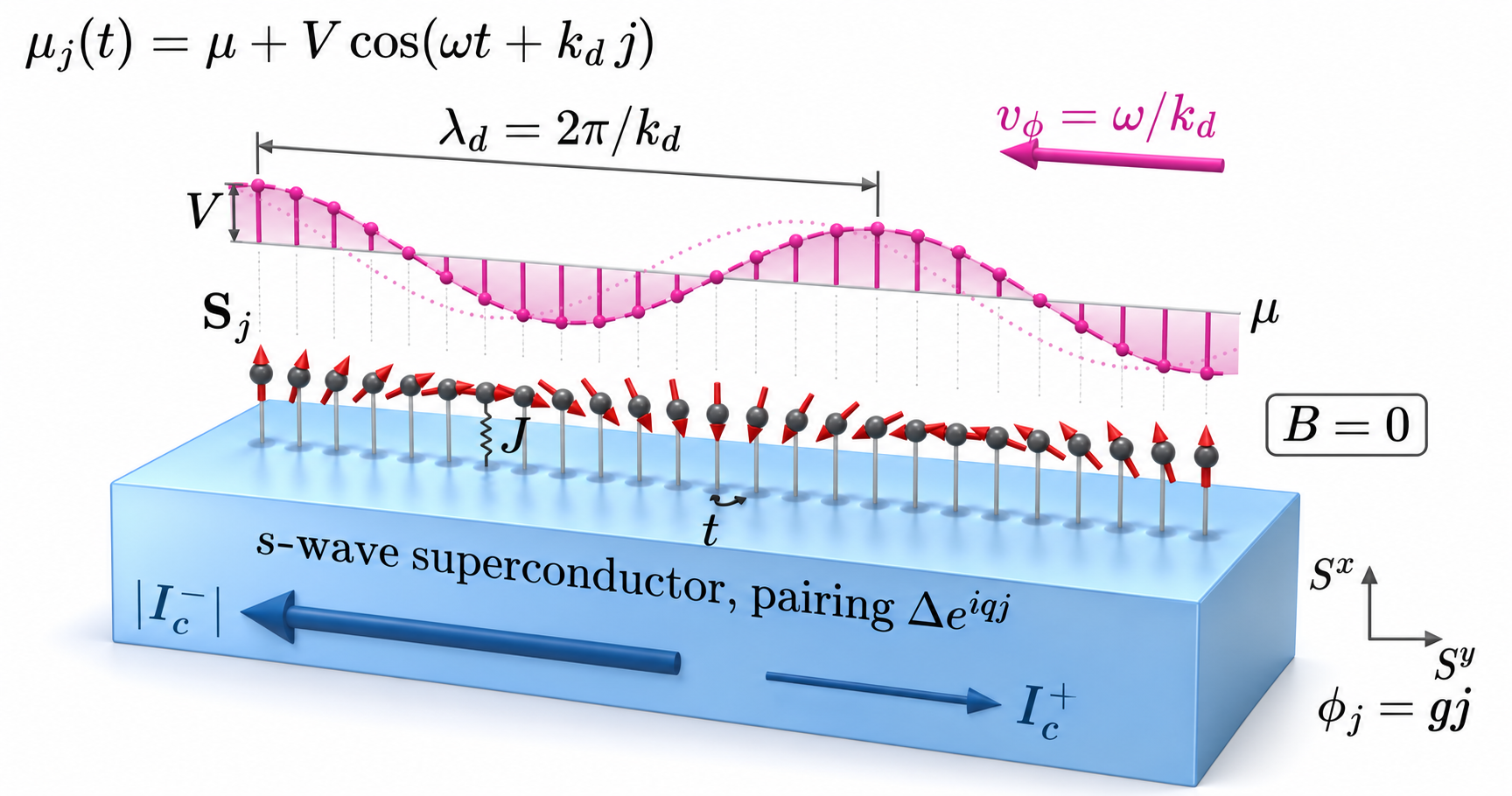}
\caption{Schematic of the driven helical Shiba chain. Magnetic adatoms with spins $\mathbf{S}_j$ rotating by $\phi_j=gj$ are exchange-coupled ($J$) to electrons hopping with amplitude $t$ on top of an $s$-wave superconductor that induces pairing $\Delta e^{iqj}$ with Cooper-pair momentum $q$. The chemical potential is modulated by a traveling wave of wavelength $2\pi/\kd$, which moves with phase velocity $v_\phi=\omega/\kd$. The wave renders the two critical currents unequal, $I_c^{+}\neq|I_c^{-}|$, without any magnetic field.}
\label{fig:1}
\end{figure}
\par\textit{Model and the traveling-wave drive.-} We consider the Hamiltonian of a chain of classical magnetic moments deposited on a conventional superconductor as sketched in Fig.~\ref{fig:1} \cite{Choy2011,NadjPerge2013,Pientka2013,Klinovaja2013,Bhowmik2025},
\begin{align}
H_0=&-t\sum_{j,\sigma}\Big(c^{\dagger}_{j+1\sigma}c_{j\sigma}+{\rm H.c.}\Big)
-\mu\sum_{j,\sigma}c^{\dagger}_{j\sigma}c_{j\sigma}\non\\
&+J\sum_{j,\sigma\sigma'}c^{\dagger}_{j\sigma}\big(\mathbf{S}_j\!\cdot\!\bm{\sigma}\big)_{\sigma\sigma'}c_{j\sigma'}
+\sum_{j}\Big(\Delta\,e^{iqj}\,c^{\dagger}_{j\ua}c^{\dagger}_{j\da}+{\rm H.c.}\Big),
\label{eq:H0}
\end{align}
where $\sigma,\sigma'=\ua,\da$ are spin indices, $t$ is the hopping, $\mu$ the chemical potential, and $\bm{\sigma}$ the Pauli matrices. The classical adatom spins form a planar spiral, $\mathbf{S}_j=(\cos\phi_j,\sin\phi_j,0)$ with $\phi_j=gj$ ($g$ denotes the spiral pitch), stabilized by RKKY interactions \cite{Klinovaja2013,Braunecker2013,Vazifeh2013,Schecter2016}, that couple to the electron spin through the exchange $J$. The last term is the $s$-wave pairing induced by the bulk superconductor that
carries a finite Cooper-pair momentum $q$ through the Fulde-Ferrell phase factor $e^{iqj}$
\cite{Fulde1964,Larkin1965}. Incidentally, the supercurrent flowing in the chain is modulated by the same $q$. Following Refs.~\cite{Legg2022,Angehrn2024}, $\Delta$ is treated as fixed, which is appropriate for a proximitized chain backed by a rigid parent superconductor. \par The spiral is what makes the chain topological. A local spin rotation $e^{-i\phi_j\sigma_z/2}$ unwinds the texture into a uniform exchange field $J\sigma_x$ and transfers the winding onto the hopping, via $t\to te^{-ig\sigma_z/2}$. At $q=0$ the Bloch Hamiltonian becomes $h(k)=-2t_g\cos k-\mu-\alpha_g\sin k\,\sigma_z+J\sigma_x$, with $t_g=t\cos(g/2)$ and $\alpha_g=2t\sin(g/2)$ \cite{Braunecker2010,Kjaergaard2012}. The spiral pitch $g$ thus acts as an SOC and $J$ as a Zeeman field perpendicular to it, replicating the Rashba-wire recipe \cite{Lutchyn2010,Oreg2010,Alicea2012} without any intrinsic SOC. The topological phase boundaries are determined by bulk-gap closings at $k=0,\pi$. Consistent with the standard prediction for Rashba nanowires, the chain hosts MZMs when $(|\mu|-2t_g)^2+\Delta^2<J^2<(|\mu|+2t_g)^2+\Delta^2$ \cite{Kitaev2001,Lutchyn2010,Oreg2010,Pientka2013}. Crucially, $H_0$ is invariant under an inversion combined with a $\pi$ spin rotation about the in-plane axis, which maps $q\to-q$. Consequently, the ground-state energy satisfies $E(q)=E(-q)$, precluding a static diode effect in the absence of an external field \cite{Bhowmik2025}.
\begin{figure*}[t]
\centering\includegraphics[width=6.6526in]{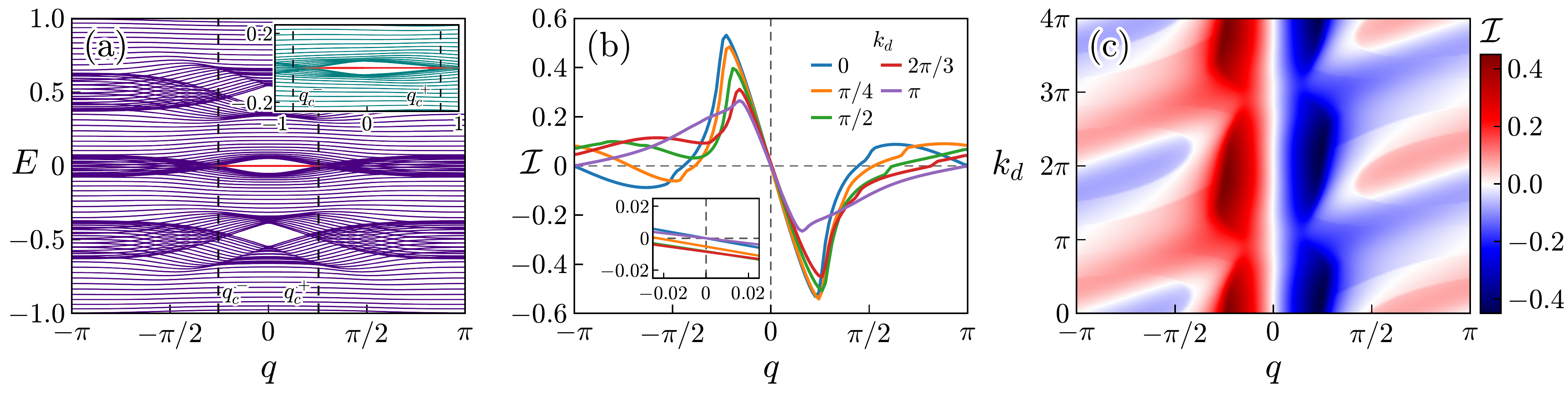}
\caption{(a) Floquet quasienergy spectrum versus Cooper-pair momentum $q$ at $\kd=0$, showing a
zero-energy branch confined to the window $[q_c^-,q_c^+]$, which is symmetric about $q=0$. (b)
Supercurrent $\I(q)$ for several $\kd$. The response is near-linear inside the gapped window and the
amplitude falls monotonically with $\kd$. Inset: magnification about
the origin showing $\I(0)\neq0$, \ie\ a finite equilibrium pair momentum $q_0$, for every
$\kd\neq n\pi$. (c) Current in the $\kd$--$q$ plane, displaying $2\pi$ periodicity in $\kd$ and the
antisymmetry $\I(q,\kd)=-\I(-q,-\kd)$. Throughout, $t=0.5$, $\Delta=0.7$, $g=\pi/2$, $V=1.5$, and
here $J=2$, $\mu=2.5$.}
\label{fig:2}
\end{figure*}
\par To break this symmetry without any field we take our cue guided by the two-tone construction of Ref.~\cite{RoyJJ2025}, where a phase mismatch between two drives generated an artificial gauge structure. A bulk analogue requires a phase difference across \textit{every} bond, i.e., a wave whose phase advances by $\kd$ from one site to another:
\begin{equation}
\mu\;\to\;\mu_j(t)=\mu+V\cos(\omega t+\kd j),
\label{eq:drive}
\end{equation}
so that $H(t)=H_0-\sum_jV\cos(\omega t+\kd j)\,c_j^{\dagger}c_j$. For $\kd=0$ all the sites oscillate in unison, and for $\kd=\pi$ neighbors oscillate out of phase, forming a staggered \textit{standing} pattern. For any other values of $\kd$, the crests \textit{travel} with phase velocity $\omega/\kd$, acting as a conveyor belt that sweeps along the chain. Such a running modulation is routinely produced by arrays of gates driven with phase-shifted ac voltages, as in conveyor-mode shuttling \cite{Seidler2022,Langrock2023}, by surface acoustic waves \cite{Hermelin2011,McNeil2011,Satzinger2018}, or by moving optical lattices \cite{Lohse2016,Nakajima2016}. These techniques are compatible with transport studies on engineered Shiba and quantum-dot Kitaev chains \cite{Kim2018,Schneider2021,Dvir2023,Samuelson2024}.
\par Since $H(t+T)=H(t)$, $T=2\pi/\omega$, the stroboscopic dynamics follow from the Floquet Hamiltonian
\cite{Shirley1965,Sambe1973},
\beq
H_{\rm eff}=\frac{i}{T}\ln\Big[\mathcal{T}\exp\Big(-i\int_{0}^{T}\!H(t)\,dt\Big)\Big],
\label{eq:Heff}
\eeq
whose eigenvalues are the quasienergies $\epsilon$, defined modulo $\omega$. The drive enters
this construction in a very particular way. Expanding Eq.~\eqref{eq:drive} in harmonics yields
$H_{\pm1}=\tfrac{V}{2}\sum_k c^{\dagger}_{k\pm \kd}c_k$. Thus absorption of one quantum $\omega$ is
\textit{inseparably} linked with a momentum kick $\kd$. The $m$-th Floquet sideband is therefore not
merely displaced in energy by $m\omega$; it is displaced in momentum by $m\kd$ as well. The
Floquet replica lattice in the extended $(k,\epsilon)$ zone is \textit{tilted} rather than rectangular, and a
tilted lattice possesses no $k\to-k$ mirror. This is the band-structure origin of everything that
follows. The same physics is transparent in a rotating frame, $\mathcal{U}(t)=\exp[i\sum_j(V/\omega)\sin(\omega t+\kd j)\,c_j^{\dagger}c_j]$, which removes the scalar potential, leaving the exchange term untouched, but dresses each bond with a \textit{traveling Peierls phase} (Sec.~S1 of the Supplemental Material (SM) \cite{SM}):
\begin{align}
t\;\to\;t\,e^{i\theta_j(t)}=t\sum_{n}i^{n}J_n(z_d)\,e^{in[\omega t+\kd(j+1/2)]},\label{eq:peierls}
\end{align}
with $\theta_j(t)=z_d\cos[\omega t+\kd(j+\tfrac12)]$ and $z_d=(2V/\omega)\sin(\kd/2)$. Equation~\eqref{eq:peierls} describes a traveling vector potential, i.e., a synthetic gauge field \cite{Peierls1933,Dalibard2011,Struck2012,Aidelsburger2013,Miyake2013}. Its static component renormalizes the hopping by the dynamical-localization factor $J_0(z_d)$ \cite{Dunlap1986}, while each harmonic $n$ transfers $(n\omega,n\kd)$ simultaneously. At $\kd=0$, $z_d=0$, which renders the drive to be inert. At $\kd=\pi$,
$z_d$ is \textit{maximal}, yet $\theta_j=-z_d (-1)^j\sin\omega t$ is a standing wave
obeying $\theta_j(-t)=-\theta_j(t)$, so $\mathcal{T}$ survives and the rectification is forbidden. For
every other $\kd$, the bond phase is a genuine traveling wave: inversion maps it to its
counter-propagating one, time reversal does the same, and only the product $\mathcal{PT}$ is
preserved. Both symmetries that forbid rectification are broken via one single drive, with no magnetic field.
\par \textit{Floquet supercurrent and nonreciprocity.-}
At equilibrium, the supercurrent follows from the free energy via, \cite{Legg2022,YuanFu2022}
\begin{equation}
\I(q)=\frac{2e}{\hbar}\,\partial_q\Omega(q),\quad \Omega(q)=-\frac{1}{\beta}\sum_{\alpha}\ln \Big[ \cosh\frac{\beta E_\alpha(q)}{2}\Big],
\label{eq:current}
\end{equation}
where $E_\alpha$ denotes the Bogoliubov-de Gennes (BdG) energies. We focus at zero temperature, where $\Omega$ reduces to the ground-state energy $E(q)=\sum_{E_\alpha<0}E_\alpha(q)$ and $\I=(2e/\hbar)\,\partial_qE$. Remarkably, the same expression remains valid for the driven chain without modification. With all negative-quasienergy Floquet states being occupied, the time-averaged current obeys a Floquet Hellmann-Feynman relation \cite{Kumari2024,RoyJJ2025} (see Appendix D of Ref.~\cite{RoyJJ2025}),
\begin{equation}
\I(q)=\frac{2e}{\hbar}\sum_{\epsilon_\alpha<0}\partial_q\,\epsilon_\alpha(q),
\label{eq:floquetcurrent}
\end{equation}
which we evaluate with $2e/\hbar=1$, the unit of which the current is measured. A diode requires $E(q)\neq E(-q)$. If $E$ is even in $q$, then $\I$ is odd, $\I(q)=-\I(-q)$, and the extrema are equal, $I_c^{+}=|I_c^{-}|$. 
\par In the numerical calculations presented below, we have kept $t=0.5$, $\Delta=0.7$, $g=\pi/2$, $V=1.5$, and $\omega=3$.  Fig.~\ref{fig:2}(a) illustrates the symmetric case at $\kd=0$. MZMs are pinned at zero quasienergy within a window $q_c^{-}<q<q_c^{+}$ with $q_c^{+}=|q_c^{-}|$, beyond which the Doppler shift of the Bogoliubov states exceeds the gap and the chain turns gapless. Once a traveling wave is switched on ($\kd\neq0,\pi$), the zero-mode window becomes lopsided, $q_c^{+}\neq|q_c^{-}|$, and accordingly $E(q)\neq E(-q)$ (see inset of Fig. \ref{fig:2}(a)). The resulting current-phase relation (CPR) is shown in Fig.~\ref{fig:2}(b). Inside the window defined by dashed vertical lines (Fig. \ref{fig:2}(a)) energy grows quadratically about the minimum, $E\simeq E_0+\tfrac{\kappa}{2}(q-q_0)^2$ with superfluid stiffness $\kappa$, and $\I$ is therefore linear in $q$. The critical currents are reached at the window edges, where quasiparticles begin to depair. Beyond these the non-self-consistent current is unphysical, since a self-consistent gap would collapse there \cite{Bhowmik2025}. So we extract $I_c^{\pm}$ strictly within $[q_c^{-},q_c^{+}]$, as in Ref.~\cite{Legg2022}.  \par Two features stand out here. First, the overall current amplitude decreases with $\kd$. The dynamical localization factor $J_0(z_d)$ of Eq.~\eqref{eq:peierls} being at work: it maximally softens the stiffness at $\kd=\pi$, where $z_d$ is largest. The suppression, however, is \textit{asymmetric}: $I_c^{+}$ is depleted faster than $|I_c^{-}|$. Second, the inset reveals $\I(0)\neq0$ for $\kd\neq0,\pi$. The equilibrium Cooper-pair momentum shifts to $q_0\neq0$, a bulk analogue of the anomalous $\varphi_0$ current of driven junctions \cite{RoyJJ2025,OrtegaTaberner2023}, generated here without any field. The global structure is displayed in the CPR map of Fig.~\ref{fig:2}(c). The current is $2\pi$ periodic in $\kd$, as dictated by the lattice, and its extrema always lie within $[q_c^{-},q_c^{+}]$. Extending the map to negative $\kd$ reveals the relation $\I(q,\kd)=-\I(-q,-\kd)$, inherited from the symmetry that maps $(q,\kd)\to(-q,-\kd)$ (Sec.~S4 \cite{SM}). Therefore, at fixed $\kd\neq0,\pi$, $\I(q)\neq-\I(-q)$.
\begin{figure}[t]
\includegraphics[width=\columnwidth]{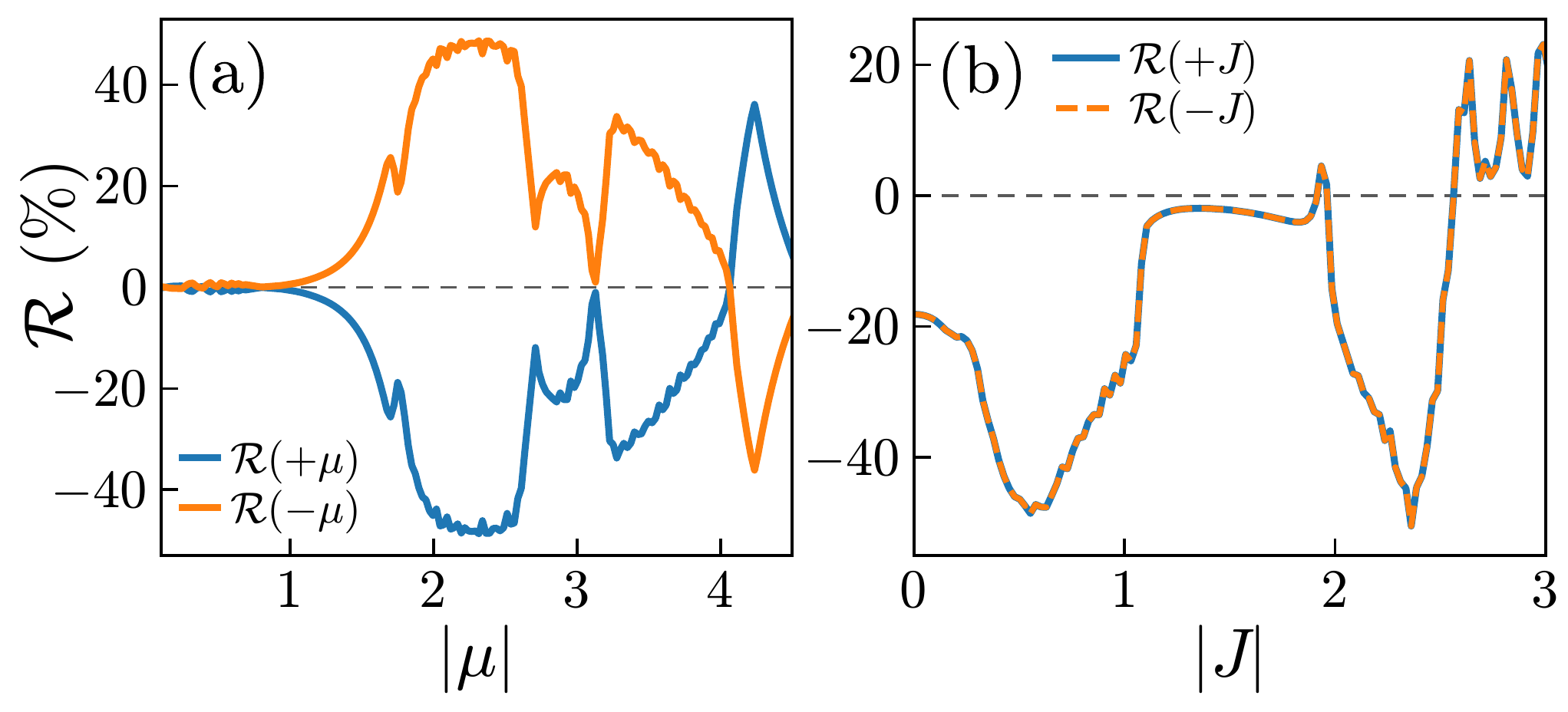}
\caption{Rectification versus (a) chemical potential at $J=0.5$ and (b) exchange coupling at
$\mu=2.5$, for $\kd=2\pi/3$. The two panels obey opposite parities: $\R(\mu)=-\R(-\mu)$, so the diode
reverses with band filling and vanishes at half filling, whereas $\R(J)=\R(-J)$, so reversing every
adatom moment leaves it unchanged. Note that $\R$ remains finite as $J\to0$.}
\label{fig:3}
\end{figure}
\par We quantify the diode by the rectification factor
\begin{equation}
\R=\frac{I_c^{+}-|I_c^{-}|}{I_c^{+}+|I_c^{-}|}.
\label{eq:R}
\end{equation}
with $I_c^{\pm}$ evaluated at $q_c^{\pm}$, inside the gapped window where the current is physical.
The antisymmetry of Fig.~\ref{fig:2}(c) immediately gives $
\R(-\kd)=-\R(\kd),
\label{eq:Rodd}$ whose meaning is simply that reversing the direction in which the wave travels reverses the diode property: the
drive's propagation direction denotes the diode's polarity. This also forces
$\R=0$ at $\kd=0$ and $\kd=\pi$, the two values at which the wave stops traveling.
\par At this stage, an important question arises: where does the response attain its maximum between these zeros? A second-order Floquet analysis (Sec.~S2
\cite{SM}) shows that the drive adds to the band an odd-in-momentum term,
\begin{equation}
\delta\epsilon_{\rm odd}(k)=-\mathcal{A}(\kd)\sin2k,\quad
\mathcal{A}(\kd)\simeq\frac{2t^2V^2}{\omega^3}\sin\kd\,(1-\cos\kd).
\label{eq:odd}
\end{equation}
Since $\R\propto\mathcal{A}$, the rectification is extremal where $\partial_{\kd}[\sin\kd(1-\cos\kd)]=0$, i.e., at $\cos\kd=-1/2$, or with $\kd=2\pi/3,4\pi/3,\dots$. Notably, this is a general, parameter-free prediction of the mechanism. 
An odd-in-$k$ band term is
exactly the magnetochiral ingredient identified in Ref.~\cite{Legg2022} as the prerequisite for
nonreciprocity, with the important difference that $\sin 2k$ is a \textit{second} harmonic and
therefore cannot be absorbed by any shift of $q$, unlike a term linear in $k$, which is a mere
Galilean boost.
% The specific form $\sin2k$ is also what makes the diode possible. A term linear in $k$ is a Galilean boost that is absorbed into $q$ and only shifts $q_0$, whereas a diode requires odd harmonics beyond linear order \cite{Legg2022,Angehrn2024,YuanFu2022}. The $\sin2k\simeq2k-\tfrac43k^3$ harmonic supplies exactly such a term, which no shift of $q$ can remove.
In real space, Eq.~\eqref{eq:odd} is a \textit{purely imaginary} next-nearest-neighbor hopping,
\begin{equation}
\delta H=\sum_{j}\Big(\frac{i\mathcal{A}}{2}\,c_j^{\dagger}c_{j+2}+{\rm H.c.}\Big),
\label{eq:haldane}
\end{equation}
a Haldane-like term in one-dimension \cite{Haldane1988,Jotzu2014} carrying a $\pi/2$ Peierls phase. It is the static fingerprint of the traveling gauge field~\eqref{eq:peierls}. Complex conjugation flips its sign ($\mathcal{T}$ broken), and so does reversing the bond direction ($\mathcal{P}$ broken).
\begin{figure}[t]
\includegraphics[width=\columnwidth]{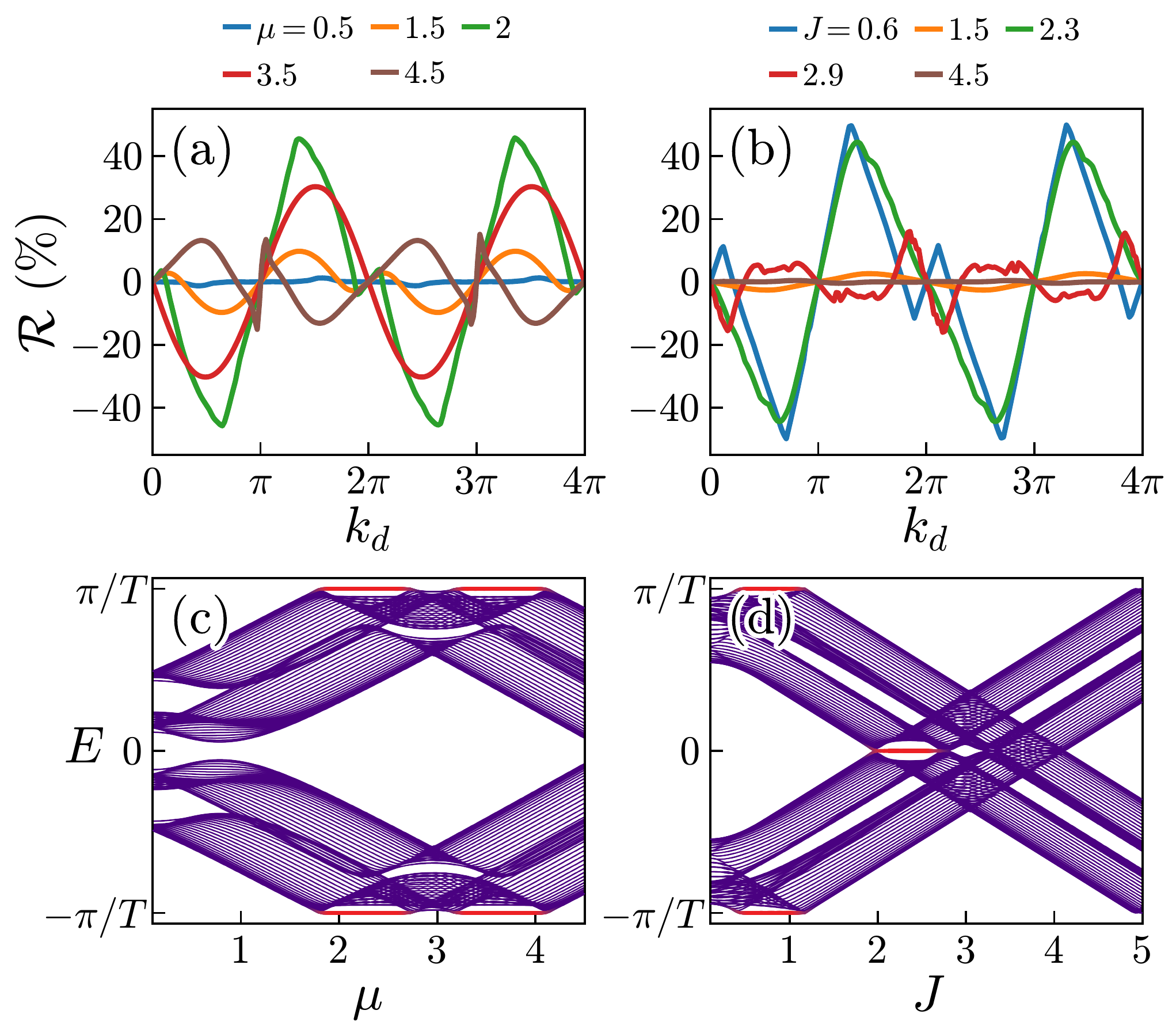}
\caption{(a),(b) Rectification versus drive wavevector for several $\mu$ at $J=0.5$ and several $J$ at
$\mu=2.5$. The response is $2\pi$-periodic with exact zeros at $\kd=n\pi$, and is extremal near
$\kd=2\pi/3$ and $4\pi/3$. Some
parameter choices reach $40$--$50\%$ while others stay below $10\%$. (c),(d) Floquet quasienergy
spectra over the full zone versus $\mu$ at $J=0.5$ and versus $J$ at $\mu=2.5$; red marks localized
branches. The large-$\R$ curves in (a) coincide with the $\pi$-mode windows of (c), and those in (b)
with the $\pi$- and zero-mode windows of (d).}
\label{fig:4}
\end{figure}
\begin{figure*}[t]
\centering\includegraphics[width=6.6526in]{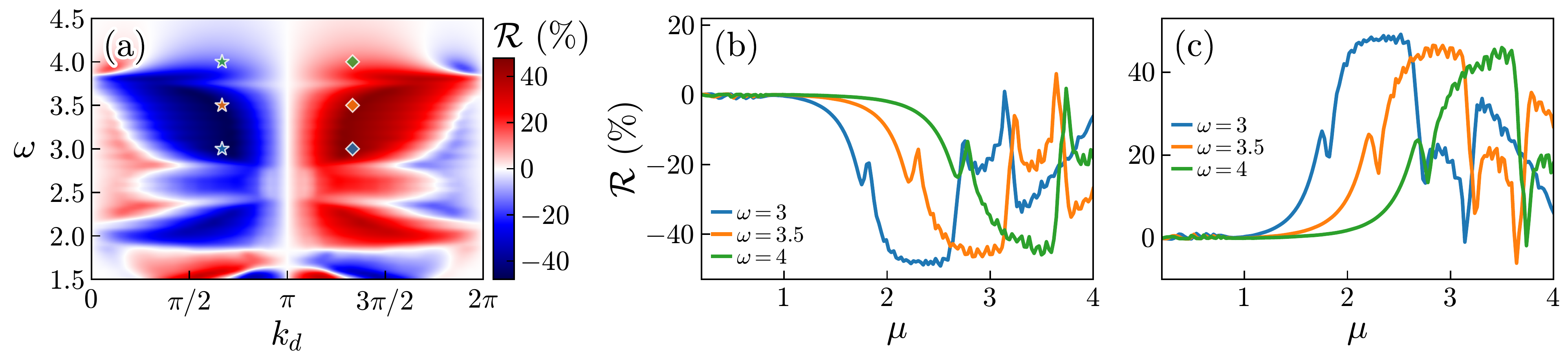}
\caption{((a) Rectification in the $\kd$--$\omega$ plane at $\mu=2.5$, $J=0.5$. Below $\omega\simeq2$
the sign alternates as single-photon resonances sweep the band; above it the sign is fixed by $\kd$
alone, deepening near $\kd=2\pi/3$ and $4\pi/3$, and fading beyond $\omega\simeq4$ towards the static
limit. (b),(c) $\R$ versus $\mu$ at $\kd=2\pi/3$ and $4\pi/3$ for $\omega=3$, $3.5$ and $4$. The
onset in $\mu$ shifts systematically with frequency, tracking the drive-tunable topological boundary.}
\label{fig:5}
\end{figure*}
\par\textit{Rectification landscape.-}
We now map $\R$ across the parameters controlling the topology of the chain. Fig. ~\ref{fig:3} reveals two exact and
opposite parities. The first, $\R(\mu)=-\R(-\mu)$ (\ref{fig:3}(a)), follows from the bipartite lattice structure and
reads cleanly: for $\mu<0$ the band is less than half filled and the carriers are electron-like, for
$\mu>0$ they are hole-like. Rewriting the electrons as holes reverses the
Cooper-pair momentum, so the convenient direction reverses accordingly; at half filling neither
description is preferred and $\R$ vanishes identically. The second, $\R(J)=\R(-J)$ (\ref{fig:3}(b)), is more unusual and more useful. Flipping every adatom moment is undone
by a rigid $\pi$ rotation of the spin frame, which leaves the spin-independent drive unharmed. The magnetization therefore does \textit{not} set the
diode direction; only ${\rm sgn}(\kd)$ and ${\rm sgn}(\mu)$ do. This is the opposite of a
field-induced diode, where $\R(B)=-\R(-B)$, and it is an experimental advantage: a chain with
reversed domains, or an unknown magnetization sign, yields the same diode effect. \par Most strikingly, $\R$ remains finite as $J\to0$ [Fig.~\ref{fig:3}(b)], where the adatoms, and with them all magnetism, are absent. The reason is that the drive is \textit{spin-blind}: it couples to the density $n_j$, not to $\bm{\sigma}$. Time-reversal breaking therefore occurs not in the spin channel but in the orbital (charge) channel, through the directionality of the traveling wave.
\par Having established the role of the system parameters, we now turn to the drive-induced parameters. Figures~\ref{fig:4}(a) and \ref{fig:4}(b)
show $\R(\kd)$ for several $\mu$ and $J$. The response is $2\pi$-periodic, vanishes exactly at
$\kd=n\pi$, and is generically extremal near $\kd=2\pi/3$ and $4\pi/3$, as Eq.~\eqref{eq:odd} predicts;
its profile ranges from smoothly sinusoidal to a sawtooth-like triangular form, reflecting whether the
critical currents are set by a smooth interior extremum or by a window edge moving linearly with
$\kd$. Most striking is the dichotomy of the scale: certain curves reach $40$--$50\%$ while the remainder
never exceed $10\%$. The spectra in Figs.~\ref{fig:4}(c) and \ref{fig:4}(d) reveal that this is no coincidence. At $J=0.5$ the $E=0$ gap is empty throughout and the chain hosts
\textit{only} Floquet $\pi$ modes, in windows $\mu\simeq1.8$--$2.7$ and $3.4$--$4.1$; precisely the
$\mu$ values at which $\R$ is large in Fig.~\ref{fig:4}(a). Similarly, at $\mu=2.5$ the chain hosts $\pi$ modes
for $J\simeq0.3$--$1.15$ and zero modes for $J\simeq2.0$--$2.75$, and both windows show up as
large-$\R$ curves in Fig.~\ref{fig:4}(b), while the trivial interval between them does not. We have confirmed the same correspondence over many other spectra: \textit{the topological phases host the largest rectification}. 
\par The origin is geometric (Sec.~S3 \cite{SM}). Because $\I\simeq-\kappa(q-q_0)$ across the window, the rectification measures how far the equilibrium point is displaced from the center of the window relative to its half-width, $\R=(q_0-q_{\rm mid})/q_c$, with $q_{\rm mid}=(q_c^{+}+q_c^{-})/2$ and $q_c=(q_c^{+}-q_c^{-})/2$. The odd harmonic~\eqref{eq:odd} displaces the window by an energy $\epsilon_{\rm asym}$ that knows nothing about the pairing gap, whereas the half-width is set by the pair-breaking gap $\Delta_{\rm eff}$ defined as the smallest excitation gap at the center of the window that the condensate must overcome. Hence
\begin{equation}
\R\simeq\frac{\epsilon_{\rm asym}}{\Delta_{\rm eff}},\qquad
\frac{\R_{\rm topo}}{\R_{\rm triv}}\simeq\frac{\Delta}{\Delta_m}>1.
\label{eq:invgap}
\end{equation}
In the trivial phase the smallest excitation corresponds to bulk quasiparticles, so $\Delta_{\rm eff}\simeq\Delta$. In a topological phase it is the minigap $\Delta_m$ that separates the Majorana branch from the continuum. This minigap is structurally small, because the phase is bounded on both sides by gap closings. A narrow window displaced by a fixed amount is largely asymmetric, while a wide one barely notices. This inverse-gap law also reconciles the apparently conflicting earlier findings. In magnetochiral nanowires the surviving asymmetry is linear in $k$, which is a Galilean boost, and the diode is suppressed \cite{Angehrn2024}. Here the drive-generated $\sin2k$ harmonic survives, while the denominator of Eq.~\eqref{eq:invgap} shrinks.
\par \par The second drive parameter, the frequency, provides an additional and independent knob [Fig.~\ref{fig:5}(a)]. At low frequencies, ($\omega\lesssim2$) $\R$ alternates in sign as a function of both $\omega$ and $\kd$, reflecting a dense sequence of multiphoton resonances sweeping the band, where the perturbative treatment behind
Eq.~\eqref{eq:odd} breaks down. Above it, the sign is fixed by $\kd$ alone; the response deepens near
$\kd=2\pi/3$ and $4\pi/3$ as predicted, and beyond $\omega\simeq4$ it fades as
$\mathcal{A}\propto\omega^{-3}$ towards the static limit. Cuts at $\omega=3$, $3.5$, and $4$ [Fig.~\ref{fig:5}(b)] reveal a remarkably rigid behavior. $\R$ vanishes at half filling, stays negligible up to a threshold value of $\mu$, and then develops a plateau of $|\R|\simeq46$--$49\%$ followed by a structured tail. The entire pattern shifts to larger $\mu$ as $\omega$ increases, the hallmark of a single-photon resonance. The drive frequency therefore \textit{relocates the rectifying window in situ}. Switching to $\kd=4\pi/3$ [Fig.~\ref{fig:5}(c)] mirrors the response onto positive values, $\R(4\pi/3)=-\R(2\pi/3)$, so that the direction of travel of the wave alone selects the diode polarity. The operational consequences are immediate: one device switches between forward, off and reverse
rectification, at $\kd=2\pi/3$, $\pi$ and $4\pi/3$, without the other alteration of parameters, and since
Eq.~\eqref{eq:invgap} ties $\R$ to the inverse minigap, the rectification doubles as a transport
read-out of the topological transition.
\par \textit{Conclusion.-}
We have shown that a single traveling wave in the chemical potential of a helical Shiba chain replaces both the magnetic field and the spin-orbit coupling usually required for a superconducting diode. The wave acts as a spin-blind synthetic gauge field. It tilts the Floquet replica lattice and imprints a Haldane-like $\sin2k$ harmonic, producing rectification up to $\sim50\%$ with a universal optimum near $\kd=2\pi/3$. \emph{The diode obeys}: $\R(\kd)=-\R(-\kd)$, $\R(\mu)=-\R(-\mu)$, and $\R(J)=\R(-J)$. Its polarity is set by the direction of the wave and the carrier type, and not by the magnetization. It survives without magnetic adatoms, is enhanced in the topological phases through an inverse-gap law, and can be relocated in situ by the drive frequency. 
% Traveling-wave Floquet engineering thus emerges as a versatile, magnet-free route to nonreciprocal topological superconductivity.
\par \textit{Acknowledgements.-} KR acknowledges the research fellowship from the MoE, Government of India. BT acknowledges support from the Scientific and Technological Research Council of Türkiye (TÜBİTAK) under Grant No.~125F435 and from the Turkish Academy of Sciences (TÜBA) under Grant No ~AD-2026. SB acknowledges support from TÜBİTAK-BİDEB.

% ---- references of the letter (compiled from refs.bib) ----
\makeatletter
\begingroup\let\addcontentsline\@gobblethree   % keep the letter's References out of the SM contents list
\makeatother

\endgroup
%

%====================================================================
%  SUPPLEMENTAL MATERIAL (same document for arXiv)
%====================================================================
\clearpage
\onecolumngrid
\setcounter{equation}{0}
\setcounter{figure}{0}
\setcounter{table}{0}
\setcounter{page}{1}
\renewcommand{\thepage}{S\arabic{page}}
\renewcommand{\theequation}{S\arabic{equation}}
\renewcommand{\thefigure}{S\arabic{figure}}
\renewcommand{\thetable}{S\arabic{table}}
\renewcommand{\theHequation}{S\arabic{equation}}
\renewcommand{\theHfigure}{S\arabic{figure}}
\renewcommand{\theHtable}{S\arabic{table}}
\renewcommand{\bibnumfmt}[1]{[S#1]}
\renewcommand{\citenumfont}[1]{S#1}

\textbf{\large Supplemental Material for ``From Traveling Waves to Synthetic Gauge Fields: A Field-Free Route to Superconducting Diode Effect''}

\medskip
\makeatletter
\def\l@section#1#2{%
  \@dottedtocline{1}{1.5em}{2.3em}{#1}{#2}%
}
\makeatother
\tableofcontents

%====================================================================
\section*{S1: The Traveling Drive as a Running Peierls Phase}

In the main text we claimed that the traveling \textit{scalar} potential of Eq.~(3) is equivalent to
a traveling \textit{vector} potential living on the bonds. Here we establish that statement exactly,
with no approximation, and extract the parameter $\K$ that controls the symmetry-breaking effect.
The Hamiltonian of the driven chain is $H(t)=H_{\rm kin}+H_{\rm loc}+H_{\rm pair}+H_{\rm dr}(t)$, with
\begin{align}
H_{\rm kin}&=-t\sum_{j,\sigma}\big(c^{\dagger}_{j+1\sigma}c_{j\sigma}+{\rm H.c.}\big),
\qquad
H_{\rm loc}=-\mu\sum_{j,\sigma}c^{\dagger}_{j\sigma}c_{j\sigma}
+J\sum_{j,\sigma\sigma'}c^{\dagger}_{j\sigma}\big(\mathbf{S}_{j}\!\cdot\!\bm{\sigma}\big)_{\sigma\sigma'}c_{j\sigma'},
\label{S:Hparts1}\\
H_{\rm pair}&=\sum_{j}\big(\Delta\,e^{iqj}\,c^{\dagger}_{j\ua}c^{\dagger}_{j\da}+{\rm H.c.}\big),
\qquad
H_{\rm dr}(t)=V\sum_{j,\sigma}\cos\!\big(\omega t+\kd j\big)\,c^{\dagger}_{j\sigma}c_{j\sigma},
\label{S:Hparts2}
\end{align}
where $\sigma,\sigma'=\ua,\da$ and $q$ is the Cooper-pair momentum of the Fulde--Ferrell state. It is
convenient to abbreviate the local density as $n_{j}=\sum_{\sigma}c^{\dagger}_{j\sigma}c_{j\sigma}$,
so that $H_{\rm dr}(t)=V\sum_{j}\cos(\omega t+\kd j)\,n_{j}$. The drive itself is diagonal in the site basis
and carries no explicit spin structure.
\par To expose the gauge structure, we move to a rotating frame. Let $|\psi(t)\rangle$ solve $i\partial_t|\psi\rangle=H(t)|\psi\rangle$ and write $|\psi\rangle=\mathcal{U}(t)|\tilde\psi\rangle$ for a unitary $\mathcal{U}(t)$. Substituting this transformation into the Schrödinger equation yields,
\begin{equation}
i(\partial_t\mathcal{U})|\tilde\psi\rangle+i\,\mathcal{U}\,\partial_t|\tilde\psi\rangle=H\,\mathcal{U}|\tilde\psi\rangle
\;\;\Longrightarrow\;\;
i\partial_t|\tilde\psi\rangle=\tilde H|\tilde\psi\rangle,
\label{S1:framep}
\end{equation}
where the effective Hamiltonian in the rotating frame is given by,
\begin{equation}
\tilde H=\mathcal{U}^\dagger H\,\mathcal{U}-i\,\mathcal{U}^\dagger\partial_t\mathcal{U}.
\label{S1:frame}
\end{equation}
We choose $\mathcal{U}(t)=e^{-i\Phi(t)}$ with $\Phi(t)=\sum_j\varphi_j(t)\,n_j$, where the real functions $\varphi_j(t)$ are fixed as,
\beq
\varphi_j(t)=\int^{t}\!V\cos\!\big(\omega t'+\kd j\big)\,dt'
= - \frac{V}{\omega}\sin\!\big(\omega t+\kd j\big),
\label{S:Phi}
\eeq
Differentiating $\mathcal{U}(t)$ and using $[n_{j},\mathcal{U}]=0$ gives
$\partial_{t}\mathcal{U}=i\sum_{j}\dot\varphi_jn_{j}\mathcal{U}$, hence
\beq
i \big(\partial_{t}\mathcal{U}\big) \mathcal{U}^{\dagger}=-\sum_{j}\dot{\varphi}_{j}n_{j}
=-V\sum_{j}\cos\!\big(\omega t+\kd j\big)n_{j}=-H_{\rm dr}(t).
\label{S:cancel}
\eeq
Since $H_{\rm dr}$ is built from $n_{j}$ alone it is invariant under conjugation,
$\mathcal{U}H_{\rm dr}\mathcal{U}^{\dagger}=H_{\rm dr}$, so the two contributions in \eqref{S1:frame} cancel identically.
The drive has thus been eliminated from the diagonal \emph{exactly}. It must therefore reappear elsewhere.  The drive dependence is thus transferred to the off-diagonal operators through the transformation identity,
\beq
e^{i\varphi_{j}n_{j}}\,c^{\dagger}_{j\sigma}\,e^{-i\varphi_{j}n_{j}}=e^{i\varphi_{j}}c^{\dagger}_{j\sigma},
\label{S:shift}
\eeq
The canonical
algebra $[n_{j},c^{\dagger}_{j\sigma}]=c^{\dagger}_{j\sigma}$ makes every commutator reproduce
$c^{\dagger}_{j\sigma}$ with an extra factor $i\varphi_{j}$, so the series resums to the exponential.
Three consequences follow.

(i) \emph{On-site terms are untouched.} Every term in $H_{\rm loc}$ has the form
$c^{\dagger}_{j\sigma}c_{j\sigma'}$ with both operators on the \emph{same} site, so it picks up
$e^{i\varphi_{j}}e^{-i\varphi_{j}}=1$. The chemical potential and the exchange coupling to the spin texture
are completely unaffected.

(ii) \emph{The hopping acquires a bond phase.} Here the two operators sit on different sites and the
phases no longer cancel,
\beq
\mathcal{U}\big(-t\,c^{\dagger}_{j+1\sigma}c_{j\sigma}\big)\mathcal{U}^{\dagger}
=-t\,e^{i\theta_{j}(t)}c^{\dagger}_{j+1\sigma}c_{j\sigma},
\qquad
\theta_{j}(t)\equiv\varphi_{j+1}(t)-\varphi_{j}(t).
\label{S:thetadef}
\eeq
This is precisely a Peierls substitution, $\theta_{j}$ is the line integral of a synthetic vector
potential across the bond $(j,j{+}1)$.

(iii) \emph{The pairing acquires a phase, which is harmless.} Both operators in
$c^{\dagger}_{j\ua}c^{\dagger}_{j\da}$ are creation operators on site $j$, so the phases \emph{add}
and the pair amplitude becomes $\Delta e^{iqj}\to\Delta e^{iqj}e^{2i\varphi_{j}(t)}$; the
Fulde--Ferrell factor is untouched and only the extra dynamical phase needs to be examined. Its period average follows from the Jacobi--Anger identity
$<e^{iz\sin \alpha}> = <e^{iz\sin(\omega t+\phi_{0})}>=\mathcal{J}_{0}(z)$, valid for \emph{any} offset
$\phi_{0}$, so with $z=2V/\omega$ and $\phi_{0}=\kd j$,
\beq
\overline{e^{2i\Phi_{j}(t)}}=\mathcal{J}_{0}\!\Big(\frac{2V}{\omega}\Big)
\qquad\text{independently of }j .
\label{S:J0}
\eeq
To leading order the pairing dressing is therefore a \emph{real}, \emph{uniform},
\emph{site-independent} renormalisation $\Delta\to\mathcal{J}_{0}(2V/\omega)\Delta$, even in $V$.
Being real and uniform it is invariant under both inversion and time reversal and cannot generate any
nonreciprocity. All directional physics thus resides in $\theta_{j}$, which we can evaluate via inserting \eqref{S:Phi} to \eqref{S:thetadef}. Let $A=\omega t+\kd j$, so that $\varphi_j=-(V/\omega)\sin A$ and $\varphi_{j+1}=-(V/\omega)\sin(A+\kd)$. Using $\sin X-\sin Y=2\cos\tfrac{X+Y}{2}\sin\tfrac{X-Y}{2}$ with $X=A$ and $Y=A+\kd$,
\begin{equation}
\theta_j(t)\equiv\varphi_j-\varphi_{j+1}=-\frac{V}{\omega}\Big[\sin A-\sin(A+\kd)\Big]
=-\frac{V}{\omega}\Big[2\cos\!\Big(A+\frac{\kd}{2}\Big)\sin\!\Big(-\frac{\kd}{2}\Big)\Big]
=z_d\cos\!\Big[\omega t+\kd\Big(j+\frac12\Big)\Big],
\label{S1:theta}
\end{equation}
with the amplitude
\begin{equation}
z_d=\frac{2V}{\omega}\sin\frac{\kd}{2}.
\label{eq:amplitude}
\end{equation}
In the rotating frame the drive is thus entirely carried by the hopping, $t\to t\,e^{i\theta_j(t)}$, which is Eq.~(4) of the main text. 
% Its time derivative is
% \begin{equation}
% \dot\theta_j=-2V\sin\frac{\kd}{2}\sin\!\Big[\omega t+\kd\Big(j+\tfrac12\Big)\Big]=\big[-V_j(t)\big]-\big[-V_{j+1}(t)\big],
% \end{equation}
% which is the drop in potential energy across the bond, i.e., the electric field (Faraday's law in lattice form). The phase $\theta_j(t)$ is therefore a lattice vector potential, and its propagating profile constitutes a traveling electric field of amplitude $2V\sin(\kd/2)$ moving with phase velocity $\omega/\kd$.

\par Three special cases follow directly from Eq.~\eqref{S1:theta}.
(i) For $\kd=0$, $z_d=0$ and $\theta_j\equiv0$. The uniform drive is a pure gauge term: it produces no field along the chain and merely adds a global time-dependent phase to $\Delta$. It therefore cannot distinguish $+q$ from $-q$.
(ii) For $\kd=\pi$, $\theta_j=(2V/\omega)\cos(\omega t+\pi j+\pi/2)=-(2V/\omega)(-1)^j\sin\omega t$. This is a \textit{standing}, staggered pattern whose crests do not move, so it has no direction.
(iii) For all other $\kd$, the profile $\cos[\omega t+\kd(j+\tfrac12)]$ propagates, and the gauge field has a direction.

\par The harmonic content of the bond phase reveals the tilted replica lattice. Using the Jacobi-Anger identity $e^{iz\cos\alpha}=\sum_{n=-\infty}^{\infty}i^nJ_n(z)e^{in\alpha}$,
\begin{equation}
t\,e^{i\theta_j(t)}=t\sum_n i^nJ_n(z_d)\,e^{in\omega t}\,e^{in\kd(j+1/2)} .
\label{S1:JA}
\end{equation}
Two conclusions follow. First, the time-averaged ($n=0$) hopping is reduced to $tJ_0(z_d)$, a $\kd$-dependent dynamical localization \cite{S-Dunlap1986}. For small $z_d$, $J_0(z_d)\simeq1-z_d^2/4=1-V^2(1-\cos\kd)/(2\omega^2)$, which is weakest at $\kd=0$ and strongest at $\kd=\pi$. This explains the overall decrease of the current amplitude with $\kd$ in Fig.~2(b). Second, the $n$-th harmonic carries the factor $e^{in\omega t}e^{in\kd j}$, so it changes the energy by $n\omega$ \textit{and} the momentum by $n\kd$ at the same time. In the extended $(k,\epsilon)$ zone, the Floquet replicas therefore sit on the tilted lattice $(k+n\kd,\ \epsilon+n\omega)$. Under $k\to-k$ this lattice maps onto $(k-n\kd,\epsilon+n\omega)$, which coincides with the original lattice only if $\kd\equiv-\kd$ (mod $2\pi$), i.e., $\kd=0$ or $\pi$.

\section*{S2: Drive-Induced Odd Harmonic and the Optimum at \texorpdfstring{$k_d=2n\pi/3$}{kd=2pi/3}}
We now compute how the traveling drive deforms the electronic band and show that it generates an odd harmonic $\propto\sin2k$ whose amplitude is maximal at $\kd=2\pi/3$. Since the drive is spin-independent, the relevant seed is a property of the kinetic band. We therefore work with the spin-degenerate tight-binding band $\varepsilon_k=-2t\cos k$, which both spin species share in the laboratory frame. 
\par We first fix the Fourier conventions. With $c_j=N^{-1/2}\sum_ke^{ikj}c_k$, $k=2\pi m/N$, and $\sum_je^{i(k'-k)j}=N\delta_{kk'}$,
\begin{equation}
\sum_jc_j^\dagger c_{j+1}=\frac1N\sum_j\sum_{k,k'}e^{-ikj}e^{ik'(j+1)}c_k^\dagger c_{k'}=\sum_ke^{ik}c_k^\dagger c_k,
\qquad\text{so}\qquad
-t\sum_j\big(c_j^\dagger c_{j+1}+{\rm H.c.}\big)=\sum_k\varepsilon_kc_k^\dagger c_k .
\end{equation}
Now for a time-periodic Hamiltonian satisfying $H(t+T)=H(t)$, we introduce its Fourier decomposition, $H(t)=\sum_m H_m e^{im\omega t}$. Floquet's theorem then maps the time-dependent Schrödinger equation onto a time-independent eigenvalue problem in the extended Sambe space \cite{S-Shirley1965,S-Sambe1973}, whose matrix elements are
\begin{equation}
\big(H_F\big)_{mn}=H_{m-n}+m\omega\,\delta_{mn},
\label{S:HF}
\end{equation}
where $m$ denotes the Floquet sideband index and $m\omega$ represents the associated quasienergy shift. Expressing the cosine modulation in exponential form, the driving Hamiltonian becomes
\begin{equation}
H_{\rm dr}(t)=e^{i\omega t}\underbrace{\frac{V}{2}\sum_{j,\sigma}e^{i\kd j}c^{\dagger}_{j\sigma}c_{j\sigma}}_{H_{+1}}
+e^{-i\omega t}\underbrace{\frac{V}{2}\sum_{j,\sigma}e^{-i\kd j}c^{\dagger}_{j\sigma}c_{j\sigma}}_{H_{-1}},
\label{S2:Hpm}
\end{equation}
indicating that the drive contains only the first harmonics, namely, $m=\pm1$. Applying the Fourier transformation $c_{j\sigma}=N^{-1/2}\sum_k e^{ikj}c_{k\sigma}$ and using the identity
\[
\frac{1}{N}\sum_j e^{i(\kd-k+k')j}=\delta_{k',k-\kd},
\]
we obtain the momentum-space representations of the two nonzero Fourier components:
\begin{equation}
H_{+1}=\frac{V}{2}\sum_{k,\sigma}c^{\dagger}_{k+\kd,\sigma}c_{k\sigma},
\qquad
H_{-1}=\frac{V}{2}\sum_{k,\sigma}c^{\dagger}_{k-\kd,\sigma}c_{k\sigma}.
\label{S:Hpm}
\end{equation}
A single-particle state $|k,m\rangle$ (momentum $k$ in sideband $m$) has unperturbed energy $\varepsilon_k+m\omega$. By Eq.~\eqref{S2:Hpm}, $|k,0\rangle$ is coupled with amplitude $-V/2$ to exactly two states:
\begin{equation}
|k,0\rangle\leftrightarrow|k+\kd,+1\rangle\ \ (\text{energy }\varepsilon_{k+\kd}+\omega),\qquad
|k,0\rangle\leftrightarrow|k-\kd,-1\rangle\ \ (\text{energy }\varepsilon_{k-\kd}-\omega).
\end{equation}
In first quantisation
$H_{\pm1}=\tfrac{V}{2}e^{\pm i\kd\hat{x}}$, so $[H_{+1},H_{-1}]=0$: a pure potential drive commutes
with itself at different times, and the leading van Vleck commutator vanishes identically. The effect must therefore arise from the interplay of the
drive with the \emph{kinetic} term, which is why the result below depends on the band structure. The unperturbed state $|k,0\rangle$ has energy $\epsilon_{k}$ and is connected by $H_{+1}$ to
$|k+\kd,+1\rangle$ at $\epsilon_{k+\kd}+\omega$, and by $H_{-1}$ to $|k-\kd,-1\rangle$ at
$\epsilon_{k-\kd}-\omega$, both with matrix element $V/2$. Therefore, the standard second-order perturbation theory gives,
\begin{equation}
\delta\varepsilon(k)=\frac{V^2}{4}\left[\frac{1}{\varepsilon_k-\varepsilon_{k+\kd}-\omega}+\frac{1}{\varepsilon_k-\varepsilon_{k-\kd}+\omega}\right].
\label{S2:shift}
\end{equation}
This expression already shows why right- and left-movers are shifted differently. For the state at $-k$, the up-shifted replica sits at $-k+\kd$, with energy $\varepsilon_{-k+\kd}=\varepsilon_{k-\kd}$, and the down-shifted one at $-k-\kd$, with energy $\varepsilon_{k+\kd}$. The left-mover thus visits the \textit{same} two levels as the right-mover, but with the labels $\pm\omega$ interchanged:
\begin{equation}
\delta\varepsilon(-k;\kd)=\frac{V^2}{4}\left[\frac{1}{\varepsilon_k-\varepsilon_{k-\kd}-\omega}+\frac{1}{\varepsilon_k-\varepsilon_{k+\kd}+\omega}\right]=\delta\varepsilon(k;-\kd).
\end{equation}
The energy denominators therefore differ, and $\delta\varepsilon(k)\neq\delta\varepsilon(-k)$. There are two exceptions. For $\kd=0$ both replicas sit at $\varepsilon_k$, and Eq.~\eqref{S2:shift} vanishes identically, $\tfrac{V^2}{4}(\tfrac{1}{-\omega}+\tfrac1\omega)=0$. For $\kd=\pi$, $\varepsilon_{k+\pi}=\varepsilon_{k-\pi}$, so the interchange has no effect.
\par We now evaluate the denominators. 
Using the dispersion relation $\epsilon_k=-2t\cos k-\mu$ and the trigonometric identity $\cos X-\cos Y=-2\sin\frac{X+Y}{2}\sin\frac{X-Y}{2}$, the energy differences can be expressed as,
\begin{align}
\epsilon_{k}-\epsilon_{k+\kd}&=2t\big[\cos(k+\kd)-\cos k\big]=-\Lambda\,\sin\!\Big(k+\frac{\kd}{2}\Big),\\
\epsilon_{k}-\xi_{k-\kd}&=2t\big[\cos(k-\kd)-\cos k\big]=+\Lambda\,\sin\!\Big(k-\frac{\kd}{2}\Big),
\end{align}
with
\beq
\Lambda\equiv4t\sin\!\Big(\frac{\kd}{2}\Big),
\qquad
S_{\pm}\equiv\sin\!\Big(k\pm\frac{\kd}{2}\Big).
\eeq
The same factor $\sin(\kd/2)$ that governed the Peierls amplitude $z_d$ in Eq.~\eqref{eq:amplitude} has
reappeared, as it must. Substituting,
\beq
\delta\epsilon_{k}=\frac{V^{2}/4}{\omega+\Lambda S_{-}}-\frac{V^{2}/4}{\omega+\Lambda S_{+}} .
\label{S:closed}
\eeq
At this point, we combine \eqref{S:closed} 
over a common
denominator. Using $\sin X-\sin Y=2\cos\frac{X+Y}{2}\sin\frac{X-Y}{2}$ for the numerator and
$\sin(A{+}B)\sin(A{-}B)=\sin^{2}A-\sin^{2}B$ for the cross term in the denominator, together with
$\Lambda\cos(\kd/2)=2t\sin \kd$, one finds
$\delta\epsilon_{k}=N_{e}/(D_{e}+D_{o})$ with
\begin{align}
N_{e}(k)&=2V^{2}t\,\sin^{2}\!\Big(\frac{\kd}{2}\Big)\cos k &&\text{(even in }k),\\
D_{e}(k)&=\omega^{2}+\Lambda^{2}\Big[\sin^{2}k-\sin^{2}\!\Big(\frac{\kd}{2}\Big)\Big]
&&\text{(even in }k),\\
D_{o}(k)&=4t\,\omega\,\sin \kd\,\sin k &&\text{(\textbf{odd} in }k).
\label{S:Do}
\end{align}
Since $\delta\epsilon_{-k}=N_{e}/(D_{e}-D_{o})$, the even and odd parts follow immediately:
\beq
\delta\epsilon^{\rm even}_{k}=\frac{N_{e}D_{e}}{D_{e}^{2}-D_{o}^{2}},
\qquad
\;\delta\epsilon^{\rm odd}_{k}=\frac{-\,N_{e}D_{o}}{D_{e}^{2}-D_{o}^{2}}\;
\label{S:odd}
\eeq
The entire odd-in-$k$ content of the drive is thus proportional to $D_{o}$, and by
\eqref{S:Do} $D_{o}\propto\sin \kd$. Hence $\delta\epsilon^{\rm odd}_{k}$ vanishes \emph{identically} at
$\kd=0,\pi,2\pi,\dots$ to all orders in the drive strength, and is \emph{odd} under $\kd\to-\kd$.
These are exact statements and the microscopic counterpart of the symmetry argument of S1.

In the high-frequency regime, $\omega\gg|\Lambda|$, the even denominator approaches $D_e\simeq\omega^2$, while $D_o^2\ll D_e^2$. Consequently,
\eqref{S:odd} reduces, using $2\sin k\cos k=\sin 2k$ and
$2\sin^{2}(\kd/2)=1-\cos \kd$, to
\beq
\;\delta\epsilon^{\rm odd}_{k}=-\A(\kd)\,\sin 2k,
\qquad
\A(\kd)=\frac{2V^{2}t^{2}}{\omega^{3}}\,\sin \kd\,\big(1-\cos \kd\big).\;
\label{S:master}
\eeq
Since $\mathcal{A}\propto\mathcal{F}(\kd)=\sin^2(\kd/2)\sin\kd$, we determine the optimal drive wave vector by introducing $u=\kd/2$, which yields
\beq
\mathcal{F}=2\sin^3u\cos u,
\qquad
\frac{d\mathcal{F}}{du}
=2\sin^2u\big(3\cos^2u-\sin^2u\big)
=2\sin^2u\big(3-4\sin^2u\big).
\eeq
Excluding the trivial solutions $\sin u=0$, the stationary condition gives $\sin^2u=3/4$, with the maximum in the interval $0<\kd<\pi$ occurring at $u=\pi/3$. Thus, the optimal drive wave vector and the corresponding maximum are
\beq
\kd^{\rm opt}=\frac{2\pi}{3},
\qquad
\mathcal{F}_{\max}=\frac{3\sqrt{3}}{8}\simeq0.6495.
\eeq
By the antisymmetry $\mathcal{F}(\kd+ \pi)=-\mathcal{F}(\kd)$, the corresponding minimum occurs at $\kd=4\pi/3$. In the long-wavelength limit, $\kd\ll1$, the function behaves as $\mathcal{F}\simeq\kd^3/4$, indicating a cubic suppression of the effect as the drive wavelength increases. This scaling imposes a practical constraint on the ratio of the lattice constant to the drive wavelength.

\par Moreover, the odd harmonic has a transparent real-space form. Writing $\sin2k=(e^{2ik}-e^{-2ik})/(2i)$ and using $\sum_k e^{2ik}c_k^\dagger c_k=\sum_jc_j^\dagger c_{j+2}$,
\begin{equation}
\sum_k\delta\varepsilon_{\rm odd}(k)\,c_k^\dagger c_k=\frac{i\mathcal{A}}{2}\sum_k\big(e^{2ik}-e^{-2ik}\big)c_k^\dagger c_k
=\sum_j\Big(\frac{i\mathcal{A}}{2}\,c_j^\dagger c_{j+2}+{\rm H.c.}\Big).
\end{equation}
This is a purely imaginary next-nearest-neighbor hopping, $\tfrac{\mathcal{A}}{2}e^{i\pi/2}$, i.e., a hopping threaded by a $\pi/2$ Peierls phase. It is the one-dimensional analogue of the complex second-neighbor term of the Haldane model \cite{S-Haldane1988,S-Jotzu2014}. Complex conjugation flips $i\to-i$, so $\mathcal{T}$ is broken; inversion reverses the bond direction, $c_j^\dagger c_{j+2}\to c_{-j}^\dagger c_{-j-2}$, which also flips the sign, so $\mathcal{P}$ is broken. This
is the real-space embodiment of the synthetic gauge flux of Eq.~\eqref{S1:theta}, and it also accounts
for the higher harmonics (non-sinusoidal structure) of the current--momentum relation once
$\kd\neq0$.

\par It is worth emphasizing why a $\sin2k$ harmonic, and not a term linear in $k$, produces a diode. Consider pairing between $k+q/2$ and $-k+q/2$ near the Fermi points $\pm k_F$ of a single band with velocity $v$. An odd term that is \textit{linear} in $k$ shifts the right-moving Fermi point up by $+\lambda_F$ and the left-moving one down by $-\lambda_F$. This is identical to a Galilean boost $k\to k+\lambda_F/v$, absorbed by $q\to q-2\lambda_F/v$. It moves the equilibrium momentum $q_0$ and the superconducting window together, leaving $I_c^{+}=|I_c^{-}|$. A diode requires odd components that \textit{cannot} be removed by such a shift, namely harmonics beyond linear order \cite{S-YuanFu2022,S-Legg2022,S-Angehrn2024}. The expansion $\sin2k=2k-\tfrac43k^3+\dots$ shows that the drive-generated term contains exactly such a cubic component. Its linear part produces the equilibrium shift $q_0\neq0$ [inset of Fig.~2(b)], while its nonlinear part displaces the window relative to $q_0$, as quantified in Sec.~S3.
\begin{figure}[t]
\includegraphics[width=0.7\textwidth]{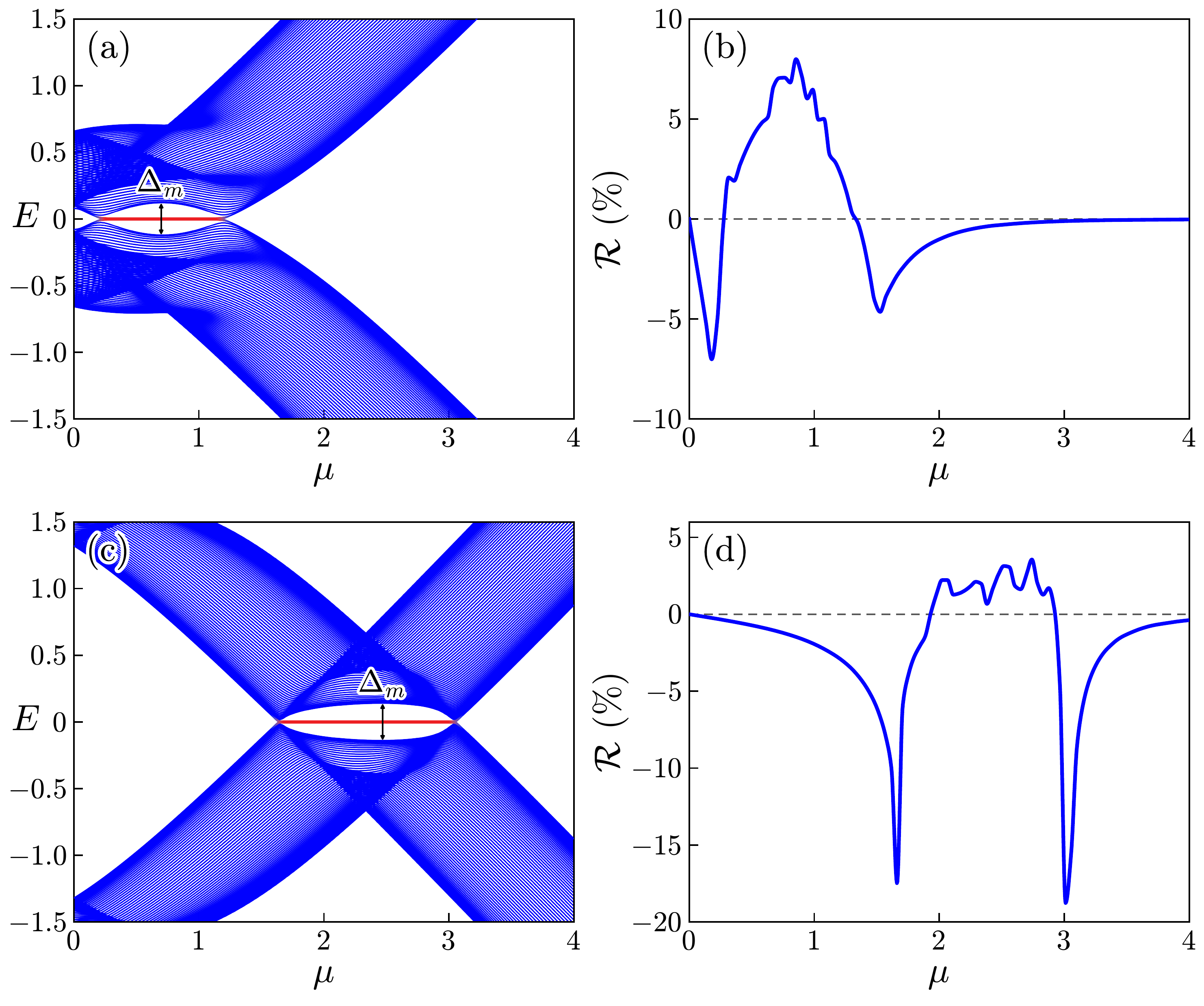}
\caption{Energy spectra (left panels) and corresponding rectification $\mathcal{R}$ (right panels) as functions of the chemical potential $\mu$ for two representative finite values of the exchange coupling $J$, in the presence of a static magnetic field $B\sigma_z$. The zero-energy modes identify the topological regions, where the rectification is comparatively large. As $\mu$ enters the trivial regime, the rectification generally decreases and approaches zero deep in the trivial phase.}
\label{fig:S1}
\end{figure}
\section*{S3: Inverse-Gap Law: Why Topological Phases Show Stronger Rectification}
In the main text, we showed that rectification is strongest in the topological regime and substantially weaker in the trivial one. This behavior is not merely a drive-induced effect but originates from the underlying topological properties of the system. To verify this, we also investigate rectification in the static limit, where both a finite exchange coupling $J$ and an external magnetic field are required. Specifically, we add a magnetic field along the $z$-direction, represented by $B\sigma_z$, to Eq.~(1) of the main text. Fig.~S1 shows the energy spectrum as a function of $\mu$ for two representative nonzero values of $J$, with zero-energy modes appearing in both cases even at fixed $B$. The corresponding rectification is also plotted as a function of $\mu$. A direct comparison reveals that rectification is largest within the topological window, gradually decreases in the trivial regime, and eventually vanishes deep in the trivial limit. In the following, we explain the physical mechanism underlying this behavior.
\par Here we show that the rectification is the ratio of a gap-independent asymmetry energy to the pair-breaking gap, $\R\simeq\epsilon_{\rm asym}/\Delta_{\rm eff}$, and the topology enters only through the denominator.

\par We start from the Bogoliubov spectrum at finite pair momentum. In an FF state with momentum $q$, an electron at $k+q/2$ pairs with one at $-k+q/2$. Let $\xi_k$ denote the effective (drive-dressed) band measured from the chemical potential, and consider a single branch for clarity. In the Nambu basis $(c_{k+q/2,\uparrow},\,c^\dagger_{-k+q/2,\downarrow})^T$ the BdG matrix is
\begin{equation}
h(k,q)=\begin{pmatrix}\xi_{k+q/2}&\Delta\\ \Delta&-\xi_{-k+q/2}\end{pmatrix}.
\end{equation}
Writing $\xi_\pm\equiv\xi_{\pm k+q/2}$, the condition $\det(h-E)=0$ reads $(\xi_+-E)(-\xi_--E)-\Delta^2=0$, i.e., $E^2-(\xi_+-\xi_-)E-\xi_+\xi_--\Delta^2=0$. Solving this quadratic and using $\tfrac14(\xi_+-\xi_-)^2+\xi_+\xi_-=\tfrac14(\xi_++\xi_-)^2$,
\begin{equation}
E_\pm(k,q)=\delta\xi\pm R,\qquad \delta\xi=\tfrac12(\xi_+-\xi_-),\qquad \bar\xi=\tfrac12(\xi_++\xi_-),\qquad R=\sqrt{\bar\xi^2+\Delta^2}\ \ge\Delta .
\label{S3:E}
\end{equation}
Here $R$ is the energy needed to break a pair, while $\delta\xi$ is the Doppler shift produced by the flow, which displaces both branches rigidly.

\par This spectrum determines the superconducting window. The condensate is stable as long as no quasiparticle is pushed through zero energy, i.e., $E_+>0>E_-$ for all $k$. By Eq.~\eqref{S3:E} this requires $|\delta\xi|<R$ everywhere (the Landau criterion):
\begin{equation}
E_{\rm min}(q)\equiv\min_k\big[R(k,q)-|\delta\xi(k,q)|\big]>0 .
\label{S3:Emin}
\end{equation}
This defines the interval $q_c^{-}<q<q_c^{+}$, bounded by the gap-closing points shown in Fig.~2(a). We characterize this interval by its half-width and midpoint,
\begin{equation}
q_c=\frac{q_c^{+}-q_c^{-}}{2},
\qquad
q_{\rm mid}=\frac{q_c^{+}+q_c^{-}}{2},
\end{equation}
and define the equilibrium pair momentum $q_0$ through $\I(q_0)=0$. The pair-breaking gap is defined as $\Delta_{\rm eff}\equiv E_{\rm min}(q_{\rm mid})$, corresponding to the maximum excitation gap attained within the condensate's stability window.
\par Within this window the rectification reduces to a geometric ratio. Inside the window the ground-state energy is smooth and minimal at $q_0$, $E\simeq E_0+\tfrac{\kappa}{2}(q-q_0)^2$, so the current (with the sign convention of Fig.~2) is linear, $\I\simeq-\kappa(q-q_0)$. Its extrema are reached at the window edges as,
\begin{equation}
I_c^{+}=\kappa\,(q_0-q_c^{-}),\qquad |I_c^{-}|=\kappa\,(q_c^{+}-q_0)
\end{equation}
with the same prefactor, which cancels in the ratio (Eq. (7) of the main text),
\begin{equation}
\R=\frac{(q_0-q_c^{-})-(q_c^{+}-q_0)}{(q_0-q_c^{-})+(q_c^{+}-q_0)}=\frac{2q_0-q_c^{+}-q_c^{-}}{q_c^{+}-q_c^{-}}=\frac{q_0-q_{\rm mid}}{q_c}.
\label{S3:geom}
\end{equation}
The rectification is thus the displacement of the equilibrium point from the center of the window, measured in units of the window's half-width. If the window is centered on $q_0$, there is no diode.
\par It remains to determine the physical mechanisms governing the width and offset of the stability window, which arise from distinct contributions to the band dispersion. To this end, we decompose the band into its even and odd components, \(\xi_k=\xi_k^e+\xi_k^o\). For the driven chain, the odd component is \(\xi_k^o=-\mathcal{A}(\kd)\sin 2k\) (Sec.~S2). Using the parity relations \(\xi^e_{-k+q/2}=\xi^e_{k-q/2}\) and \(\xi^o_{-k+q/2}=-\xi^o_{k-q/2}\), we obtain
\begin{align}
\delta\xi
&=\frac{1}{2}\left[\xi^e_{k+q/2}-\xi^e_{k-q/2}\right]
+\frac{1}{2}\left[\xi^o_{k+q/2}+\xi^o_{k-q/2}\right],
\label{S3:dxi}\\
\bar\xi
&=\frac{1}{2}\left[\xi^e_{k+q/2}+\xi^e_{k-q/2}\right]
+\frac{1}{2}\left[\xi^o_{k+q/2}-\xi^o_{k-q/2}\right].
\end{align}
Expanding each term in a Taylor series about \(q=0\),
\[
f\left(k\pm\frac{q}{2}\right)
=f(k)\pm\frac{q}{2}f'(k)
+\frac{q^2}{8}f''(k)
\pm\frac{q^3}{48}f'''(k)+\cdots,
\]
the symmetric and antisymmetric combinations yield
\[
\begin{aligned}
\frac{1}{2}\left[f\left(k+\frac{q}{2}\right)
-f\left(k-\frac{q}{2}\right)\right]
=\frac{q}{2}f'(k)+O(q^3), \qquad 
\frac{1}{2}\left[f\left(k+\frac{q}{2}\right)
+f\left(k-\frac{q}{2}\right)\right]
=f(k)+O(q^2).
\end{aligned}
\label{Eq:Taylor}
\]
Applying these expansions to Eq.~\eqref{S3:dxi}, we find, to leading order in \(q\),
\begin{align}
\delta\xi
=\underbrace{\xi_k^o}_{\text{independent of }q}
+\underbrace{\frac{q}{2}\partial_k\xi_k^e}_{\text{linear in }q}
+O(q^2),
\label{S3:dxi2}.
\end{align}
This is the whole mechanism. The even band enters the Doppler shift only in proportion to $q$, while the odd harmonic enters as a \textit{constant offset}, already present at $q=0$. Because the window edges satisfy $|\delta\xi|=R$, the constant shifts
both roots in the \emph{same} direction while the linear term pushes them in \emph{opposite}
directions, so the window is displaced without being widened.
\par This separation becomes quantitative near the weakest link of the condensate. Let $k^*$ be the momentum at which the minimum in Eq.~\eqref{S3:Emin} is attained, i.e., the weakest link of the condensate.
Near $q_{\rm mid}$, $R(k^*,q)\simeq\Delta_{\rm eff}$ to leading order. With $\gamma=\tfrac12\partial_k\xi^e|_{k^*}>0$ and $\epsilon=\xi^o_{k^*}$, Eq.~\eqref{S3:dxi} turns Eq.~\eqref{S3:Emin} into a tent-shaped function,
\begin{equation}
E_{\rm min}(q)\simeq\Delta_{\rm eff}-|\gamma q+\epsilon| .
\end{equation}
Its zeros follow from $\gamma q+\epsilon=\pm\Delta_{\rm eff}$, namely $q_c^{\pm}=(\pm\Delta_{\rm eff}-\epsilon)/\gamma$. Taking the sum and difference of these critical momenta, we obtain,
\begin{equation}
q_{\rm mid}=-\frac{\epsilon}{\gamma},\qquad q_c=\frac{\Delta_{\rm eff}}{\gamma}.
\label{S3:two}
\end{equation}
The half-width is controlled by the gap, while the center is controlled by $\epsilon=\xi^o_{k^*}$, a property of the \textit{normal-state} band in which $\Delta$ does not appear. Substituting Eq.~\eqref{S3} into Eq.~\eqref{S3}, the factor ($\gamma$) cancels, yielding
\begin{equation}
\R\simeq\frac{\epsilon_{\rm asym}}{\Delta_{\rm eff}},
\qquad
\epsilon_{\rm asym}\equiv\epsilon+\gamma q_0.
\label{S3}
\end{equation}
Here $\epsilon_{\rm asym}$ is the net energy by which the odd harmonic displaces the pair-breaking threshold relative to the equilibrium state. It is proportional to $\mathcal{A}(\kd)$, which also explains the optimum at $\kd=2\pi/3$ (Sec.~S2), and it is independent of the gap. Topology enters only through $\Delta_{\rm eff}$. In a \textit{trivial} phase the weakest excitation is a bulk quasiparticle, so by Eq.~\eqref{S3:E} $\Delta_{\rm eff}\simeq\Delta$ (or even larger away from a phase transition, with exact equality at $(J=0)$). In a \textit{topological} phase a Majorana branch lies inside the bulk gap, at zero quasienergy or at $\pm\pi/T$. The weakest link is then the minigap $\Delta_m$ separating that branch from the continuum. That $\Delta_m\ll\Delta$ is structural rather than accidental. A topological window is bounded on both sides by gap closings, so the gap rises from zero and returns to zero across it. By contrast, a trivial phase extends over wide parameter ranges without closings, and its gap stays comparable to or larger than $\Delta$ except in the vicinity of a transition.
Consequently,
\begin{equation}
\frac{\R_{\rm topo}}{\R_{\rm triv}}\simeq\frac{\Delta}{\Delta_m}>1 .
\label{S3:inverse_gap}
\end{equation}
In summary, the drive displaces the superconducting window by a fixed amount that knows nothing about the gap. What topology changes is not the displacement but the \textit{width} of the window being displaced. A narrow window shifted by a fixed amount is strongly lopsided, while a wide one barely notices. This is consistent with the results for the static model shown in Fig.~\ref{fig:S1}. A similar phenomenology emerges even in the driven scenario, where $\Delta_{\rm eff}$ is governed by the drive-induced pairing gap. Nevertheless, the inverse-gap dependence of the rectification factor remains valid, explaining why rectification is strongest in the topological windows hosting either zero or $\pi$ modes, while remaining weakest in the trivial regime.
\begin{figure}[t]
\includegraphics[width=0.7\textwidth]{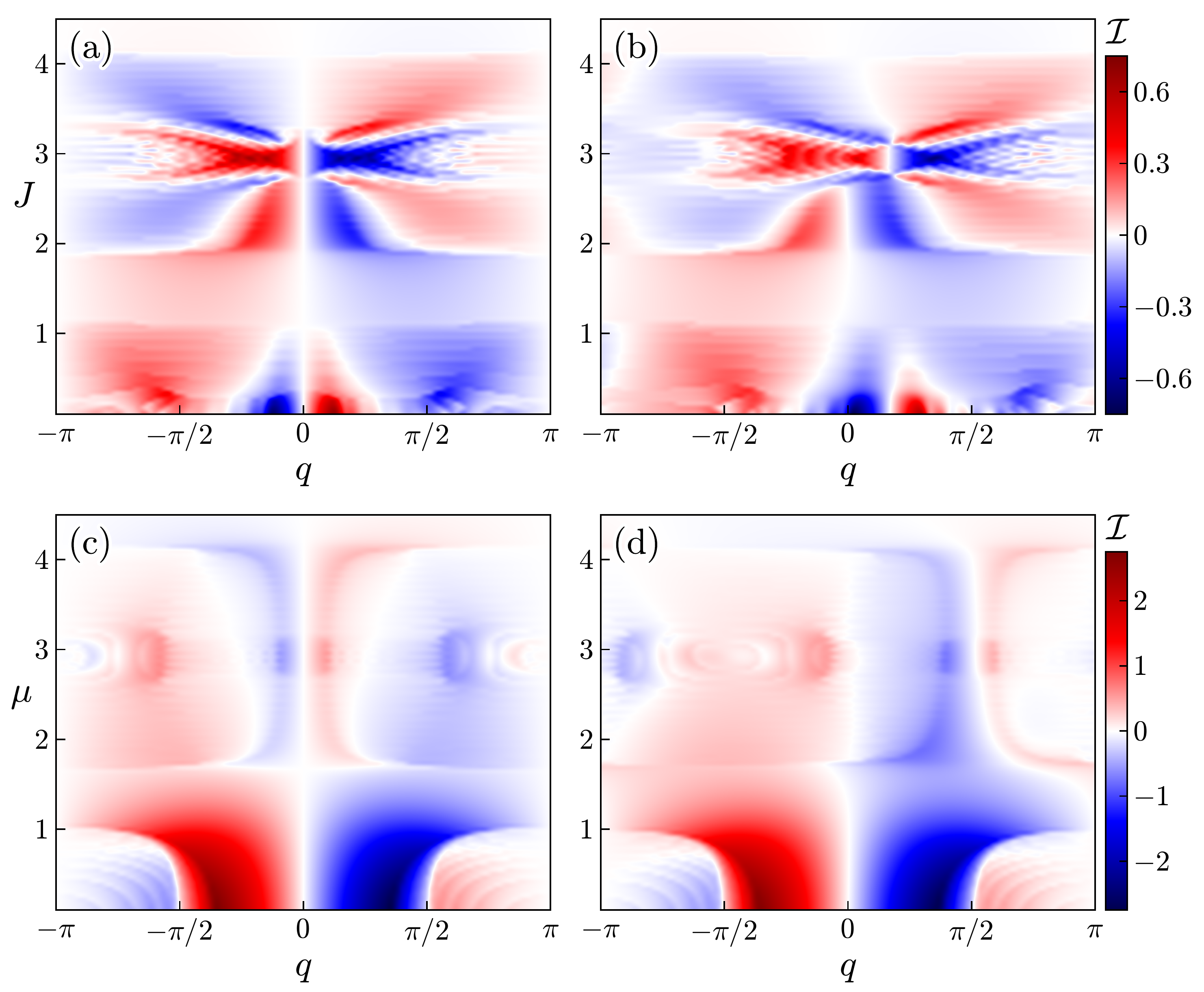}
\caption{Supercurrent $\I$ in the $(q,J)$ plane at fixed $\mu=2.5$, for (a) the standing drive $\kd=0$ and (b) the traveling drive $\kd=2\pi/3$. In (a) the map is exactly antisymmetric under $q\to-q$, as required by the combined inversion and spin-rotation symmetry, and the zero-current contour is a sharp vertical line at $q=0$. In (b) this contour is visibly displaced from $q=0$, which is a direct image of the equilibrium Cooper-pair momentum $q_0\neq0$, and the positive and negative current lobes become unequal in both magnitude and extent. The distortion is strongest inside the topological windows, in agreement with the inverse-gap law of Eq.~\eqref{S3:inverse_gap}. Other parameters: $t=0.5$, $\Delta=0.7$, $g=\pi/2$, $V=1.5$, $\omega=3$.}
\label{fig:S2}
\end{figure}
\section*{S4: Current Maps in the $(q,J)$ and $(q,\mu)$ Planes}
Fig.~2 of the main text shows the current--momentum relation at fixed $\mu$ and $J$. Here we map it
over the system parameters themselves, which exposes the geometry of Sec.~S3 directly.
\par Fig.~\ref{fig:S2} display $\I$ in the $(q,J)$ plane at $\mu=2.5$ and in the $(q,\mu)$ plane at $J=0.5$, respectively, for $\kd=0$ and $\kd=2\pi/3$. The reference panels [$\kd=0$, Figs.~\ref{fig:S2}(a) and (c)] serve as a reference. They are exactly antisymmetric under $q\to-q$, and the zero-current contour is a sharp vertical line at $q=0$. With a traveling wave [$\kd=2\pi/3$, panels (b) and (d)] the zero-current contour is visibly displaced from $q=0$, giving a direct image of the equilibrium momentum $q_0\neq0$. In addition, the positive and negative current lobes become unequal. The distortion is strongest within the topological windows, i.e., the $\pi$-mode window $0.3\lesssim J\lesssim1.2$ and the zero-mode window $2.0\lesssim J\lesssim2.8$ in Fig.~\ref{fig:S2}(a) and (b) (see Fig. 4(d) of the main text), and the $\pi$-mode windows $1.8\lesssim\mu\lesssim2.7$ and $3.2\lesssim\mu\lesssim4.1$ in Fig.~\ref{fig:S2}(c) and (d) (see Fig. 4(c) of the main text). This is consistent with the inverse-gap law of Sec.~S3.
\par Further, we note that at large $J$ the exchange field eventually exceeds every other scale and destroys the paired state
altogether; above $J\simeq4.2$ the gapped window closes, $q_{c}^{\pm}$ cease to exist, and the
rectification is identically zero rather than merely small. At large $\mu$ the chemical potential
moves away from the region supporting low-energy states, the condensate loses the spectral weight that
carries current, and the superfluid stiffness $\kappa$ collapses; since the current amplitude is
proportional to $\kappa$, $\I\to0$. The first is a loss of pairing, the second a loss of carriers.
\par Several additional features are noteworthy. Panel (c) develops a horizontal near-null strip at
$\mu\simeq1.7$, where the current nearly vanishes across the \emph{entire} $q$ range, just below the
onset of the first $\pi$-mode window at $\mu\simeq1.8$; this is where the superconducting window is
pinching shut at the transition, consistent with $q_{c}\to0$ in Eq.~\eqref{S3:two}. Above
$\mu\simeq4.2$, coinciding with the closing of the second $\pi$-mode window, the entire current
pattern reverses sign, a phenomenon which we call as the $0-\pi$ phase transition in diode current.
\par We close the discussion with a remark on the numerics. The quasienergies are obtained by exact diagonalization of $H_{\rm eff}=(i/T)\ln U(T)$, Eq.~(3) of the main text, with the propagator constructed by time slicing. The current is evaluated from Eq.~(6) of the main text by numerical differentiation with respect to $q$. Following Refs.~\cite{S-Legg2022,S-Angehrn2024}, the pairing amplitude is held fixed, and the critical currents $I_c^{\pm}$ are extracted within the gapped window $[q_c^{-},q_c^{+}]$ only, since beyond it the non-self-consistent current is unphysical.

% ---- references of the Supplemental Material ----


\begin{thebibliography}{98}%
\makeatletter
\providecommand \@ifxundefined [1]{%
 \@ifx{#1\undefined}
}%
\providecommand \@ifnum [1]{%
 \ifnum #1\expandafter \@firstoftwo
 \else \expandafter \@secondoftwo
 \fi
}%
\providecommand \@ifx [1]{%
 \ifx #1\expandafter \@firstoftwo
 \else \expandafter \@secondoftwo
 \fi
}%
\providecommand \natexlab [1]{#1}%
\providecommand \enquote  [1]{``#1''}%
\providecommand \bibnamefont  [1]{#1}%
\providecommand \bibfnamefont [1]{#1}%
\providecommand \citenamefont [1]{#1}%
\providecommand \href@noop [0]{\@secondoftwo}%
\providecommand \href [0]{\begingroup \@sanitize@url \@href}%
\providecommand \@href[1]{\@@startlink{#1}\@@href}%
\providecommand \@@href[1]{\endgroup#1\@@endlink}%
\providecommand \@sanitize@url [0]{\catcode `\\12\catcode `\$12\catcode
  `\&12\catcode `\#12\catcode `\^12\catcode `\_12\catcode `\%12\relax}%
\providecommand \@@startlink[1]{}%
\providecommand \@@endlink[0]{}%
\providecommand \url  [0]{\begingroup\@sanitize@url \@url }%
\providecommand \@url [1]{\endgroup\@href {#1}{\urlprefix }}%
\providecommand \urlprefix  [0]{URL }%
\providecommand \Eprint [0]{\href }%
\providecommand \doibase [0]{https://doi.org/}%
\providecommand \selectlanguage [0]{\@gobble}%
\providecommand \bibinfo  [0]{\@secondoftwo}%
\providecommand \bibfield  [0]{\@secondoftwo}%
\providecommand \translation [1]{[#1]}%
\providecommand \BibitemOpen [0]{}%
\providecommand \bibitemStop [0]{}%
\providecommand \bibitemNoStop [0]{.\EOS\space}%
\providecommand \EOS [0]{\spacefactor3000\relax}%
\providecommand \BibitemShut  [1]{\csname bibitem#1\endcsname}%
\let\auto@bib@innerbib\@empty
%</preamble>
\bibitem [{\citenamefont {Ando}\ \emph {et~al.}(2020)\citenamefont {Ando},
  \citenamefont {Miyasaka}, \citenamefont {Li}, \citenamefont {Ishizuka},
  \citenamefont {Arakawa}, \citenamefont {Shiota}, \citenamefont {Moriyama},
  \citenamefont {Yanase},\ and\ \citenamefont {Ono}}]{Ando2020}%
  \BibitemOpen
  \bibfield  {author} {\bibinfo {author} {\bibfnamefont {F.}~\bibnamefont
  {Ando}}, \bibinfo {author} {\bibfnamefont {Y.}~\bibnamefont {Miyasaka}},
  \bibinfo {author} {\bibfnamefont {T.}~\bibnamefont {Li}}, \bibinfo {author}
  {\bibfnamefont {J.}~\bibnamefont {Ishizuka}}, \bibinfo {author}
  {\bibfnamefont {T.}~\bibnamefont {Arakawa}}, \bibinfo {author} {\bibfnamefont
  {Y.}~\bibnamefont {Shiota}}, \bibinfo {author} {\bibfnamefont
  {T.}~\bibnamefont {Moriyama}}, \bibinfo {author} {\bibfnamefont
  {Y.}~\bibnamefont {Yanase}},\ and\ \bibinfo {author} {\bibfnamefont
  {T.}~\bibnamefont {Ono}},\ }\bibfield  {title} {\bibinfo {title} {Observation
  of superconducting diode effect},\ }\href
  {https://doi.org/10.1038/s41586-020-2590-4} {\bibfield  {journal} {\bibinfo
  {journal} {Nature (London)}\ }\textbf {\bibinfo {volume} {584}},\ \bibinfo
  {pages} {373} (\bibinfo {year} {2020})}\BibitemShut {NoStop}%
\bibitem [{\citenamefont {Nadeem}\ \emph {et~al.}(2023)\citenamefont {Nadeem},
  \citenamefont {Fuhrer},\ and\ \citenamefont {Wang}}]{Nadeem2023}%
  \BibitemOpen
  \bibfield  {author} {\bibinfo {author} {\bibfnamefont {M.}~\bibnamefont
  {Nadeem}}, \bibinfo {author} {\bibfnamefont {M.~S.}\ \bibnamefont {Fuhrer}},\
  and\ \bibinfo {author} {\bibfnamefont {X.}~\bibnamefont {Wang}},\ }\bibfield
  {title} {\bibinfo {title} {The superconducting diode effect},\ }\href
  {https://doi.org/10.1038/s42254-023-00632-w} {\bibfield  {journal} {\bibinfo
  {journal} {Nat. Rev. Phys.}\ }\textbf {\bibinfo {volume} {5}},\ \bibinfo
  {pages} {558} (\bibinfo {year} {2023})}\BibitemShut {NoStop}%
\bibitem [{\citenamefont {Baumgartner}\ \emph {et~al.}(2022)\citenamefont
  {Baumgartner}, \citenamefont {Fuchs}, \citenamefont {Costa}, \citenamefont
  {Reinhardt}, \citenamefont {Gronin}, \citenamefont {Gardner}, \citenamefont
  {Lindemann}, \citenamefont {Manfra}, \citenamefont {{Faria Junior}},
  \citenamefont {Kochan} \emph {et~al.}}]{Baumgartner2022}%
  \BibitemOpen
  \bibfield  {author} {\bibinfo {author} {\bibfnamefont {C.}~\bibnamefont
  {Baumgartner}}, \bibinfo {author} {\bibfnamefont {L.}~\bibnamefont {Fuchs}},
  \bibinfo {author} {\bibfnamefont {A.}~\bibnamefont {Costa}}, \bibinfo
  {author} {\bibfnamefont {S.}~\bibnamefont {Reinhardt}}, \bibinfo {author}
  {\bibfnamefont {S.}~\bibnamefont {Gronin}}, \bibinfo {author} {\bibfnamefont
  {G.~C.}\ \bibnamefont {Gardner}}, \bibinfo {author} {\bibfnamefont
  {T.}~\bibnamefont {Lindemann}}, \bibinfo {author} {\bibfnamefont {M.~J.}\
  \bibnamefont {Manfra}}, \bibinfo {author} {\bibfnamefont {P.~E.}\
  \bibnamefont {{Faria Junior}}}, \bibinfo {author} {\bibfnamefont
  {D.}~\bibnamefont {Kochan}}, \emph {et~al.},\ }\bibfield  {title} {\bibinfo
  {title} {Supercurrent rectification and magnetochiral effects in symmetric
  {Josephson} junctions},\ }\href {https://doi.org/10.1038/s41565-021-01009-9}
  {\bibfield  {journal} {\bibinfo  {journal} {Nat. Nanotechnol.}\ }\textbf
  {\bibinfo {volume} {17}},\ \bibinfo {pages} {39} (\bibinfo {year}
  {2022})}\BibitemShut {NoStop}%
\bibitem [{\citenamefont {Pal}\ \emph {et~al.}(2022)\citenamefont {Pal},
  \citenamefont {Chakraborty}, \citenamefont {Sivakumar}, \citenamefont
  {Davydova}, \citenamefont {Gopi}, \citenamefont {Pandeya}, \citenamefont
  {Krieger}, \citenamefont {Zhang}, \citenamefont {Date}, \citenamefont {Ju}
  \emph {et~al.}}]{Pal2022}%
  \BibitemOpen
  \bibfield  {author} {\bibinfo {author} {\bibfnamefont {B.}~\bibnamefont
  {Pal}}, \bibinfo {author} {\bibfnamefont {A.}~\bibnamefont {Chakraborty}},
  \bibinfo {author} {\bibfnamefont {P.~K.}\ \bibnamefont {Sivakumar}}, \bibinfo
  {author} {\bibfnamefont {M.}~\bibnamefont {Davydova}}, \bibinfo {author}
  {\bibfnamefont {A.~K.}\ \bibnamefont {Gopi}}, \bibinfo {author}
  {\bibfnamefont {A.~K.}\ \bibnamefont {Pandeya}}, \bibinfo {author}
  {\bibfnamefont {J.~A.}\ \bibnamefont {Krieger}}, \bibinfo {author}
  {\bibfnamefont {Y.}~\bibnamefont {Zhang}}, \bibinfo {author} {\bibfnamefont
  {M.}~\bibnamefont {Date}}, \bibinfo {author} {\bibfnamefont {S.}~\bibnamefont
  {Ju}}, \emph {et~al.},\ }\bibfield  {title} {\bibinfo {title} {Josephson
  diode effect from {Cooper} pair momentum in a topological semimetal},\ }\href
  {https://doi.org/10.1038/s41567-022-01699-5} {\bibfield  {journal} {\bibinfo
  {journal} {Nat. Phys.}\ }\textbf {\bibinfo {volume} {18}},\ \bibinfo {pages}
  {1228} (\bibinfo {year} {2022})}\BibitemShut {NoStop}%
\bibitem [{\citenamefont {Wu}\ \emph {et~al.}(2022)\citenamefont {Wu},
  \citenamefont {Wang}, \citenamefont {Xu}, \citenamefont {Sivakumar},
  \citenamefont {Pasco}, \citenamefont {Filippozzi}, \citenamefont {Parkin},
  \citenamefont {Zeng}, \citenamefont {McQueen},\ and\ \citenamefont
  {Ali}}]{Wu2022}%
  \BibitemOpen
  \bibfield  {author} {\bibinfo {author} {\bibfnamefont {H.}~\bibnamefont
  {Wu}}, \bibinfo {author} {\bibfnamefont {Y.}~\bibnamefont {Wang}}, \bibinfo
  {author} {\bibfnamefont {Y.}~\bibnamefont {Xu}}, \bibinfo {author}
  {\bibfnamefont {P.~K.}\ \bibnamefont {Sivakumar}}, \bibinfo {author}
  {\bibfnamefont {C.}~\bibnamefont {Pasco}}, \bibinfo {author} {\bibfnamefont
  {U.}~\bibnamefont {Filippozzi}}, \bibinfo {author} {\bibfnamefont {S.~S.~P.}\
  \bibnamefont {Parkin}}, \bibinfo {author} {\bibfnamefont {Y.-J.}\
  \bibnamefont {Zeng}}, \bibinfo {author} {\bibfnamefont {T.}~\bibnamefont
  {McQueen}},\ and\ \bibinfo {author} {\bibfnamefont {M.~N.}\ \bibnamefont
  {Ali}},\ }\bibfield  {title} {\bibinfo {title} {The field-free {Josephson}
  diode in a van der {Waals} heterostructure},\ }\href
  {https://doi.org/10.1038/s41586-022-04504-8} {\bibfield  {journal} {\bibinfo
  {journal} {Nature (London)}\ }\textbf {\bibinfo {volume} {604}},\ \bibinfo
  {pages} {653} (\bibinfo {year} {2022})}\BibitemShut {NoStop}%
\bibitem [{\citenamefont {Trahms}\ \emph {et~al.}(2023)\citenamefont {Trahms},
  \citenamefont {Melischek}, \citenamefont {Steiner}, \citenamefont {Mahendru},
  \citenamefont {Tamir}, \citenamefont {Bogdanoff}, \citenamefont {Peters},
  \citenamefont {Reecht}, \citenamefont {Winkelmann}, \citenamefont {von Oppen}
  \emph {et~al.}}]{Trahms2023}%
  \BibitemOpen
  \bibfield  {author} {\bibinfo {author} {\bibfnamefont {M.}~\bibnamefont
  {Trahms}}, \bibinfo {author} {\bibfnamefont {L.}~\bibnamefont {Melischek}},
  \bibinfo {author} {\bibfnamefont {J.~F.}\ \bibnamefont {Steiner}}, \bibinfo
  {author} {\bibfnamefont {B.}~\bibnamefont {Mahendru}}, \bibinfo {author}
  {\bibfnamefont {I.}~\bibnamefont {Tamir}}, \bibinfo {author} {\bibfnamefont
  {N.}~\bibnamefont {Bogdanoff}}, \bibinfo {author} {\bibfnamefont
  {O.}~\bibnamefont {Peters}}, \bibinfo {author} {\bibfnamefont
  {G.}~\bibnamefont {Reecht}}, \bibinfo {author} {\bibfnamefont {C.~B.}\
  \bibnamefont {Winkelmann}}, \bibinfo {author} {\bibfnamefont
  {F.}~\bibnamefont {von Oppen}}, \emph {et~al.},\ }\bibfield  {title}
  {\bibinfo {title} {Diode effect in {Josephson} junctions with a single
  magnetic atom},\ }\href {https://doi.org/10.1038/s41586-023-05743-z}
  {\bibfield  {journal} {\bibinfo  {journal} {Nature (London)}\ }\textbf
  {\bibinfo {volume} {615}},\ \bibinfo {pages} {628} (\bibinfo {year}
  {2023})}\BibitemShut {NoStop}%
\bibitem [{\citenamefont {Bauriedl}\ \emph {et~al.}(2022)\citenamefont
  {Bauriedl}, \citenamefont {B\"{a}uml}, \citenamefont {Fuchs}, \citenamefont
  {Baumgartner}, \citenamefont {Paulik}, \citenamefont {Bauer}, \citenamefont
  {Lin}, \citenamefont {Lupton}, \citenamefont {Taniguchi}, \citenamefont
  {Watanabe} \emph {et~al.}}]{Bauriedl2022}%
  \BibitemOpen
  \bibfield  {author} {\bibinfo {author} {\bibfnamefont {L.}~\bibnamefont
  {Bauriedl}}, \bibinfo {author} {\bibfnamefont {C.}~\bibnamefont {B\"{a}uml}},
  \bibinfo {author} {\bibfnamefont {L.}~\bibnamefont {Fuchs}}, \bibinfo
  {author} {\bibfnamefont {C.}~\bibnamefont {Baumgartner}}, \bibinfo {author}
  {\bibfnamefont {N.}~\bibnamefont {Paulik}}, \bibinfo {author} {\bibfnamefont
  {J.~M.}\ \bibnamefont {Bauer}}, \bibinfo {author} {\bibfnamefont {K.-Q.}\
  \bibnamefont {Lin}}, \bibinfo {author} {\bibfnamefont {J.~M.}\ \bibnamefont
  {Lupton}}, \bibinfo {author} {\bibfnamefont {T.}~\bibnamefont {Taniguchi}},
  \bibinfo {author} {\bibfnamefont {K.}~\bibnamefont {Watanabe}}, \emph
  {et~al.},\ }\bibfield  {title} {\bibinfo {title} {Supercurrent diode effect
  and magnetochiral anisotropy in few-layer {NbSe}$_2$},\ }\href
  {https://doi.org/10.1038/s41467-022-31954-5} {\bibfield  {journal} {\bibinfo
  {journal} {Nat. Commun.}\ }\textbf {\bibinfo {volume} {13}},\ \bibinfo
  {pages} {4266} (\bibinfo {year} {2022})}\BibitemShut {NoStop}%
\bibitem [{\citenamefont {Lin}\ \emph {et~al.}(2022)\citenamefont {Lin},
  \citenamefont {Siriviboon}, \citenamefont {Scammell}, \citenamefont {Liu},
  \citenamefont {Rhodes}, \citenamefont {Watanabe}, \citenamefont {Taniguchi},
  \citenamefont {Hone}, \citenamefont {Scheurer},\ and\ \citenamefont
  {Li}}]{Lin2022}%
  \BibitemOpen
  \bibfield  {author} {\bibinfo {author} {\bibfnamefont {J.-X.}\ \bibnamefont
  {Lin}}, \bibinfo {author} {\bibfnamefont {P.}~\bibnamefont {Siriviboon}},
  \bibinfo {author} {\bibfnamefont {H.~D.}\ \bibnamefont {Scammell}}, \bibinfo
  {author} {\bibfnamefont {S.}~\bibnamefont {Liu}}, \bibinfo {author}
  {\bibfnamefont {D.}~\bibnamefont {Rhodes}}, \bibinfo {author} {\bibfnamefont
  {K.}~\bibnamefont {Watanabe}}, \bibinfo {author} {\bibfnamefont
  {T.}~\bibnamefont {Taniguchi}}, \bibinfo {author} {\bibfnamefont
  {J.}~\bibnamefont {Hone}}, \bibinfo {author} {\bibfnamefont {M.~S.}\
  \bibnamefont {Scheurer}},\ and\ \bibinfo {author} {\bibfnamefont {J.~I.~A.}\
  \bibnamefont {Li}},\ }\bibfield  {title} {\bibinfo {title} {Zero-field
  superconducting diode effect in small-twist-angle trilayer graphene},\ }\href
  {https://doi.org/10.1038/s41567-022-01700-1} {\bibfield  {journal} {\bibinfo
  {journal} {Nat. Phys.}\ }\textbf {\bibinfo {volume} {18}},\ \bibinfo {pages}
  {1221} (\bibinfo {year} {2022})}\BibitemShut {NoStop}%
\bibitem [{\citenamefont {Strambini}\ \emph {et~al.}(2022)\citenamefont
  {Strambini}, \citenamefont {Spies}, \citenamefont {Ligato}, \citenamefont
  {Ili\'{c}}, \citenamefont {Rouco}, \citenamefont {Gonz\'{a}lez-Orellana},
  \citenamefont {Ilyn}, \citenamefont {Rogero}, \citenamefont {Bergeret},
  \citenamefont {Moodera} \emph {et~al.}}]{Strambini2022}%
  \BibitemOpen
  \bibfield  {author} {\bibinfo {author} {\bibfnamefont {E.}~\bibnamefont
  {Strambini}}, \bibinfo {author} {\bibfnamefont {M.}~\bibnamefont {Spies}},
  \bibinfo {author} {\bibfnamefont {N.}~\bibnamefont {Ligato}}, \bibinfo
  {author} {\bibfnamefont {S.}~\bibnamefont {Ili\'{c}}}, \bibinfo {author}
  {\bibfnamefont {M.}~\bibnamefont {Rouco}}, \bibinfo {author} {\bibfnamefont
  {C.}~\bibnamefont {Gonz\'{a}lez-Orellana}}, \bibinfo {author} {\bibfnamefont
  {M.}~\bibnamefont {Ilyn}}, \bibinfo {author} {\bibfnamefont {C.}~\bibnamefont
  {Rogero}}, \bibinfo {author} {\bibfnamefont {F.~S.}\ \bibnamefont
  {Bergeret}}, \bibinfo {author} {\bibfnamefont {J.~S.}\ \bibnamefont
  {Moodera}}, \emph {et~al.},\ }\bibfield  {title} {\bibinfo {title}
  {Superconducting spintronic tunnel diode},\ }\href
  {https://doi.org/10.1038/s41467-022-29990-2} {\bibfield  {journal} {\bibinfo
  {journal} {Nat. Commun.}\ }\textbf {\bibinfo {volume} {13}},\ \bibinfo
  {pages} {2431} (\bibinfo {year} {2022})}\BibitemShut {NoStop}%
\bibitem [{\citenamefont {Narita}\ \emph {et~al.}(2022)\citenamefont {Narita},
  \citenamefont {Ishizuka}, \citenamefont {Kawarazaki}, \citenamefont {Kan},
  \citenamefont {Shiota}, \citenamefont {Moriyama}, \citenamefont {Shimakawa},
  \citenamefont {Ognev}, \citenamefont {Samardak}, \citenamefont {Yanase} \emph
  {et~al.}}]{Narita2022}%
  \BibitemOpen
  \bibfield  {author} {\bibinfo {author} {\bibfnamefont {H.}~\bibnamefont
  {Narita}}, \bibinfo {author} {\bibfnamefont {J.}~\bibnamefont {Ishizuka}},
  \bibinfo {author} {\bibfnamefont {R.}~\bibnamefont {Kawarazaki}}, \bibinfo
  {author} {\bibfnamefont {D.}~\bibnamefont {Kan}}, \bibinfo {author}
  {\bibfnamefont {Y.}~\bibnamefont {Shiota}}, \bibinfo {author} {\bibfnamefont
  {T.}~\bibnamefont {Moriyama}}, \bibinfo {author} {\bibfnamefont
  {Y.}~\bibnamefont {Shimakawa}}, \bibinfo {author} {\bibfnamefont {A.~V.}\
  \bibnamefont {Ognev}}, \bibinfo {author} {\bibfnamefont {A.~S.}\ \bibnamefont
  {Samardak}}, \bibinfo {author} {\bibfnamefont {Y.}~\bibnamefont {Yanase}},
  \emph {et~al.},\ }\bibfield  {title} {\bibinfo {title} {Field-free
  superconducting diode effect in noncentrosymmetric superconductor/ferromagnet
  multilayers},\ }\href {https://doi.org/10.1038/s41565-022-01159-4} {\bibfield
   {journal} {\bibinfo  {journal} {Nat. Nanotechnol.}\ }\textbf {\bibinfo
  {volume} {17}},\ \bibinfo {pages} {823} (\bibinfo {year} {2022})}\BibitemShut
  {NoStop}%
\bibitem [{\citenamefont {Jeon}\ \emph {et~al.}(2022)\citenamefont {Jeon},
  \citenamefont {Kim}, \citenamefont {Yoon}, \citenamefont {Jeon},
  \citenamefont {Han}, \citenamefont {Cottet}, \citenamefont {Kontos},\ and\
  \citenamefont {Parkin}}]{Jeon2022}%
  \BibitemOpen
  \bibfield  {author} {\bibinfo {author} {\bibfnamefont {K.-R.}\ \bibnamefont
  {Jeon}}, \bibinfo {author} {\bibfnamefont {J.-K.}\ \bibnamefont {Kim}},
  \bibinfo {author} {\bibfnamefont {J.}~\bibnamefont {Yoon}}, \bibinfo {author}
  {\bibfnamefont {J.-C.}\ \bibnamefont {Jeon}}, \bibinfo {author}
  {\bibfnamefont {H.}~\bibnamefont {Han}}, \bibinfo {author} {\bibfnamefont
  {A.}~\bibnamefont {Cottet}}, \bibinfo {author} {\bibfnamefont
  {T.}~\bibnamefont {Kontos}},\ and\ \bibinfo {author} {\bibfnamefont
  {S.~S.~P.}\ \bibnamefont {Parkin}},\ }\bibfield  {title} {\bibinfo {title}
  {Zero-field polarity-reversible {Josephson} supercurrent diodes enabled by a
  proximity-magnetized {Pt} barrier},\ }\href
  {https://doi.org/10.1038/s41563-022-01300-7} {\bibfield  {journal} {\bibinfo
  {journal} {Nat. Mater.}\ }\textbf {\bibinfo {volume} {21}},\ \bibinfo {pages}
  {1008} (\bibinfo {year} {2022})}\BibitemShut {NoStop}%
\bibitem [{\citenamefont {Rikken}\ \emph {et~al.}(2001)\citenamefont {Rikken},
  \citenamefont {F\"{o}lling},\ and\ \citenamefont {Wyder}}]{Rikken2001}%
  \BibitemOpen
  \bibfield  {author} {\bibinfo {author} {\bibfnamefont {G.~L. J.~A.}\
  \bibnamefont {Rikken}}, \bibinfo {author} {\bibfnamefont {J.}~\bibnamefont
  {F\"{o}lling}},\ and\ \bibinfo {author} {\bibfnamefont {P.}~\bibnamefont
  {Wyder}},\ }\bibfield  {title} {\bibinfo {title} {Electrical magnetochiral
  anisotropy},\ }\href {https://doi.org/10.1103/physrevlett.87.236602}
  {\bibfield  {journal} {\bibinfo  {journal} {Phys. Rev. Lett.}\ }\textbf
  {\bibinfo {volume} {87}},\ \bibinfo {pages} {236602} (\bibinfo {year}
  {2001})}\BibitemShut {NoStop}%
\bibitem [{\citenamefont {Wakatsuki}\ \emph {et~al.}(2017)\citenamefont
  {Wakatsuki}, \citenamefont {Saito}, \citenamefont {Hoshino}, \citenamefont
  {Itahashi}, \citenamefont {Ideue}, \citenamefont {Ezawa}, \citenamefont
  {Iwasa},\ and\ \citenamefont {Nagaosa}}]{Wakatsuki2017}%
  \BibitemOpen
  \bibfield  {author} {\bibinfo {author} {\bibfnamefont {R.}~\bibnamefont
  {Wakatsuki}}, \bibinfo {author} {\bibfnamefont {Y.}~\bibnamefont {Saito}},
  \bibinfo {author} {\bibfnamefont {S.}~\bibnamefont {Hoshino}}, \bibinfo
  {author} {\bibfnamefont {Y.~M.}\ \bibnamefont {Itahashi}}, \bibinfo {author}
  {\bibfnamefont {T.}~\bibnamefont {Ideue}}, \bibinfo {author} {\bibfnamefont
  {M.}~\bibnamefont {Ezawa}}, \bibinfo {author} {\bibfnamefont
  {Y.}~\bibnamefont {Iwasa}},\ and\ \bibinfo {author} {\bibfnamefont
  {N.}~\bibnamefont {Nagaosa}},\ }\bibfield  {title} {\bibinfo {title}
  {Nonreciprocal charge transport in noncentrosymmetric superconductors},\
  }\href {https://doi.org/10.1126/sciadv.1602390} {\bibfield  {journal}
  {\bibinfo  {journal} {Sci. Adv.}\ }\textbf {\bibinfo {volume} {3}},\ \bibinfo
  {pages} {e1602390} (\bibinfo {year} {2017})}\BibitemShut {NoStop}%
\bibitem [{\citenamefont {Tokura}\ and\ \citenamefont
  {Nagaosa}(2018)}]{Tokura2018}%
  \BibitemOpen
  \bibfield  {author} {\bibinfo {author} {\bibfnamefont {Y.}~\bibnamefont
  {Tokura}}\ and\ \bibinfo {author} {\bibfnamefont {N.}~\bibnamefont
  {Nagaosa}},\ }\bibfield  {title} {\bibinfo {title} {Nonreciprocal responses
  from non-centrosymmetric quantum materials},\ }\href
  {https://doi.org/10.1038/s41467-018-05759-4} {\bibfield  {journal} {\bibinfo
  {journal} {Nat. Commun.}\ }\textbf {\bibinfo {volume} {9}},\ \bibinfo {pages}
  {3740} (\bibinfo {year} {2018})}\BibitemShut {NoStop}%
\bibitem [{\citenamefont {Yuan}\ and\ \citenamefont {Fu}(2022)}]{YuanFu2022}%
  \BibitemOpen
  \bibfield  {author} {\bibinfo {author} {\bibfnamefont {N.~F.~Q.}\
  \bibnamefont {Yuan}}\ and\ \bibinfo {author} {\bibfnamefont {L.}~\bibnamefont
  {Fu}},\ }\bibfield  {title} {\bibinfo {title} {Supercurrent diode effect and
  finite-momentum superconductors},\ }\href
  {https://doi.org/10.1073/pnas.2119548119} {\bibfield  {journal} {\bibinfo
  {journal} {Proc. Natl. Acad. Sci. U.S.A.}\ }\textbf {\bibinfo {volume}
  {119}},\ \bibinfo {pages} {e2119548119} (\bibinfo {year} {2022})}\BibitemShut
  {NoStop}%
\bibitem [{\citenamefont {Daido}\ \emph {et~al.}(2022)\citenamefont {Daido},
  \citenamefont {Ikeda},\ and\ \citenamefont {Yanase}}]{Daido2022}%
  \BibitemOpen
  \bibfield  {author} {\bibinfo {author} {\bibfnamefont {A.}~\bibnamefont
  {Daido}}, \bibinfo {author} {\bibfnamefont {Y.}~\bibnamefont {Ikeda}},\ and\
  \bibinfo {author} {\bibfnamefont {Y.}~\bibnamefont {Yanase}},\ }\bibfield
  {title} {\bibinfo {title} {Intrinsic superconducting diode effect},\ }\href
  {https://doi.org/10.1103/physrevlett.128.037001} {\bibfield  {journal}
  {\bibinfo  {journal} {Phys. Rev. Lett.}\ }\textbf {\bibinfo {volume} {128}},\
  \bibinfo {pages} {037001} (\bibinfo {year} {2022})}\BibitemShut {NoStop}%
\bibitem [{\citenamefont {Daido}\ and\ \citenamefont
  {Yanase}(2022)}]{DaidoYanase2022}%
  \BibitemOpen
  \bibfield  {author} {\bibinfo {author} {\bibfnamefont {A.}~\bibnamefont
  {Daido}}\ and\ \bibinfo {author} {\bibfnamefont {Y.}~\bibnamefont {Yanase}},\
  }\bibfield  {title} {\bibinfo {title} {Superconducting diode effect and
  nonreciprocal transition lines},\ }\href
  {https://doi.org/10.1103/physrevb.106.205206} {\bibfield  {journal} {\bibinfo
   {journal} {Phys. Rev. B}\ }\textbf {\bibinfo {volume} {106}},\ \bibinfo
  {pages} {205206} (\bibinfo {year} {2022})}\BibitemShut {NoStop}%
\bibitem [{\citenamefont {Ili\'{c}}\ and\ \citenamefont
  {Bergeret}(2022)}]{Ilic2022}%
  \BibitemOpen
  \bibfield  {author} {\bibinfo {author} {\bibfnamefont {S.}~\bibnamefont
  {Ili\'{c}}}\ and\ \bibinfo {author} {\bibfnamefont {F.~S.}\ \bibnamefont
  {Bergeret}},\ }\bibfield  {title} {\bibinfo {title} {Theory of the
  supercurrent diode effect in {Rashba} superconductors with arbitrary
  disorder},\ }\href {https://doi.org/10.1103/physrevlett.128.177001}
  {\bibfield  {journal} {\bibinfo  {journal} {Phys. Rev. Lett.}\ }\textbf
  {\bibinfo {volume} {128}},\ \bibinfo {pages} {177001} (\bibinfo {year}
  {2022})}\BibitemShut {NoStop}%
\bibitem [{\citenamefont {He}\ \emph {et~al.}(2022)\citenamefont {He},
  \citenamefont {Tanaka},\ and\ \citenamefont {Nagaosa}}]{He2022}%
  \BibitemOpen
  \bibfield  {author} {\bibinfo {author} {\bibfnamefont {J.~J.}\ \bibnamefont
  {He}}, \bibinfo {author} {\bibfnamefont {Y.}~\bibnamefont {Tanaka}},\ and\
  \bibinfo {author} {\bibfnamefont {N.}~\bibnamefont {Nagaosa}},\ }\bibfield
  {title} {\bibinfo {title} {A phenomenological theory of superconductor
  diodes},\ }\href {https://doi.org/10.1088/1367-2630/ac6766} {\bibfield
  {journal} {\bibinfo  {journal} {New J. Phys.}\ }\textbf {\bibinfo {volume}
  {24}},\ \bibinfo {pages} {053014} (\bibinfo {year} {2022})}\BibitemShut
  {NoStop}%
\bibitem [{\citenamefont {Davydova}\ \emph {et~al.}(2022)\citenamefont
  {Davydova}, \citenamefont {Prembabu},\ and\ \citenamefont
  {Fu}}]{Davydova2022}%
  \BibitemOpen
  \bibfield  {author} {\bibinfo {author} {\bibfnamefont {M.}~\bibnamefont
  {Davydova}}, \bibinfo {author} {\bibfnamefont {S.}~\bibnamefont {Prembabu}},\
  and\ \bibinfo {author} {\bibfnamefont {L.}~\bibnamefont {Fu}},\ }\bibfield
  {title} {\bibinfo {title} {Universal {Josephson} diode effect},\ }\href
  {https://doi.org/10.1126/sciadv.abo0309} {\bibfield  {journal} {\bibinfo
  {journal} {Sci. Adv.}\ }\textbf {\bibinfo {volume} {8}},\ \bibinfo {pages}
  {eabo0309} (\bibinfo {year} {2022})}\BibitemShut {NoStop}%
\bibitem [{\citenamefont {Zhang}\ \emph {et~al.}(2022)\citenamefont {Zhang},
  \citenamefont {Gu}, \citenamefont {Li}, \citenamefont {Hu},\ and\
  \citenamefont {Jiang}}]{Zhang2022}%
  \BibitemOpen
  \bibfield  {author} {\bibinfo {author} {\bibfnamefont {Y.}~\bibnamefont
  {Zhang}}, \bibinfo {author} {\bibfnamefont {Y.}~\bibnamefont {Gu}}, \bibinfo
  {author} {\bibfnamefont {P.}~\bibnamefont {Li}}, \bibinfo {author}
  {\bibfnamefont {J.}~\bibnamefont {Hu}},\ and\ \bibinfo {author}
  {\bibfnamefont {K.}~\bibnamefont {Jiang}},\ }\bibfield  {title} {\bibinfo
  {title} {General theory of {Josephson} diodes},\ }\href
  {https://doi.org/10.1103/physrevx.12.041013} {\bibfield  {journal} {\bibinfo
  {journal} {Phys. Rev. X}\ }\textbf {\bibinfo {volume} {12}},\ \bibinfo
  {pages} {041013} (\bibinfo {year} {2022})}\BibitemShut {NoStop}%
\bibitem [{\citenamefont {Misaki}\ and\ \citenamefont
  {Nagaosa}(2021)}]{Misaki2021}%
  \BibitemOpen
  \bibfield  {author} {\bibinfo {author} {\bibfnamefont {K.}~\bibnamefont
  {Misaki}}\ and\ \bibinfo {author} {\bibfnamefont {N.}~\bibnamefont
  {Nagaosa}},\ }\bibfield  {title} {\bibinfo {title} {Theory of the
  nonreciprocal {Josephson} effect},\ }\href
  {https://doi.org/10.1103/physrevb.103.245302} {\bibfield  {journal} {\bibinfo
   {journal} {Phys. Rev. B}\ }\textbf {\bibinfo {volume} {103}},\ \bibinfo
  {pages} {245302} (\bibinfo {year} {2021})}\BibitemShut {NoStop}%
\bibitem [{\citenamefont {Smidman}\ \emph {et~al.}(2017)\citenamefont
  {Smidman}, \citenamefont {Salamon}, \citenamefont {Yuan},\ and\ \citenamefont
  {Agterberg}}]{Smidman2017}%
  \BibitemOpen
  \bibfield  {author} {\bibinfo {author} {\bibfnamefont {M.}~\bibnamefont
  {Smidman}}, \bibinfo {author} {\bibfnamefont {M.~B.}\ \bibnamefont
  {Salamon}}, \bibinfo {author} {\bibfnamefont {H.~Q.}\ \bibnamefont {Yuan}},\
  and\ \bibinfo {author} {\bibfnamefont {D.~F.}\ \bibnamefont {Agterberg}},\
  }\bibfield  {title} {\bibinfo {title} {Superconductivity and spin-orbit
  coupling in non-centrosymmetric materials: a review},\ }\href
  {https://doi.org/10.1088/1361-6633/80/3/036501} {\bibfield  {journal}
  {\bibinfo  {journal} {Rep. Prog. Phys.}\ }\textbf {\bibinfo {volume} {80}},\
  \bibinfo {pages} {036501} (\bibinfo {year} {2017})}\BibitemShut {NoStop}%
\bibitem [{\citenamefont {Legg}\ \emph {et~al.}(2022)\citenamefont {Legg},
  \citenamefont {Loss},\ and\ \citenamefont {Klinovaja}}]{Legg2022}%
  \BibitemOpen
  \bibfield  {author} {\bibinfo {author} {\bibfnamefont {H.~F.}\ \bibnamefont
  {Legg}}, \bibinfo {author} {\bibfnamefont {D.}~\bibnamefont {Loss}},\ and\
  \bibinfo {author} {\bibfnamefont {J.}~\bibnamefont {Klinovaja}},\ }\bibfield
  {title} {\bibinfo {title} {Superconducting diode effect due to magnetochiral
  anisotropy in topological insulator and {Rashba} nanowires},\ }\href
  {https://doi.org/10.1103/physrevb.106.104501} {\bibfield  {journal} {\bibinfo
   {journal} {Phys. Rev. B}\ }\textbf {\bibinfo {volume} {106}},\ \bibinfo
  {pages} {104501} (\bibinfo {year} {2022})}\BibitemShut {NoStop}%
\bibitem [{\citenamefont {Angehrn}\ \emph {et~al.}(2024)\citenamefont
  {Angehrn}, \citenamefont {Legg}, \citenamefont {Loss},\ and\ \citenamefont
  {Klinovaja}}]{Angehrn2024}%
  \BibitemOpen
  \bibfield  {author} {\bibinfo {author} {\bibfnamefont {L.}~\bibnamefont
  {Angehrn}}, \bibinfo {author} {\bibfnamefont {H.~F.}\ \bibnamefont {Legg}},
  \bibinfo {author} {\bibfnamefont {D.}~\bibnamefont {Loss}},\ and\ \bibinfo
  {author} {\bibfnamefont {J.}~\bibnamefont {Klinovaja}},\ }\bibfield  {title}
  {\bibinfo {title} {Relations between normal state nonreciprocal transport and
  the superconducting diode effect in the trivial and topological phases},\
  }\href {https://doi.org/10.1063/5.0216679} {\bibfield  {journal} {\bibinfo
  {journal} {Appl. Phys. Lett.}\ }\textbf {\bibinfo {volume} {125}},\ \bibinfo
  {pages} {023507} (\bibinfo {year} {2024})}\BibitemShut {NoStop}%
\bibitem [{\citenamefont {de~Picoli}\ \emph {et~al.}(2023)\citenamefont
  {de~Picoli}, \citenamefont {Blood}, \citenamefont {Lyanda-Geller},\ and\
  \citenamefont {V\"{a}yrynen}}]{dePicoli2023}%
  \BibitemOpen
  \bibfield  {author} {\bibinfo {author} {\bibfnamefont {T.}~\bibnamefont
  {de~Picoli}}, \bibinfo {author} {\bibfnamefont {Z.}~\bibnamefont {Blood}},
  \bibinfo {author} {\bibfnamefont {Y.}~\bibnamefont {Lyanda-Geller}},\ and\
  \bibinfo {author} {\bibfnamefont {J.~I.}\ \bibnamefont {V\"{a}yrynen}},\
  }\bibfield  {title} {\bibinfo {title} {Superconducting diode effect in
  quasi-one-dimensional systems},\ }\href
  {https://doi.org/10.1103/physrevb.107.224518} {\bibfield  {journal} {\bibinfo
   {journal} {Phys. Rev. B}\ }\textbf {\bibinfo {volume} {107}},\ \bibinfo
  {pages} {224518} (\bibinfo {year} {2023})}\BibitemShut {NoStop}%
\bibitem [{\citenamefont {Huang}\ \emph {et~al.}(2024)\citenamefont {Huang},
  \citenamefont {de~Picoli},\ and\ \citenamefont {V\"{a}yrynen}}]{Huang2024}%
  \BibitemOpen
  \bibfield  {author} {\bibinfo {author} {\bibfnamefont {J.}~\bibnamefont
  {Huang}}, \bibinfo {author} {\bibfnamefont {T.}~\bibnamefont {de~Picoli}},\
  and\ \bibinfo {author} {\bibfnamefont {J.~I.}\ \bibnamefont {V\"{a}yrynen}},\
  }\bibfield  {title} {\bibinfo {title} {Superconducting diode effect in
  two-dimensional topological insulator edges and {Josephson} junctions},\
  }\href {https://doi.org/10.1063/5.0213137} {\bibfield  {journal} {\bibinfo
  {journal} {Appl. Phys. Lett.}\ }\textbf {\bibinfo {volume} {125}},\ \bibinfo
  {pages} {032602} (\bibinfo {year} {2024})}\BibitemShut {NoStop}%
\bibitem [{\citenamefont {Lu}\ \emph {et~al.}(2023)\citenamefont {Lu},
  \citenamefont {Ikegaya}, \citenamefont {Burset}, \citenamefont {Tanaka},\
  and\ \citenamefont {Nagaosa}}]{Lu2023}%
  \BibitemOpen
  \bibfield  {author} {\bibinfo {author} {\bibfnamefont {B.}~\bibnamefont
  {Lu}}, \bibinfo {author} {\bibfnamefont {S.}~\bibnamefont {Ikegaya}},
  \bibinfo {author} {\bibfnamefont {P.}~\bibnamefont {Burset}}, \bibinfo
  {author} {\bibfnamefont {Y.}~\bibnamefont {Tanaka}},\ and\ \bibinfo {author}
  {\bibfnamefont {N.}~\bibnamefont {Nagaosa}},\ }\bibfield  {title} {\bibinfo
  {title} {Tunable {Josephson} diode effect on the surface of topological
  insulators},\ }\href {https://doi.org/10.1103/physrevlett.131.096001}
  {\bibfield  {journal} {\bibinfo  {journal} {Phys. Rev. Lett.}\ }\textbf
  {\bibinfo {volume} {131}},\ \bibinfo {pages} {096001} (\bibinfo {year}
  {2023})}\BibitemShut {NoStop}%
\bibitem [{\citenamefont {Cayao}\ \emph {et~al.}(2024)\citenamefont {Cayao},
  \citenamefont {Nagaosa},\ and\ \citenamefont {Tanaka}}]{Cayao2024}%
  \BibitemOpen
  \bibfield  {author} {\bibinfo {author} {\bibfnamefont {J.}~\bibnamefont
  {Cayao}}, \bibinfo {author} {\bibfnamefont {N.}~\bibnamefont {Nagaosa}},\
  and\ \bibinfo {author} {\bibfnamefont {Y.}~\bibnamefont {Tanaka}},\
  }\bibfield  {title} {\bibinfo {title} {Enhancing the {Josephson} diode effect
  with {Majorana} bound states},\ }\href
  {https://doi.org/10.1103/physrevb.109.l081405} {\bibfield  {journal}
  {\bibinfo  {journal} {Phys. Rev. B}\ }\textbf {\bibinfo {volume} {109}},\
  \bibinfo {pages} {L081405} (\bibinfo {year} {2024})}\BibitemShut {NoStop}%
\bibitem [{\citenamefont {Liu}\ \emph {et~al.}(2024)\citenamefont {Liu},
  \citenamefont {Huang},\ and\ \citenamefont {Wang}}]{LiuHuangWang2024}%
  \BibitemOpen
  \bibfield  {author} {\bibinfo {author} {\bibfnamefont {Z.}~\bibnamefont
  {Liu}}, \bibinfo {author} {\bibfnamefont {L.}~\bibnamefont {Huang}},\ and\
  \bibinfo {author} {\bibfnamefont {J.}~\bibnamefont {Wang}},\ }\bibfield
  {title} {\bibinfo {title} {Josephson diode effect in topological
  superconductors},\ }\href {https://doi.org/10.1103/physrevb.110.014519}
  {\bibfield  {journal} {\bibinfo  {journal} {Phys. Rev. B}\ }\textbf {\bibinfo
  {volume} {110}},\ \bibinfo {pages} {014519} (\bibinfo {year}
  {2024})}\BibitemShut {NoStop}%
\bibitem [{\citenamefont {Yu}(1965)}]{Yu1965}%
  \BibitemOpen
  \bibfield  {author} {\bibinfo {author} {\bibfnamefont {L.}~\bibnamefont
  {Yu}},\ }\bibfield  {title} {\bibinfo {title} {Bound state in superconductors
  with paramagnetic impurities},\ }\href {https://doi.org/10.7498/aps.21.75}
  {\bibfield  {journal} {\bibinfo  {journal} {Acta Phys. Sin.}\ }\textbf
  {\bibinfo {volume} {21}},\ \bibinfo {pages} {75} (\bibinfo {year}
  {1965})}\BibitemShut {NoStop}%
\bibitem [{\citenamefont {Shiba}(1968)}]{Shiba1968}%
  \BibitemOpen
  \bibfield  {author} {\bibinfo {author} {\bibfnamefont {H.}~\bibnamefont
  {Shiba}},\ }\bibfield  {title} {\bibinfo {title} {Classical spins in
  superconductors},\ }\href {https://doi.org/10.1143/ptp.40.435} {\bibfield
  {journal} {\bibinfo  {journal} {Prog. Theor. Phys.}\ }\textbf {\bibinfo
  {volume} {40}},\ \bibinfo {pages} {435} (\bibinfo {year} {1968})}\BibitemShut
  {NoStop}%
\bibitem [{\citenamefont {Rusinov}(1969)}]{Rusinov1969}%
  \BibitemOpen
  \bibfield  {author} {\bibinfo {author} {\bibfnamefont {A.~I.}\ \bibnamefont
  {Rusinov}},\ }\bibfield  {title} {\bibinfo {title} {Superconductivity near a
  paramagnetic impurity},\ }\href@noop {} {\bibfield  {journal} {\bibinfo
  {journal} {JETP Lett.}\ }\textbf {\bibinfo {volume} {9}},\ \bibinfo {pages}
  {85} (\bibinfo {year} {1969})}\BibitemShut {NoStop}%
\bibitem [{\citenamefont {Choy}\ \emph {et~al.}(2011)\citenamefont {Choy},
  \citenamefont {Edge}, \citenamefont {Akhmerov},\ and\ \citenamefont
  {Beenakker}}]{Choy2011}%
  \BibitemOpen
  \bibfield  {author} {\bibinfo {author} {\bibfnamefont {T.-P.}\ \bibnamefont
  {Choy}}, \bibinfo {author} {\bibfnamefont {J.~M.}\ \bibnamefont {Edge}},
  \bibinfo {author} {\bibfnamefont {A.~R.}\ \bibnamefont {Akhmerov}},\ and\
  \bibinfo {author} {\bibfnamefont {C.~W.~J.}\ \bibnamefont {Beenakker}},\
  }\bibfield  {title} {\bibinfo {title} {Majorana fermions emerging from
  magnetic nanoparticles on a superconductor without spin-orbit coupling},\
  }\href {https://doi.org/10.1103/physrevb.84.195442} {\bibfield  {journal}
  {\bibinfo  {journal} {Phys. Rev. B}\ }\textbf {\bibinfo {volume} {84}},\
  \bibinfo {pages} {195442} (\bibinfo {year} {2011})}\BibitemShut {NoStop}%
\bibitem [{\citenamefont {Nadj-Perge}\ \emph {et~al.}(2013)\citenamefont
  {Nadj-Perge}, \citenamefont {Drozdov}, \citenamefont {Bernevig},\ and\
  \citenamefont {Yazdani}}]{NadjPerge2013}%
  \BibitemOpen
  \bibfield  {author} {\bibinfo {author} {\bibfnamefont {S.}~\bibnamefont
  {Nadj-Perge}}, \bibinfo {author} {\bibfnamefont {I.~K.}\ \bibnamefont
  {Drozdov}}, \bibinfo {author} {\bibfnamefont {B.~A.}\ \bibnamefont
  {Bernevig}},\ and\ \bibinfo {author} {\bibfnamefont {A.}~\bibnamefont
  {Yazdani}},\ }\bibfield  {title} {\bibinfo {title} {Proposal for realizing
  {Majorana} fermions in chains of magnetic atoms on a superconductor},\ }\href
  {https://doi.org/10.1103/physrevb.88.020407} {\bibfield  {journal} {\bibinfo
  {journal} {Phys. Rev. B}\ }\textbf {\bibinfo {volume} {88}},\ \bibinfo
  {pages} {020407} (\bibinfo {year} {2013})}\BibitemShut {NoStop}%
\bibitem [{\citenamefont {Klinovaja}\ \emph {et~al.}(2013)\citenamefont
  {Klinovaja}, \citenamefont {Stano}, \citenamefont {Yazdani},\ and\
  \citenamefont {Loss}}]{Klinovaja2013}%
  \BibitemOpen
  \bibfield  {author} {\bibinfo {author} {\bibfnamefont {J.}~\bibnamefont
  {Klinovaja}}, \bibinfo {author} {\bibfnamefont {P.}~\bibnamefont {Stano}},
  \bibinfo {author} {\bibfnamefont {A.}~\bibnamefont {Yazdani}},\ and\ \bibinfo
  {author} {\bibfnamefont {D.}~\bibnamefont {Loss}},\ }\bibfield  {title}
  {\bibinfo {title} {Topological superconductivity and {Majorana} fermions in
  {RKKY} systems},\ }\href {https://doi.org/10.1103/physrevlett.111.186805}
  {\bibfield  {journal} {\bibinfo  {journal} {Phys. Rev. Lett.}\ }\textbf
  {\bibinfo {volume} {111}},\ \bibinfo {pages} {186805} (\bibinfo {year}
  {2013})}\BibitemShut {NoStop}%
\bibitem [{\citenamefont {Braunecker}\ and\ \citenamefont
  {Simon}(2013)}]{Braunecker2013}%
  \BibitemOpen
  \bibfield  {author} {\bibinfo {author} {\bibfnamefont {B.}~\bibnamefont
  {Braunecker}}\ and\ \bibinfo {author} {\bibfnamefont {P.}~\bibnamefont
  {Simon}},\ }\bibfield  {title} {\bibinfo {title} {Interplay between classical
  magnetic moments and superconductivity in quantum one-dimensional conductors:
  toward a self-sustained topological {Majorana} phase},\ }\href
  {https://doi.org/10.1103/physrevlett.111.147202} {\bibfield  {journal}
  {\bibinfo  {journal} {Phys. Rev. Lett.}\ }\textbf {\bibinfo {volume} {111}},\
  \bibinfo {pages} {147202} (\bibinfo {year} {2013})}\BibitemShut {NoStop}%
\bibitem [{\citenamefont {Vazifeh}\ and\ \citenamefont
  {Franz}(2013)}]{Vazifeh2013}%
  \BibitemOpen
  \bibfield  {author} {\bibinfo {author} {\bibfnamefont {M.~M.}\ \bibnamefont
  {Vazifeh}}\ and\ \bibinfo {author} {\bibfnamefont {M.}~\bibnamefont
  {Franz}},\ }\bibfield  {title} {\bibinfo {title} {Self-organized topological
  state with {Majorana} fermions},\ }\href
  {https://doi.org/10.1103/physrevlett.111.206802} {\bibfield  {journal}
  {\bibinfo  {journal} {Phys. Rev. Lett.}\ }\textbf {\bibinfo {volume} {111}},\
  \bibinfo {pages} {206802} (\bibinfo {year} {2013})}\BibitemShut {NoStop}%
\bibitem [{\citenamefont {Pientka}\ \emph {et~al.}(2013)\citenamefont
  {Pientka}, \citenamefont {Glazman},\ and\ \citenamefont {von
  Oppen}}]{Pientka2013}%
  \BibitemOpen
  \bibfield  {author} {\bibinfo {author} {\bibfnamefont {F.}~\bibnamefont
  {Pientka}}, \bibinfo {author} {\bibfnamefont {L.~I.}\ \bibnamefont
  {Glazman}},\ and\ \bibinfo {author} {\bibfnamefont {F.}~\bibnamefont {von
  Oppen}},\ }\bibfield  {title} {\bibinfo {title} {Topological superconducting
  phase in helical {Shiba} chains},\ }\href
  {https://doi.org/10.1103/physrevb.88.155420} {\bibfield  {journal} {\bibinfo
  {journal} {Phys. Rev. B}\ }\textbf {\bibinfo {volume} {88}},\ \bibinfo
  {pages} {155420} (\bibinfo {year} {2013})}\BibitemShut {NoStop}%
\bibitem [{\citenamefont {Nadj-Perge}\ \emph {et~al.}(2014)\citenamefont
  {Nadj-Perge}, \citenamefont {Drozdov}, \citenamefont {Li}, \citenamefont
  {Chen}, \citenamefont {Jeon}, \citenamefont {Seo}, \citenamefont {MacDonald},
  \citenamefont {Bernevig},\ and\ \citenamefont {Yazdani}}]{NadjPerge2014}%
  \BibitemOpen
  \bibfield  {author} {\bibinfo {author} {\bibfnamefont {S.}~\bibnamefont
  {Nadj-Perge}}, \bibinfo {author} {\bibfnamefont {I.~K.}\ \bibnamefont
  {Drozdov}}, \bibinfo {author} {\bibfnamefont {J.}~\bibnamefont {Li}},
  \bibinfo {author} {\bibfnamefont {H.}~\bibnamefont {Chen}}, \bibinfo {author}
  {\bibfnamefont {S.}~\bibnamefont {Jeon}}, \bibinfo {author} {\bibfnamefont
  {J.}~\bibnamefont {Seo}}, \bibinfo {author} {\bibfnamefont {A.~H.}\
  \bibnamefont {MacDonald}}, \bibinfo {author} {\bibfnamefont {B.~A.}\
  \bibnamefont {Bernevig}},\ and\ \bibinfo {author} {\bibfnamefont
  {A.}~\bibnamefont {Yazdani}},\ }\bibfield  {title} {\bibinfo {title}
  {Observation of {Majorana} fermions in ferromagnetic atomic chains on a
  superconductor},\ }\href {https://doi.org/10.1126/science.1259327} {\bibfield
   {journal} {\bibinfo  {journal} {Science}\ }\textbf {\bibinfo {volume}
  {346}},\ \bibinfo {pages} {602} (\bibinfo {year} {2014})}\BibitemShut
  {NoStop}%
\bibitem [{\citenamefont {Ruby}\ \emph {et~al.}(2017)\citenamefont {Ruby},
  \citenamefont {Heinrich}, \citenamefont {Peng}, \citenamefont {von Oppen},\
  and\ \citenamefont {Franke}}]{Ruby2017}%
  \BibitemOpen
  \bibfield  {author} {\bibinfo {author} {\bibfnamefont {M.}~\bibnamefont
  {Ruby}}, \bibinfo {author} {\bibfnamefont {B.~W.}\ \bibnamefont {Heinrich}},
  \bibinfo {author} {\bibfnamefont {Y.}~\bibnamefont {Peng}}, \bibinfo {author}
  {\bibfnamefont {F.}~\bibnamefont {von Oppen}},\ and\ \bibinfo {author}
  {\bibfnamefont {K.~J.}\ \bibnamefont {Franke}},\ }\bibfield  {title}
  {\bibinfo {title} {Exploring a proximity-coupled {Co} chain on {Pb}(110) as a
  possible {Majorana} platform},\ }\href
  {https://doi.org/10.1021/acs.nanolett.7b01728} {\bibfield  {journal}
  {\bibinfo  {journal} {Nano Lett.}\ }\textbf {\bibinfo {volume} {17}},\
  \bibinfo {pages} {4473} (\bibinfo {year} {2017})}\BibitemShut {NoStop}%
\bibitem [{\citenamefont {Kim}\ \emph {et~al.}(2018)\citenamefont {Kim},
  \citenamefont {Palacio-Morales}, \citenamefont {Posske}, \citenamefont
  {R\'{o}zsa}, \citenamefont {Palot\'{a}s}, \citenamefont {Szunyogh},
  \citenamefont {Thorwart},\ and\ \citenamefont {Wiesendanger}}]{Kim2018}%
  \BibitemOpen
  \bibfield  {author} {\bibinfo {author} {\bibfnamefont {H.}~\bibnamefont
  {Kim}}, \bibinfo {author} {\bibfnamefont {A.}~\bibnamefont
  {Palacio-Morales}}, \bibinfo {author} {\bibfnamefont {T.}~\bibnamefont
  {Posske}}, \bibinfo {author} {\bibfnamefont {L.}~\bibnamefont {R\'{o}zsa}},
  \bibinfo {author} {\bibfnamefont {K.}~\bibnamefont {Palot\'{a}s}}, \bibinfo
  {author} {\bibfnamefont {L.}~\bibnamefont {Szunyogh}}, \bibinfo {author}
  {\bibfnamefont {M.}~\bibnamefont {Thorwart}},\ and\ \bibinfo {author}
  {\bibfnamefont {R.}~\bibnamefont {Wiesendanger}},\ }\bibfield  {title}
  {\bibinfo {title} {Toward tailoring {Majorana} bound states in artificially
  constructed magnetic atom chains on elemental superconductors},\ }\href
  {https://doi.org/10.1126/sciadv.aar5251} {\bibfield  {journal} {\bibinfo
  {journal} {Sci. Adv.}\ }\textbf {\bibinfo {volume} {4}},\ \bibinfo {pages}
  {eaar5251} (\bibinfo {year} {2018})}\BibitemShut {NoStop}%
\bibitem [{\citenamefont {Schneider}\ \emph {et~al.}(2021)\citenamefont
  {Schneider}, \citenamefont {Beck}, \citenamefont {Posske}, \citenamefont
  {Crawford}, \citenamefont {Mascot}, \citenamefont {Rachel}, \citenamefont
  {Wiesendanger},\ and\ \citenamefont {Wiebe}}]{Schneider2021}%
  \BibitemOpen
  \bibfield  {author} {\bibinfo {author} {\bibfnamefont {L.}~\bibnamefont
  {Schneider}}, \bibinfo {author} {\bibfnamefont {P.}~\bibnamefont {Beck}},
  \bibinfo {author} {\bibfnamefont {T.}~\bibnamefont {Posske}}, \bibinfo
  {author} {\bibfnamefont {D.}~\bibnamefont {Crawford}}, \bibinfo {author}
  {\bibfnamefont {E.}~\bibnamefont {Mascot}}, \bibinfo {author} {\bibfnamefont
  {S.}~\bibnamefont {Rachel}}, \bibinfo {author} {\bibfnamefont
  {R.}~\bibnamefont {Wiesendanger}},\ and\ \bibinfo {author} {\bibfnamefont
  {J.}~\bibnamefont {Wiebe}},\ }\bibfield  {title} {\bibinfo {title}
  {Topological {Shiba} bands in artificial spin chains on superconductors},\
  }\href {https://doi.org/10.1038/s41567-021-01234-y} {\bibfield  {journal}
  {\bibinfo  {journal} {Nat. Phys.}\ }\textbf {\bibinfo {volume} {17}},\
  \bibinfo {pages} {943} (\bibinfo {year} {2021})}\BibitemShut {NoStop}%
\bibitem [{\citenamefont {Fulde}\ and\ \citenamefont
  {Ferrell}(1964)}]{Fulde1964}%
  \BibitemOpen
  \bibfield  {author} {\bibinfo {author} {\bibfnamefont {P.}~\bibnamefont
  {Fulde}}\ and\ \bibinfo {author} {\bibfnamefont {R.~A.}\ \bibnamefont
  {Ferrell}},\ }\bibfield  {title} {\bibinfo {title} {Superconductivity in a
  strong spin-exchange field},\ }\href
  {https://doi.org/10.1103/physrev.135.a550} {\bibfield  {journal} {\bibinfo
  {journal} {Phys. Rev.}\ }\textbf {\bibinfo {volume} {135}},\ \bibinfo {pages}
  {A550} (\bibinfo {year} {1964})}\BibitemShut {NoStop}%
\bibitem [{\citenamefont {Larkin}\ and\ \citenamefont
  {Ovchinnikov}(1965)}]{Larkin1965}%
  \BibitemOpen
  \bibfield  {author} {\bibinfo {author} {\bibfnamefont {A.~I.}\ \bibnamefont
  {Larkin}}\ and\ \bibinfo {author} {\bibfnamefont {Y.~N.}\ \bibnamefont
  {Ovchinnikov}},\ }\bibfield  {title} {\bibinfo {title} {Nonuniform state of
  superconductors},\ }\href@noop {} {\bibfield  {journal} {\bibinfo  {journal}
  {Sov. Phys. JETP}\ }\textbf {\bibinfo {volume} {20}},\ \bibinfo {pages} {762}
  (\bibinfo {year} {1965})}\BibitemShut {NoStop}%
\bibitem [{\citenamefont {Bhowmik}\ and\ \citenamefont
  {Saha}(2025)}]{Bhowmik2025}%
  \BibitemOpen
  \bibfield  {author} {\bibinfo {author} {\bibfnamefont {S.}~\bibnamefont
  {Bhowmik}}\ and\ \bibinfo {author} {\bibfnamefont {A.}~\bibnamefont {Saha}},\
  }\bibfield  {title} {\bibinfo {title} {Topological {Majorana} zero modes and
  superconducting diode effect driven by {Fulde-Ferrell-Larkin-Ovchinnikov}
  pairing in a helical {Shiba} chain},\ }\href
  {https://doi.org/10.1103/physrevb.111.l161402} {\bibfield  {journal}
  {\bibinfo  {journal} {Phys. Rev. B}\ }\textbf {\bibinfo {volume} {111}},\
  \bibinfo {pages} {L161402} (\bibinfo {year} {2025})}\BibitemShut {NoStop}%
\bibitem [{\citenamefont {Bhowmik}\ \emph {et~al.}(2025)\citenamefont
  {Bhowmik}, \citenamefont {Samanta}, \citenamefont {Nandy}, \citenamefont
  {Saha},\ and\ \citenamefont {Ghosh}}]{Bhowmik2025b}%
  \BibitemOpen
  \bibfield  {author} {\bibinfo {author} {\bibfnamefont {S.}~\bibnamefont
  {Bhowmik}}, \bibinfo {author} {\bibfnamefont {S.}~\bibnamefont {Samanta}},
  \bibinfo {author} {\bibfnamefont {A.}~\bibnamefont {Nandy}}, \bibinfo
  {author} {\bibfnamefont {A.}~\bibnamefont {Saha}},\ and\ \bibinfo {author}
  {\bibfnamefont {A.}~\bibnamefont {Ghosh}},\ }\bibfield  {title} {\bibinfo
  {title} {Optimizing one-dimensional superconducting diodes: interplay of
  {Rashba} spin-orbit coupling and magnetic field},\ }\href
  {https://doi.org/10.1038/s42005-025-02044-x} {\bibfield  {journal} {\bibinfo
  {journal} {Commun. Phys.}\ }\textbf {\bibinfo {volume} {8}},\ \bibinfo
  {pages} {260} (\bibinfo {year} {2025})}\BibitemShut {NoStop}%
\bibitem [{\citenamefont {Samanta}\ and\ \citenamefont
  {Ghosh}(2025)}]{Samanta2025}%
  \BibitemOpen
  \bibfield  {author} {\bibinfo {author} {\bibfnamefont {S.}~\bibnamefont
  {Samanta}}\ and\ \bibinfo {author} {\bibfnamefont {A.}~\bibnamefont
  {Ghosh}},\ }\bibfield  {title} {\bibinfo {title} {Field-free superconducting
  diode effect and topological {Fulde-Ferrell-Larkin-Ovchinnikov}
  superconductivity in altermagnetic {Shiba} chains},\ }\href@noop {}
  {\bibfield  {journal} {\bibinfo  {journal}
  {\href{https://arxiv.org/abs/2507.21446}{arXiv:2507.21446}}\ } (\bibinfo
  {year} {2025})}\BibitemShut {NoStop}%
\bibitem [{\citenamefont {Pal}\ \emph {et~al.}(2025)\citenamefont {Pal},
  \citenamefont {Mondal}, \citenamefont {Nag},\ and\ \citenamefont
  {Saha}}]{Pal2025}%
  \BibitemOpen
  \bibfield  {author} {\bibinfo {author} {\bibfnamefont {S.}~\bibnamefont
  {Pal}}, \bibinfo {author} {\bibfnamefont {S.}~\bibnamefont {Mondal}},
  \bibinfo {author} {\bibfnamefont {T.}~\bibnamefont {Nag}},\ and\ \bibinfo
  {author} {\bibfnamefont {A.}~\bibnamefont {Saha}},\ }\bibfield  {title}
  {\bibinfo {title} {Topological superconductivity and superconducting diode
  effect mediated via unconventional magnet and {Ising} spin-orbit coupling},\
  }\href@noop {} {\bibfield  {journal} {\bibinfo  {journal}
  {\href{https://arxiv.org/abs/2512.01266}{arXiv:2512.01266}}\ } (\bibinfo
  {year} {2025})}\BibitemShut {NoStop}%
\bibitem [{\citenamefont {Santra}\ \emph {et~al.}(2026)\citenamefont {Santra},
  \citenamefont {Samanta},\ and\ \citenamefont {Ghosh}}]{Santra2026}%
  \BibitemOpen
  \bibfield  {author} {\bibinfo {author} {\bibfnamefont {S.}~\bibnamefont
  {Santra}}, \bibinfo {author} {\bibfnamefont {S.}~\bibnamefont {Samanta}},\
  and\ \bibinfo {author} {\bibfnamefont {A.}~\bibnamefont {Ghosh}},\ }\bibfield
   {title} {\bibinfo {title} {Superconducting diode effect in multichannel
  {Majorana} wires},\ }\href@noop {} {\bibfield  {journal} {\bibinfo  {journal}
  {\href{https://arxiv.org/abs/2603.01207}{arXiv:2603.01207}}\ } (\bibinfo
  {year} {2026})}\BibitemShut {NoStop}%
\bibitem [{\citenamefont {Bhowmik}\ \emph {et~al.}(2026)\citenamefont
  {Bhowmik}, \citenamefont {Saha},\ and\ \citenamefont {Nag}}]{Bhowmik2026}%
  \BibitemOpen
  \bibfield  {author} {\bibinfo {author} {\bibfnamefont {S.}~\bibnamefont
  {Bhowmik}}, \bibinfo {author} {\bibfnamefont {A.}~\bibnamefont {Saha}},\ and\
  \bibinfo {author} {\bibfnamefont {T.}~\bibnamefont {Nag}},\ }\bibfield
  {title} {\bibinfo {title} {Superconducting diode effect via {Floquet}
  topological {Fulde-Ferrell} phase in a driven {Rashba} nanowire},\
  }\href@noop {} {\bibfield  {journal} {\bibinfo  {journal}
  {\href{https://arxiv.org/abs/2606.16459}{arXiv:2606.16459}}\ } (\bibinfo
  {year} {2026})}\BibitemShut {NoStop}%
\bibitem [{\citenamefont {Oka}\ and\ \citenamefont {Aoki}(2009)}]{Oka2009}%
  \BibitemOpen
  \bibfield  {author} {\bibinfo {author} {\bibfnamefont {T.}~\bibnamefont
  {Oka}}\ and\ \bibinfo {author} {\bibfnamefont {H.}~\bibnamefont {Aoki}},\
  }\bibfield  {title} {\bibinfo {title} {Photovoltaic {Hall} effect in
  graphene},\ }\href {https://doi.org/10.1103/physrevb.79.081406} {\bibfield
  {journal} {\bibinfo  {journal} {Phys. Rev. B}\ }\textbf {\bibinfo {volume}
  {79}},\ \bibinfo {pages} {081406} (\bibinfo {year} {2009})}\BibitemShut
  {NoStop}%
\bibitem [{\citenamefont {Lindner}\ \emph {et~al.}(2011)\citenamefont
  {Lindner}, \citenamefont {Refael},\ and\ \citenamefont
  {Galitski}}]{Lindner2011}%
  \BibitemOpen
  \bibfield  {author} {\bibinfo {author} {\bibfnamefont {N.~H.}\ \bibnamefont
  {Lindner}}, \bibinfo {author} {\bibfnamefont {G.}~\bibnamefont {Refael}},\
  and\ \bibinfo {author} {\bibfnamefont {V.}~\bibnamefont {Galitski}},\
  }\bibfield  {title} {\bibinfo {title} {Floquet topological insulator in
  semiconductor quantum wells},\ }\href {https://doi.org/10.1038/nphys1926}
  {\bibfield  {journal} {\bibinfo  {journal} {Nat. Phys.}\ }\textbf {\bibinfo
  {volume} {7}},\ \bibinfo {pages} {490} (\bibinfo {year} {2011})}\BibitemShut
  {NoStop}%
\bibitem [{\citenamefont {Kitagawa}\ \emph {et~al.}(2010)\citenamefont
  {Kitagawa}, \citenamefont {Berg}, \citenamefont {Rudner},\ and\ \citenamefont
  {Demler}}]{Kitagawa2010}%
  \BibitemOpen
  \bibfield  {author} {\bibinfo {author} {\bibfnamefont {T.}~\bibnamefont
  {Kitagawa}}, \bibinfo {author} {\bibfnamefont {E.}~\bibnamefont {Berg}},
  \bibinfo {author} {\bibfnamefont {M.}~\bibnamefont {Rudner}},\ and\ \bibinfo
  {author} {\bibfnamefont {E.}~\bibnamefont {Demler}},\ }\bibfield  {title}
  {\bibinfo {title} {Topological characterization of periodically driven
  quantum systems},\ }\href {https://doi.org/10.1103/physrevb.82.235114}
  {\bibfield  {journal} {\bibinfo  {journal} {Phys. Rev. B}\ }\textbf {\bibinfo
  {volume} {82}},\ \bibinfo {pages} {235114} (\bibinfo {year}
  {2010})}\BibitemShut {NoStop}%
\bibitem [{\citenamefont {Rudner}\ and\ \citenamefont
  {Lindner}(2020)}]{Rudner2020}%
  \BibitemOpen
  \bibfield  {author} {\bibinfo {author} {\bibfnamefont {M.~S.}\ \bibnamefont
  {Rudner}}\ and\ \bibinfo {author} {\bibfnamefont {N.~H.}\ \bibnamefont
  {Lindner}},\ }\bibfield  {title} {\bibinfo {title} {Band structure
  engineering and non-equilibrium dynamics in {Floquet} topological
  insulators},\ }\href {https://doi.org/10.1038/s42254-020-0170-z} {\bibfield
  {journal} {\bibinfo  {journal} {Nat. Rev. Phys.}\ }\textbf {\bibinfo {volume}
  {2}},\ \bibinfo {pages} {229} (\bibinfo {year} {2020})}\BibitemShut {NoStop}%
\bibitem [{\citenamefont {Eckardt}(2017)}]{Eckardt2017}%
  \BibitemOpen
  \bibfield  {author} {\bibinfo {author} {\bibfnamefont {A.}~\bibnamefont
  {Eckardt}},\ }\bibfield  {title} {\bibinfo {title} {Colloquium: Atomic
  quantum gases in periodically driven optical lattices},\ }\href
  {https://doi.org/10.1103/revmodphys.89.011004} {\bibfield  {journal}
  {\bibinfo  {journal} {Rev. Mod. Phys.}\ }\textbf {\bibinfo {volume} {89}},\
  \bibinfo {pages} {011004} (\bibinfo {year} {2017})}\BibitemShut {NoStop}%
\bibitem [{\citenamefont {Bukov}\ \emph {et~al.}(2015)\citenamefont {Bukov},
  \citenamefont {D'Alessio},\ and\ \citenamefont {Polkovnikov}}]{Bukov2015}%
  \BibitemOpen
  \bibfield  {author} {\bibinfo {author} {\bibfnamefont {M.}~\bibnamefont
  {Bukov}}, \bibinfo {author} {\bibfnamefont {L.}~\bibnamefont {D'Alessio}},\
  and\ \bibinfo {author} {\bibfnamefont {A.}~\bibnamefont {Polkovnikov}},\
  }\bibfield  {title} {\bibinfo {title} {Universal high-frequency behavior of
  periodically driven systems: from dynamical stabilization to {Floquet}
  engineering},\ }\href {https://doi.org/10.1080/00018732.2015.1055918}
  {\bibfield  {journal} {\bibinfo  {journal} {Adv. Phys.}\ }\textbf {\bibinfo
  {volume} {64}},\ \bibinfo {pages} {139} (\bibinfo {year} {2015})}\BibitemShut
  {NoStop}%
\bibitem [{\citenamefont {Goldman}\ and\ \citenamefont
  {Dalibard}(2014)}]{Goldman2014}%
  \BibitemOpen
  \bibfield  {author} {\bibinfo {author} {\bibfnamefont {N.}~\bibnamefont
  {Goldman}}\ and\ \bibinfo {author} {\bibfnamefont {J.}~\bibnamefont
  {Dalibard}},\ }\bibfield  {title} {\bibinfo {title} {Periodically driven
  quantum systems: effective {Hamiltonians} and engineered gauge fields},\
  }\href {https://doi.org/10.1103/physrevx.4.031027} {\bibfield  {journal}
  {\bibinfo  {journal} {Phys. Rev. X}\ }\textbf {\bibinfo {volume} {4}},\
  \bibinfo {pages} {031027} (\bibinfo {year} {2014})}\BibitemShut {NoStop}%
\bibitem [{\citenamefont {Roy}\ and\ \citenamefont
  {Basu}(2023)}]{Roy2023creutz}%
  \BibitemOpen
  \bibfield  {author} {\bibinfo {author} {\bibfnamefont {K.}~\bibnamefont
  {Roy}}\ and\ \bibinfo {author} {\bibfnamefont {S.}~\bibnamefont {Basu}},\
  }\bibfield  {title} {\bibinfo {title} {Topological properties of a
  periodically driven creutz ladder},\ }\href
  {https://doi.org/10.1103/PhysRevB.108.045415} {\bibfield  {journal} {\bibinfo
   {journal} {Phys. Rev. B}\ }\textbf {\bibinfo {volume} {108}},\ \bibinfo
  {pages} {045415} (\bibinfo {year} {2023})}\BibitemShut {NoStop}%
\bibitem [{\citenamefont {Roy}\ \emph {et~al.}(2025{\natexlab{a}})\citenamefont
  {Roy}, \citenamefont {Shahab},\ and\ \citenamefont
  {Basu}}]{Roy2025dissipative}%
  \BibitemOpen
  \bibfield  {author} {\bibinfo {author} {\bibfnamefont {K.}~\bibnamefont
  {Roy}}, \bibinfo {author} {\bibfnamefont {S.}~\bibnamefont {Shahab}},\ and\
  \bibinfo {author} {\bibfnamefont {S.}~\bibnamefont {Basu}},\ }\bibfield
  {title} {\bibinfo {title} {Controlling dissipative topology through floquet
  driving: From transient diagnostics to boundary states isolation},\
  }\href@noop {} {\bibfield  {journal} {\bibinfo  {journal}
  {\href{https://arxiv.org/abs/2511.23229}{arXiv:2511.23229}}\ } (\bibinfo
  {year} {2025}{\natexlab{a}})}\BibitemShut {NoStop}%
\bibitem [{\citenamefont {Roy}\ \emph {et~al.}(2025{\natexlab{b}})\citenamefont
  {Roy}, \citenamefont {Gogoi},\ and\ \citenamefont {Basu}}]{Roy2025NHcreutz}%
  \BibitemOpen
  \bibfield  {author} {\bibinfo {author} {\bibfnamefont {K.}~\bibnamefont
  {Roy}}, \bibinfo {author} {\bibfnamefont {K.}~\bibnamefont {Gogoi}},\ and\
  \bibinfo {author} {\bibfnamefont {S.}~\bibnamefont {Basu}},\ }\bibfield
  {title} {\bibinfo {title} {Topological characterization of a non-hermitian
  ladder via floquet non-bloch theory},\ }\href
  {https://doi.org/10.1103/PhysRevB.111.115424} {\bibfield  {journal} {\bibinfo
   {journal} {Phys. Rev. B}\ }\textbf {\bibinfo {volume} {111}},\ \bibinfo
  {pages} {115424} (\bibinfo {year} {2025}{\natexlab{b}})}\BibitemShut
  {NoStop}%
\bibitem [{\citenamefont {Roy}\ \emph {et~al.}(2025{\natexlab{c}})\citenamefont
  {Roy}, \citenamefont {Halder}, \citenamefont {Gogoi}, \citenamefont
  {Tanatar},\ and\ \citenamefont {Basu}}]{Roy2025ckt}%
  \BibitemOpen
  \bibfield  {author} {\bibinfo {author} {\bibfnamefont {K.}~\bibnamefont
  {Roy}}, \bibinfo {author} {\bibfnamefont {D.}~\bibnamefont {Halder}},
  \bibinfo {author} {\bibfnamefont {K.}~\bibnamefont {Gogoi}}, \bibinfo
  {author} {\bibfnamefont {B.}~\bibnamefont {Tanatar}},\ and\ \bibinfo {author}
  {\bibfnamefont {S.}~\bibnamefont {Basu}},\ }\bibfield  {title} {\bibinfo
  {title} {Floquet non-bloch formalism for a non-hermitian ladder: From
  theoretical framework to topolectrical circuits},\ }\href
  {https://doi.org/10.1103/s3b6-wz16} {\bibfield  {journal} {\bibinfo
  {journal} {Phys. Rev. Res.}\ }\textbf {\bibinfo {volume} {7}},\ \bibinfo
  {pages} {043331} (\bibinfo {year} {2025}{\natexlab{c}})}\BibitemShut
  {NoStop}%
\bibitem [{\citenamefont {Roy}\ \emph {et~al.}(2026)\citenamefont {Roy},
  \citenamefont {Kalita}, \citenamefont {Tanatar},\ and\ \citenamefont
  {Basu}}]{Roy2026hybrid}%
  \BibitemOpen
  \bibfield  {author} {\bibinfo {author} {\bibfnamefont {K.}~\bibnamefont
  {Roy}}, \bibinfo {author} {\bibfnamefont {L.}~\bibnamefont {Kalita}},
  \bibinfo {author} {\bibfnamefont {B.}~\bibnamefont {Tanatar}},\ and\ \bibinfo
  {author} {\bibfnamefont {S.}~\bibnamefont {Basu}},\ }\bibfield  {title}
  {\bibinfo {title} {Floquet generation of hybrid-order topology and
  {$\mathbb{Z}_2$}-like bipolar localization},\ }\href@noop {} {\bibfield
  {journal} {\bibinfo  {journal}
  {\href{https://arxiv.org/abs/2603.21954}{arXiv:2603.21954}}\ } (\bibinfo
  {year} {2026})}\BibitemShut {NoStop}%
\bibitem [{\citenamefont {Jiang}\ \emph {et~al.}(2011)\citenamefont {Jiang},
  \citenamefont {Kitagawa}, \citenamefont {Alicea}, \citenamefont {Akhmerov},
  \citenamefont {Pekker}, \citenamefont {Refael}, \citenamefont {Cirac},
  \citenamefont {Demler}, \citenamefont {Lukin},\ and\ \citenamefont
  {Zoller}}]{Jiang2011}%
  \BibitemOpen
  \bibfield  {author} {\bibinfo {author} {\bibfnamefont {L.}~\bibnamefont
  {Jiang}}, \bibinfo {author} {\bibfnamefont {T.}~\bibnamefont {Kitagawa}},
  \bibinfo {author} {\bibfnamefont {J.}~\bibnamefont {Alicea}}, \bibinfo
  {author} {\bibfnamefont {A.~R.}\ \bibnamefont {Akhmerov}}, \bibinfo {author}
  {\bibfnamefont {D.}~\bibnamefont {Pekker}}, \bibinfo {author} {\bibfnamefont
  {G.}~\bibnamefont {Refael}}, \bibinfo {author} {\bibfnamefont {J.~I.}\
  \bibnamefont {Cirac}}, \bibinfo {author} {\bibfnamefont {E.}~\bibnamefont
  {Demler}}, \bibinfo {author} {\bibfnamefont {M.~D.}\ \bibnamefont {Lukin}},\
  and\ \bibinfo {author} {\bibfnamefont {P.}~\bibnamefont {Zoller}},\
  }\bibfield  {title} {\bibinfo {title} {Majorana fermions in equilibrium and
  in driven cold-atom quantum wires},\ }\href
  {https://doi.org/10.1103/physrevlett.106.220402} {\bibfield  {journal}
  {\bibinfo  {journal} {Phys. Rev. Lett.}\ }\textbf {\bibinfo {volume} {106}},\
  \bibinfo {pages} {220402} (\bibinfo {year} {2011})}\BibitemShut {NoStop}%
\bibitem [{\citenamefont {Thakurathi}\ \emph {et~al.}(2013)\citenamefont
  {Thakurathi}, \citenamefont {Patel}, \citenamefont {Sen},\ and\ \citenamefont
  {Dutta}}]{Thakurathi2013}%
  \BibitemOpen
  \bibfield  {author} {\bibinfo {author} {\bibfnamefont {M.}~\bibnamefont
  {Thakurathi}}, \bibinfo {author} {\bibfnamefont {A.~A.}\ \bibnamefont
  {Patel}}, \bibinfo {author} {\bibfnamefont {D.}~\bibnamefont {Sen}},\ and\
  \bibinfo {author} {\bibfnamefont {A.}~\bibnamefont {Dutta}},\ }\bibfield
  {title} {\bibinfo {title} {Floquet generation of {Majorana} end modes and
  topological invariants},\ }\href {https://doi.org/10.1103/physrevb.88.155133}
  {\bibfield  {journal} {\bibinfo  {journal} {Phys. Rev. B}\ }\textbf {\bibinfo
  {volume} {88}},\ \bibinfo {pages} {155133} (\bibinfo {year}
  {2013})}\BibitemShut {NoStop}%
\bibitem [{\citenamefont {Roy}\ \emph {et~al.}(2024)\citenamefont {Roy},
  \citenamefont {Roy},\ and\ \citenamefont {Basu}}]{Roy2024DKC}%
  \BibitemOpen
  \bibfield  {author} {\bibinfo {author} {\bibfnamefont {K.}~\bibnamefont
  {Roy}}, \bibinfo {author} {\bibfnamefont {S.}~\bibnamefont {Roy}},\ and\
  \bibinfo {author} {\bibfnamefont {S.}~\bibnamefont {Basu}},\ }\bibfield
  {title} {\bibinfo {title} {Quasiperiodic disorder induced critical phases in
  a periodically driven dimerized {p}-wave kitaev chain},\ }\href
  {https://doi.org/10.1038/s41598-024-70995-2} {\bibfield  {journal} {\bibinfo
  {journal} {Sci. Rep.}\ }\textbf {\bibinfo {volume} {14}},\ \bibinfo {pages}
  {20603} (\bibinfo {year} {2024})}\BibitemShut {NoStop}%
\bibitem [{\citenamefont {Roy}\ and\ \citenamefont
  {Basu}(2024)}]{Roy2024rashba}%
  \BibitemOpen
  \bibfield  {author} {\bibinfo {author} {\bibfnamefont {K.}~\bibnamefont
  {Roy}}\ and\ \bibinfo {author} {\bibfnamefont {S.}~\bibnamefont {Basu}},\
  }\bibfield  {title} {\bibinfo {title} {Single and multifrequency driving
  protocols in a rashba nanowire proximitized to an {$s$}-wave
  superconductor},\ }\href {https://doi.org/10.1103/PhysRevB.110.165403}
  {\bibfield  {journal} {\bibinfo  {journal} {Phys. Rev. B}\ }\textbf {\bibinfo
  {volume} {110}},\ \bibinfo {pages} {165403} (\bibinfo {year}
  {2024})}\BibitemShut {NoStop}%
\bibitem [{\citenamefont {Roy}\ \emph {et~al.}(2025{\natexlab{d}})\citenamefont
  {Roy}, \citenamefont {Paul}, \citenamefont {Debnath}, \citenamefont
  {Bhattacharyya},\ and\ \citenamefont {Basu}}]{RoyJJ2025}%
  \BibitemOpen
  \bibfield  {author} {\bibinfo {author} {\bibfnamefont {K.}~\bibnamefont
  {Roy}}, \bibinfo {author} {\bibfnamefont {G.}~\bibnamefont {Paul}}, \bibinfo
  {author} {\bibfnamefont {D.}~\bibnamefont {Debnath}}, \bibinfo {author}
  {\bibfnamefont {K.}~\bibnamefont {Bhattacharyya}},\ and\ \bibinfo {author}
  {\bibfnamefont {S.}~\bibnamefont {Basu}},\ }\bibfield  {title} {\bibinfo
  {title} {{F}loquet-engineered diode performance of a topological {Josephson}
  junction composed of two {Kitaev} chains coupled via a quantum dot},\ }\href
  {https://doi.org/10.1103/69jq-rcsb} {\bibfield  {journal} {\bibinfo
  {journal} {Phys. Rev. B}\ }\textbf {\bibinfo {volume} {112}},\ \bibinfo
  {pages} {125407} (\bibinfo {year} {2025}{\natexlab{d}})}\BibitemShut
  {NoStop}%
\bibitem [{\citenamefont {Soori}(2023)}]{Soori2023}%
  \BibitemOpen
  \bibfield  {author} {\bibinfo {author} {\bibfnamefont {A.}~\bibnamefont
  {Soori}},\ }\bibfield  {title} {\bibinfo {title} {Anomalous {Josephson}
  effect and rectification in junctions between {Floquet} topological
  superconductors},\ }\href {https://doi.org/10.1016/j.physe.2022.115545}
  {\bibfield  {journal} {\bibinfo  {journal} {Physica E}\ }\textbf {\bibinfo
  {volume} {146}},\ \bibinfo {pages} {115545} (\bibinfo {year}
  {2023})}\BibitemShut {NoStop}%
\bibitem [{\citenamefont {Ortega-Taberner}\ \emph {et~al.}(2023)\citenamefont
  {Ortega-Taberner}, \citenamefont {Jauho},\ and\ \citenamefont
  {Paaske}}]{OrtegaTaberner2023}%
  \BibitemOpen
  \bibfield  {author} {\bibinfo {author} {\bibfnamefont {C.}~\bibnamefont
  {Ortega-Taberner}}, \bibinfo {author} {\bibfnamefont {A.-P.}\ \bibnamefont
  {Jauho}},\ and\ \bibinfo {author} {\bibfnamefont {J.}~\bibnamefont
  {Paaske}},\ }\bibfield  {title} {\bibinfo {title} {Anomalous {Josephson}
  current through a driven double quantum dot},\ }\href
  {https://doi.org/10.1103/physrevb.107.115165} {\bibfield  {journal} {\bibinfo
   {journal} {Phys. Rev. B}\ }\textbf {\bibinfo {volume} {107}},\ \bibinfo
  {pages} {115165} (\bibinfo {year} {2023})}\BibitemShut {NoStop}%
\bibitem [{\citenamefont {Schecter}\ \emph {et~al.}(2016)\citenamefont
  {Schecter}, \citenamefont {Flensberg}, \citenamefont {Christensen},
  \citenamefont {Andersen},\ and\ \citenamefont {Paaske}}]{Schecter2016}%
  \BibitemOpen
  \bibfield  {author} {\bibinfo {author} {\bibfnamefont {M.}~\bibnamefont
  {Schecter}}, \bibinfo {author} {\bibfnamefont {K.}~\bibnamefont {Flensberg}},
  \bibinfo {author} {\bibfnamefont {M.~H.}\ \bibnamefont {Christensen}},
  \bibinfo {author} {\bibfnamefont {B.~M.}\ \bibnamefont {Andersen}},\ and\
  \bibinfo {author} {\bibfnamefont {J.}~\bibnamefont {Paaske}},\ }\bibfield
  {title} {\bibinfo {title} {Self-organized topological superconductivity in a
  {Yu-Shiba-Rusinov} chain},\ }\href
  {https://doi.org/10.1103/physrevb.93.140503} {\bibfield  {journal} {\bibinfo
  {journal} {Phys. Rev. B}\ }\textbf {\bibinfo {volume} {93}},\ \bibinfo
  {pages} {140503} (\bibinfo {year} {2016})}\BibitemShut {NoStop}%
\bibitem [{\citenamefont {Braunecker}\ \emph {et~al.}(2010)\citenamefont
  {Braunecker}, \citenamefont {Japaridze}, \citenamefont {Klinovaja},\ and\
  \citenamefont {Loss}}]{Braunecker2010}%
  \BibitemOpen
  \bibfield  {author} {\bibinfo {author} {\bibfnamefont {B.}~\bibnamefont
  {Braunecker}}, \bibinfo {author} {\bibfnamefont {G.~I.}\ \bibnamefont
  {Japaridze}}, \bibinfo {author} {\bibfnamefont {J.}~\bibnamefont
  {Klinovaja}},\ and\ \bibinfo {author} {\bibfnamefont {D.}~\bibnamefont
  {Loss}},\ }\bibfield  {title} {\bibinfo {title} {Spin-selective {Peierls}
  transition in interacting one-dimensional conductors with spin-orbit
  interaction},\ }\href {https://doi.org/10.1103/physrevb.82.045127} {\bibfield
   {journal} {\bibinfo  {journal} {Phys. Rev. B}\ }\textbf {\bibinfo {volume}
  {82}},\ \bibinfo {pages} {045127} (\bibinfo {year} {2010})}\BibitemShut
  {NoStop}%
\bibitem [{\citenamefont {Kjaergaard}\ \emph {et~al.}(2012)\citenamefont
  {Kjaergaard}, \citenamefont {W\"{o}lms},\ and\ \citenamefont
  {Flensberg}}]{Kjaergaard2012}%
  \BibitemOpen
  \bibfield  {author} {\bibinfo {author} {\bibfnamefont {M.}~\bibnamefont
  {Kjaergaard}}, \bibinfo {author} {\bibfnamefont {K.}~\bibnamefont
  {W\"{o}lms}},\ and\ \bibinfo {author} {\bibfnamefont {K.}~\bibnamefont
  {Flensberg}},\ }\bibfield  {title} {\bibinfo {title} {Majorana fermions in
  superconducting nanowires without spin-orbit coupling},\ }\href
  {https://doi.org/10.1103/physrevb.85.020503} {\bibfield  {journal} {\bibinfo
  {journal} {Phys. Rev. B}\ }\textbf {\bibinfo {volume} {85}},\ \bibinfo
  {pages} {020503} (\bibinfo {year} {2012})}\BibitemShut {NoStop}%
\bibitem [{\citenamefont {Lutchyn}\ \emph {et~al.}(2010)\citenamefont
  {Lutchyn}, \citenamefont {Sau},\ and\ \citenamefont {{Das
  Sarma}}}]{Lutchyn2010}%
  \BibitemOpen
  \bibfield  {author} {\bibinfo {author} {\bibfnamefont {R.~M.}\ \bibnamefont
  {Lutchyn}}, \bibinfo {author} {\bibfnamefont {J.~D.}\ \bibnamefont {Sau}},\
  and\ \bibinfo {author} {\bibfnamefont {S.}~\bibnamefont {{Das Sarma}}},\
  }\bibfield  {title} {\bibinfo {title} {Majorana fermions and a topological
  phase transition in semiconductor-superconductor heterostructures},\ }\href
  {https://doi.org/10.1103/physrevlett.105.077001} {\bibfield  {journal}
  {\bibinfo  {journal} {Phys. Rev. Lett.}\ }\textbf {\bibinfo {volume} {105}},\
  \bibinfo {pages} {077001} (\bibinfo {year} {2010})}\BibitemShut {NoStop}%
\bibitem [{\citenamefont {Oreg}\ \emph {et~al.}(2010)\citenamefont {Oreg},
  \citenamefont {Refael},\ and\ \citenamefont {von Oppen}}]{Oreg2010}%
  \BibitemOpen
  \bibfield  {author} {\bibinfo {author} {\bibfnamefont {Y.}~\bibnamefont
  {Oreg}}, \bibinfo {author} {\bibfnamefont {G.}~\bibnamefont {Refael}},\ and\
  \bibinfo {author} {\bibfnamefont {F.}~\bibnamefont {von Oppen}},\ }\bibfield
  {title} {\bibinfo {title} {Helical liquids and {Majorana} bound states in
  quantum wires},\ }\href {https://doi.org/10.1103/physrevlett.105.177002}
  {\bibfield  {journal} {\bibinfo  {journal} {Phys. Rev. Lett.}\ }\textbf
  {\bibinfo {volume} {105}},\ \bibinfo {pages} {177002} (\bibinfo {year}
  {2010})}\BibitemShut {NoStop}%
\bibitem [{\citenamefont {Alicea}(2012)}]{Alicea2012}%
  \BibitemOpen
  \bibfield  {author} {\bibinfo {author} {\bibfnamefont {J.}~\bibnamefont
  {Alicea}},\ }\bibfield  {title} {\bibinfo {title} {New directions in the
  pursuit of {Majorana} fermions in solid state systems},\ }\href
  {https://doi.org/10.1088/0034-4885/75/7/076501} {\bibfield  {journal}
  {\bibinfo  {journal} {Rep. Prog. Phys.}\ }\textbf {\bibinfo {volume} {75}},\
  \bibinfo {pages} {076501} (\bibinfo {year} {2012})}\BibitemShut {NoStop}%
\bibitem [{\citenamefont {Kitaev}(2001)}]{Kitaev2001}%
  \BibitemOpen
  \bibfield  {author} {\bibinfo {author} {\bibfnamefont {A.~Y.}\ \bibnamefont
  {Kitaev}},\ }\bibfield  {title} {\bibinfo {title} {Unpaired {Majorana}
  fermions in quantum wires},\ }\href
  {https://doi.org/10.1070/1063-7869/44/10s/s29} {\bibfield  {journal}
  {\bibinfo  {journal} {Phys.-Usp.}\ }\textbf {\bibinfo {volume} {44}},\
  \bibinfo {pages} {131} (\bibinfo {year} {2001})}\BibitemShut {NoStop}%
\bibitem [{\citenamefont {Seidler}\ \emph {et~al.}(2022)\citenamefont
  {Seidler}, \citenamefont {Struck}, \citenamefont {Xue}, \citenamefont
  {Focke}, \citenamefont {Trellenkamp}, \citenamefont {Bluhm},\ and\
  \citenamefont {Schreiber}}]{Seidler2022}%
  \BibitemOpen
  \bibfield  {author} {\bibinfo {author} {\bibfnamefont {I.}~\bibnamefont
  {Seidler}}, \bibinfo {author} {\bibfnamefont {T.}~\bibnamefont {Struck}},
  \bibinfo {author} {\bibfnamefont {R.}~\bibnamefont {Xue}}, \bibinfo {author}
  {\bibfnamefont {N.}~\bibnamefont {Focke}}, \bibinfo {author} {\bibfnamefont
  {S.}~\bibnamefont {Trellenkamp}}, \bibinfo {author} {\bibfnamefont
  {H.}~\bibnamefont {Bluhm}},\ and\ \bibinfo {author} {\bibfnamefont {L.~R.}\
  \bibnamefont {Schreiber}},\ }\bibfield  {title} {\bibinfo {title}
  {Conveyor-mode single-electron shuttling in {Si/SiGe} for a scalable quantum
  computing architecture},\ }\href {https://doi.org/10.1038/s41534-022-00615-2}
  {\bibfield  {journal} {\bibinfo  {journal} {npj Quantum Inf.}\ }\textbf
  {\bibinfo {volume} {8}},\ \bibinfo {pages} {100} (\bibinfo {year}
  {2022})}\BibitemShut {NoStop}%
\bibitem [{\citenamefont {Langrock}\ \emph {et~al.}(2023)\citenamefont
  {Langrock}, \citenamefont {Krzywda}, \citenamefont {Focke}, \citenamefont
  {Seidler}, \citenamefont {Schreiber},\ and\ \citenamefont
  {Cywi\'{n}ski}}]{Langrock2023}%
  \BibitemOpen
  \bibfield  {author} {\bibinfo {author} {\bibfnamefont {V.}~\bibnamefont
  {Langrock}}, \bibinfo {author} {\bibfnamefont {J.~A.}\ \bibnamefont
  {Krzywda}}, \bibinfo {author} {\bibfnamefont {N.}~\bibnamefont {Focke}},
  \bibinfo {author} {\bibfnamefont {I.}~\bibnamefont {Seidler}}, \bibinfo
  {author} {\bibfnamefont {L.~R.}\ \bibnamefont {Schreiber}},\ and\ \bibinfo
  {author} {\bibfnamefont {{\L}.}~\bibnamefont {Cywi\'{n}ski}},\ }\bibfield
  {title} {\bibinfo {title} {Blueprint of a scalable spin qubit shuttle device
  for coherent mid-range qubit transfer in disordered {Si/SiGe/SiO}$_2$},\
  }\href {https://doi.org/10.1103/prxquantum.4.020305} {\bibfield  {journal}
  {\bibinfo  {journal} {PRX Quantum}\ }\textbf {\bibinfo {volume} {4}},\
  \bibinfo {pages} {020305} (\bibinfo {year} {2023})}\BibitemShut {NoStop}%
\bibitem [{\citenamefont {Hermelin}\ \emph {et~al.}(2011)\citenamefont
  {Hermelin}, \citenamefont {Takada}, \citenamefont {Yamamoto}, \citenamefont
  {Tarucha}, \citenamefont {Wieck}, \citenamefont {Saminadayar}, \citenamefont
  {B\"{a}uerle},\ and\ \citenamefont {Meunier}}]{Hermelin2011}%
  \BibitemOpen
  \bibfield  {author} {\bibinfo {author} {\bibfnamefont {S.}~\bibnamefont
  {Hermelin}}, \bibinfo {author} {\bibfnamefont {S.}~\bibnamefont {Takada}},
  \bibinfo {author} {\bibfnamefont {M.}~\bibnamefont {Yamamoto}}, \bibinfo
  {author} {\bibfnamefont {S.}~\bibnamefont {Tarucha}}, \bibinfo {author}
  {\bibfnamefont {A.~D.}\ \bibnamefont {Wieck}}, \bibinfo {author}
  {\bibfnamefont {L.}~\bibnamefont {Saminadayar}}, \bibinfo {author}
  {\bibfnamefont {C.}~\bibnamefont {B\"{a}uerle}},\ and\ \bibinfo {author}
  {\bibfnamefont {T.}~\bibnamefont {Meunier}},\ }\bibfield  {title} {\bibinfo
  {title} {Electrons surfing on a sound wave as a platform for quantum optics
  with flying electrons},\ }\href {https://doi.org/10.1038/nature10416}
  {\bibfield  {journal} {\bibinfo  {journal} {Nature (London)}\ }\textbf
  {\bibinfo {volume} {477}},\ \bibinfo {pages} {435} (\bibinfo {year}
  {2011})}\BibitemShut {NoStop}%
\bibitem [{\citenamefont {McNeil}\ \emph {et~al.}(2011)\citenamefont {McNeil},
  \citenamefont {Kataoka}, \citenamefont {Ford}, \citenamefont {Barnes},
  \citenamefont {Anderson}, \citenamefont {Jones}, \citenamefont {Farrer},\
  and\ \citenamefont {Ritchie}}]{McNeil2011}%
  \BibitemOpen
  \bibfield  {author} {\bibinfo {author} {\bibfnamefont {R.~P.~G.}\
  \bibnamefont {McNeil}}, \bibinfo {author} {\bibfnamefont {M.}~\bibnamefont
  {Kataoka}}, \bibinfo {author} {\bibfnamefont {C.~J.~B.}\ \bibnamefont
  {Ford}}, \bibinfo {author} {\bibfnamefont {C.~H.~W.}\ \bibnamefont {Barnes}},
  \bibinfo {author} {\bibfnamefont {D.}~\bibnamefont {Anderson}}, \bibinfo
  {author} {\bibfnamefont {G.~A.~C.}\ \bibnamefont {Jones}}, \bibinfo {author}
  {\bibfnamefont {I.}~\bibnamefont {Farrer}},\ and\ \bibinfo {author}
  {\bibfnamefont {D.~A.}\ \bibnamefont {Ritchie}},\ }\bibfield  {title}
  {\bibinfo {title} {On-demand single-electron transfer between distant quantum
  dots},\ }\href {https://doi.org/10.1038/nature10444} {\bibfield  {journal}
  {\bibinfo  {journal} {Nature (London)}\ }\textbf {\bibinfo {volume} {477}},\
  \bibinfo {pages} {439} (\bibinfo {year} {2011})}\BibitemShut {NoStop}%
\bibitem [{\citenamefont {Satzinger}\ \emph {et~al.}(2018)\citenamefont
  {Satzinger}, \citenamefont {Zhong}, \citenamefont {Chang}, \citenamefont
  {Peairs}, \citenamefont {Bienfait}, \citenamefont {Chou}, \citenamefont
  {Cleland}, \citenamefont {Conner}, \citenamefont {Dumur}, \citenamefont
  {Grebel} \emph {et~al.}}]{Satzinger2018}%
  \BibitemOpen
  \bibfield  {author} {\bibinfo {author} {\bibfnamefont {K.~J.}\ \bibnamefont
  {Satzinger}}, \bibinfo {author} {\bibfnamefont {Y.~P.}\ \bibnamefont
  {Zhong}}, \bibinfo {author} {\bibfnamefont {H.-S.}\ \bibnamefont {Chang}},
  \bibinfo {author} {\bibfnamefont {G.~A.}\ \bibnamefont {Peairs}}, \bibinfo
  {author} {\bibfnamefont {A.}~\bibnamefont {Bienfait}}, \bibinfo {author}
  {\bibfnamefont {M.-H.}\ \bibnamefont {Chou}}, \bibinfo {author}
  {\bibfnamefont {A.~Y.}\ \bibnamefont {Cleland}}, \bibinfo {author}
  {\bibfnamefont {C.~R.}\ \bibnamefont {Conner}}, \bibinfo {author}
  {\bibfnamefont {E.}~\bibnamefont {Dumur}}, \bibinfo {author} {\bibfnamefont
  {J.}~\bibnamefont {Grebel}}, \emph {et~al.},\ }\bibfield  {title} {\bibinfo
  {title} {Quantum control of surface acoustic-wave phonons},\ }\href
  {https://doi.org/10.1038/s41586-018-0719-5} {\bibfield  {journal} {\bibinfo
  {journal} {Nature (London)}\ }\textbf {\bibinfo {volume} {563}},\ \bibinfo
  {pages} {661} (\bibinfo {year} {2018})}\BibitemShut {NoStop}%
\bibitem [{\citenamefont {Lohse}\ \emph {et~al.}(2016)\citenamefont {Lohse},
  \citenamefont {Schweizer}, \citenamefont {Zilberberg}, \citenamefont
  {Aidelsburger},\ and\ \citenamefont {Bloch}}]{Lohse2016}%
  \BibitemOpen
  \bibfield  {author} {\bibinfo {author} {\bibfnamefont {M.}~\bibnamefont
  {Lohse}}, \bibinfo {author} {\bibfnamefont {C.}~\bibnamefont {Schweizer}},
  \bibinfo {author} {\bibfnamefont {O.}~\bibnamefont {Zilberberg}}, \bibinfo
  {author} {\bibfnamefont {M.}~\bibnamefont {Aidelsburger}},\ and\ \bibinfo
  {author} {\bibfnamefont {I.}~\bibnamefont {Bloch}},\ }\bibfield  {title}
  {\bibinfo {title} {A {Thouless} quantum pump with ultracold bosonic atoms in
  an optical superlattice},\ }\href {https://doi.org/10.1038/nphys3584}
  {\bibfield  {journal} {\bibinfo  {journal} {Nat. Phys.}\ }\textbf {\bibinfo
  {volume} {12}},\ \bibinfo {pages} {350} (\bibinfo {year} {2016})}\BibitemShut
  {NoStop}%
\bibitem [{\citenamefont {Nakajima}\ \emph {et~al.}(2016)\citenamefont
  {Nakajima}, \citenamefont {Tomita}, \citenamefont {Taie}, \citenamefont
  {Ichinose}, \citenamefont {Ozawa}, \citenamefont {Wang}, \citenamefont
  {Troyer},\ and\ \citenamefont {Takahashi}}]{Nakajima2016}%
  \BibitemOpen
  \bibfield  {author} {\bibinfo {author} {\bibfnamefont {S.}~\bibnamefont
  {Nakajima}}, \bibinfo {author} {\bibfnamefont {T.}~\bibnamefont {Tomita}},
  \bibinfo {author} {\bibfnamefont {S.}~\bibnamefont {Taie}}, \bibinfo {author}
  {\bibfnamefont {T.}~\bibnamefont {Ichinose}}, \bibinfo {author}
  {\bibfnamefont {H.}~\bibnamefont {Ozawa}}, \bibinfo {author} {\bibfnamefont
  {L.}~\bibnamefont {Wang}}, \bibinfo {author} {\bibfnamefont {M.}~\bibnamefont
  {Troyer}},\ and\ \bibinfo {author} {\bibfnamefont {Y.}~\bibnamefont
  {Takahashi}},\ }\bibfield  {title} {\bibinfo {title} {Topological {Thouless}
  pumping of ultracold fermions},\ }\href {https://doi.org/10.1038/nphys3622}
  {\bibfield  {journal} {\bibinfo  {journal} {Nat. Phys.}\ }\textbf {\bibinfo
  {volume} {12}},\ \bibinfo {pages} {296} (\bibinfo {year} {2016})}\BibitemShut
  {NoStop}%
\bibitem [{\citenamefont {Dvir}\ \emph {et~al.}(2023)\citenamefont {Dvir},
  \citenamefont {Wang}, \citenamefont {van Loo}, \citenamefont {Liu},
  \citenamefont {Mazur}, \citenamefont {Bordin}, \citenamefont {ten Haaf},
  \citenamefont {Wang}, \citenamefont {van Driel}, \citenamefont {Zatelli}
  \emph {et~al.}}]{Dvir2023}%
  \BibitemOpen
  \bibfield  {author} {\bibinfo {author} {\bibfnamefont {T.}~\bibnamefont
  {Dvir}}, \bibinfo {author} {\bibfnamefont {G.}~\bibnamefont {Wang}}, \bibinfo
  {author} {\bibfnamefont {N.}~\bibnamefont {van Loo}}, \bibinfo {author}
  {\bibfnamefont {C.-X.}\ \bibnamefont {Liu}}, \bibinfo {author} {\bibfnamefont
  {G.~P.}\ \bibnamefont {Mazur}}, \bibinfo {author} {\bibfnamefont
  {A.}~\bibnamefont {Bordin}}, \bibinfo {author} {\bibfnamefont {S.~L.~D.}\
  \bibnamefont {ten Haaf}}, \bibinfo {author} {\bibfnamefont {J.-Y.}\
  \bibnamefont {Wang}}, \bibinfo {author} {\bibfnamefont {D.}~\bibnamefont {van
  Driel}}, \bibinfo {author} {\bibfnamefont {F.}~\bibnamefont {Zatelli}}, \emph
  {et~al.},\ }\bibfield  {title} {\bibinfo {title} {Realization of a minimal
  {Kitaev} chain in coupled quantum dots},\ }\href
  {https://doi.org/10.1038/s41586-022-05585-1} {\bibfield  {journal} {\bibinfo
  {journal} {Nature (London)}\ }\textbf {\bibinfo {volume} {614}},\ \bibinfo
  {pages} {445} (\bibinfo {year} {2023})}\BibitemShut {NoStop}%
\bibitem [{\citenamefont {Samuelson}\ \emph {et~al.}(2024)\citenamefont
  {Samuelson}, \citenamefont {Svensson},\ and\ \citenamefont
  {Leijnse}}]{Samuelson2024}%
  \BibitemOpen
  \bibfield  {author} {\bibinfo {author} {\bibfnamefont {W.}~\bibnamefont
  {Samuelson}}, \bibinfo {author} {\bibfnamefont {V.}~\bibnamefont
  {Svensson}},\ and\ \bibinfo {author} {\bibfnamefont {M.}~\bibnamefont
  {Leijnse}},\ }\bibfield  {title} {\bibinfo {title} {Minimal quantum dot based
  {Kitaev} chain with only local superconducting proximity effect},\ }\href
  {https://doi.org/10.1103/physrevb.109.035415} {\bibfield  {journal} {\bibinfo
   {journal} {Phys. Rev. B}\ }\textbf {\bibinfo {volume} {109}},\ \bibinfo
  {pages} {035415} (\bibinfo {year} {2024})}\BibitemShut {NoStop}%
\bibitem [{\citenamefont {Shirley}(1965)}]{Shirley1965}%
  \BibitemOpen
  \bibfield  {author} {\bibinfo {author} {\bibfnamefont {J.~H.}\ \bibnamefont
  {Shirley}},\ }\bibfield  {title} {\bibinfo {title} {Solution of the
  {Schr\"{o}dinger} equation with a {Hamiltonian} periodic in time},\ }\href
  {https://doi.org/10.1103/physrev.138.b979} {\bibfield  {journal} {\bibinfo
  {journal} {Phys. Rev.}\ }\textbf {\bibinfo {volume} {138}},\ \bibinfo {pages}
  {B979} (\bibinfo {year} {1965})}\BibitemShut {NoStop}%
\bibitem [{\citenamefont {Sambe}(1973)}]{Sambe1973}%
  \BibitemOpen
  \bibfield  {author} {\bibinfo {author} {\bibfnamefont {H.}~\bibnamefont
  {Sambe}},\ }\bibfield  {title} {\bibinfo {title} {Steady states and
  quasienergies of a quantum-mechanical system in an oscillating field},\
  }\href {https://doi.org/10.1103/physreva.7.2203} {\bibfield  {journal}
  {\bibinfo  {journal} {Phys. Rev. A}\ }\textbf {\bibinfo {volume} {7}},\
  \bibinfo {pages} {2203} (\bibinfo {year} {1973})}\BibitemShut {NoStop}%
\bibitem [{SM()}]{SM}%
  \BibitemOpen
  \href@noop {} {}\bibinfo {note} {See Supplemental Material at [URL will be
  inserted by publisher] for (S1) the exact mapping of the traveling drive onto
  a running Peierls phase, (S2) the second-order Floquet band correction and
  the optimum at $k_d=2\pi/3$, (S3) the inverse-gap law for the rectification,
  and (S4) the symmetry relations and current maps in the $(q,J)$ and $(q,\mu)$
  planes.}\BibitemShut {Stop}%
\bibitem [{\citenamefont {Peierls}(1933)}]{Peierls1933}%
  \BibitemOpen
  \bibfield  {author} {\bibinfo {author} {\bibfnamefont {R.}~\bibnamefont
  {Peierls}},\ }\bibfield  {title} {\bibinfo {title} {Zur {Theorie} des
  {Diamagnetismus} von {Leitungselektronen}},\ }\href
  {https://doi.org/10.1007/bf01342591} {\bibfield  {journal} {\bibinfo
  {journal} {Z. Phys.}\ }\textbf {\bibinfo {volume} {80}},\ \bibinfo {pages}
  {763} (\bibinfo {year} {1933})}\BibitemShut {NoStop}%
\bibitem [{\citenamefont {Dalibard}\ \emph {et~al.}(2011)\citenamefont
  {Dalibard}, \citenamefont {Gerbier}, \citenamefont {Juzeli\=unas},\ and\
  \citenamefont {\"{O}hberg}}]{Dalibard2011}%
  \BibitemOpen
  \bibfield  {author} {\bibinfo {author} {\bibfnamefont {J.}~\bibnamefont
  {Dalibard}}, \bibinfo {author} {\bibfnamefont {F.}~\bibnamefont {Gerbier}},
  \bibinfo {author} {\bibfnamefont {G.}~\bibnamefont {Juzeli\=unas}},\ and\
  \bibinfo {author} {\bibfnamefont {P.}~\bibnamefont {\"{O}hberg}},\ }\bibfield
   {title} {\bibinfo {title} {Colloquium: Artificial gauge potentials for
  neutral atoms},\ }\href {https://doi.org/10.1103/revmodphys.83.1523}
  {\bibfield  {journal} {\bibinfo  {journal} {Rev. Mod. Phys.}\ }\textbf
  {\bibinfo {volume} {83}},\ \bibinfo {pages} {1523} (\bibinfo {year}
  {2011})}\BibitemShut {NoStop}%
\bibitem [{\citenamefont {Struck}\ \emph {et~al.}(2012)\citenamefont {Struck},
  \citenamefont {\"{O}lschl\"{a}ger}, \citenamefont {Weinberg}, \citenamefont
  {Hauke}, \citenamefont {Simonet}, \citenamefont {Eckardt}, \citenamefont
  {Lewenstein}, \citenamefont {Sengstock},\ and\ \citenamefont
  {Windpassinger}}]{Struck2012}%
  \BibitemOpen
  \bibfield  {author} {\bibinfo {author} {\bibfnamefont {J.}~\bibnamefont
  {Struck}}, \bibinfo {author} {\bibfnamefont {C.}~\bibnamefont
  {\"{O}lschl\"{a}ger}}, \bibinfo {author} {\bibfnamefont {M.}~\bibnamefont
  {Weinberg}}, \bibinfo {author} {\bibfnamefont {P.}~\bibnamefont {Hauke}},
  \bibinfo {author} {\bibfnamefont {J.}~\bibnamefont {Simonet}}, \bibinfo
  {author} {\bibfnamefont {A.}~\bibnamefont {Eckardt}}, \bibinfo {author}
  {\bibfnamefont {M.}~\bibnamefont {Lewenstein}}, \bibinfo {author}
  {\bibfnamefont {K.}~\bibnamefont {Sengstock}},\ and\ \bibinfo {author}
  {\bibfnamefont {P.}~\bibnamefont {Windpassinger}},\ }\bibfield  {title}
  {\bibinfo {title} {Tunable gauge potential for neutral and spinless particles
  in driven optical lattices},\ }\href
  {https://doi.org/10.1103/physrevlett.108.225304} {\bibfield  {journal}
  {\bibinfo  {journal} {Phys. Rev. Lett.}\ }\textbf {\bibinfo {volume} {108}},\
  \bibinfo {pages} {225304} (\bibinfo {year} {2012})}\BibitemShut {NoStop}%
\bibitem [{\citenamefont {Aidelsburger}\ \emph {et~al.}(2013)\citenamefont
  {Aidelsburger}, \citenamefont {Atala}, \citenamefont {Lohse}, \citenamefont
  {Barreiro}, \citenamefont {Paredes},\ and\ \citenamefont
  {Bloch}}]{Aidelsburger2013}%
  \BibitemOpen
  \bibfield  {author} {\bibinfo {author} {\bibfnamefont {M.}~\bibnamefont
  {Aidelsburger}}, \bibinfo {author} {\bibfnamefont {M.}~\bibnamefont {Atala}},
  \bibinfo {author} {\bibfnamefont {M.}~\bibnamefont {Lohse}}, \bibinfo
  {author} {\bibfnamefont {J.~T.}\ \bibnamefont {Barreiro}}, \bibinfo {author}
  {\bibfnamefont {B.}~\bibnamefont {Paredes}},\ and\ \bibinfo {author}
  {\bibfnamefont {I.}~\bibnamefont {Bloch}},\ }\bibfield  {title} {\bibinfo
  {title} {Realization of the {Hofstadter} {Hamiltonian} with ultracold atoms
  in optical lattices},\ }\href
  {https://doi.org/10.1103/physrevlett.111.185301} {\bibfield  {journal}
  {\bibinfo  {journal} {Phys. Rev. Lett.}\ }\textbf {\bibinfo {volume} {111}},\
  \bibinfo {pages} {185301} (\bibinfo {year} {2013})}\BibitemShut {NoStop}%
\bibitem [{\citenamefont {Miyake}\ \emph {et~al.}(2013)\citenamefont {Miyake},
  \citenamefont {Siviloglou}, \citenamefont {Kennedy}, \citenamefont {Burton},\
  and\ \citenamefont {Ketterle}}]{Miyake2013}%
  \BibitemOpen
  \bibfield  {author} {\bibinfo {author} {\bibfnamefont {H.}~\bibnamefont
  {Miyake}}, \bibinfo {author} {\bibfnamefont {G.~A.}\ \bibnamefont
  {Siviloglou}}, \bibinfo {author} {\bibfnamefont {C.~J.}\ \bibnamefont
  {Kennedy}}, \bibinfo {author} {\bibfnamefont {W.~C.}\ \bibnamefont
  {Burton}},\ and\ \bibinfo {author} {\bibfnamefont {W.}~\bibnamefont
  {Ketterle}},\ }\bibfield  {title} {\bibinfo {title} {Realizing the {Harper}
  {Hamiltonian} with laser-assisted tunneling in optical lattices},\ }\href
  {https://doi.org/10.1103/physrevlett.111.185302} {\bibfield  {journal}
  {\bibinfo  {journal} {Phys. Rev. Lett.}\ }\textbf {\bibinfo {volume} {111}},\
  \bibinfo {pages} {185302} (\bibinfo {year} {2013})}\BibitemShut {NoStop}%
\bibitem [{\citenamefont {Dunlap}\ and\ \citenamefont
  {Kenkre}(1986)}]{Dunlap1986}%
  \BibitemOpen
  \bibfield  {author} {\bibinfo {author} {\bibfnamefont {D.~H.}\ \bibnamefont
  {Dunlap}}\ and\ \bibinfo {author} {\bibfnamefont {V.~M.}\ \bibnamefont
  {Kenkre}},\ }\bibfield  {title} {\bibinfo {title} {Dynamic localization of a
  charged particle moving under the influence of an electric field},\ }\href
  {https://doi.org/10.1103/physrevb.34.3625} {\bibfield  {journal} {\bibinfo
  {journal} {Phys. Rev. B}\ }\textbf {\bibinfo {volume} {34}},\ \bibinfo
  {pages} {3625} (\bibinfo {year} {1986})}\BibitemShut {NoStop}%
\bibitem [{\citenamefont {Kumari}\ \emph {et~al.}(2024)\citenamefont {Kumari},
  \citenamefont {Seradjeh},\ and\ \citenamefont {Kundu}}]{Kumari2024}%
  \BibitemOpen
  \bibfield  {author} {\bibinfo {author} {\bibfnamefont {R.}~\bibnamefont
  {Kumari}}, \bibinfo {author} {\bibfnamefont {B.}~\bibnamefont {Seradjeh}},\
  and\ \bibinfo {author} {\bibfnamefont {A.}~\bibnamefont {Kundu}},\ }\bibfield
   {title} {\bibinfo {title} {Josephson-current signatures of unpaired
  {Floquet} {Majorana} fermions},\ }\href
  {https://doi.org/10.1103/physrevlett.133.196601} {\bibfield  {journal}
  {\bibinfo  {journal} {Phys. Rev. Lett.}\ }\textbf {\bibinfo {volume} {133}},\
  \bibinfo {pages} {196601} (\bibinfo {year} {2024})}\BibitemShut {NoStop}%
\bibitem [{\citenamefont {Haldane}(1988)}]{Haldane1988}%
  \BibitemOpen
  \bibfield  {author} {\bibinfo {author} {\bibfnamefont {F.~D.~M.}\
  \bibnamefont {Haldane}},\ }\bibfield  {title} {\bibinfo {title} {Model for a
  quantum {Hall} effect without {Landau} levels: condensed-matter realization
  of the parity anomaly},\ }\href {https://doi.org/10.1103/physrevlett.61.2015}
  {\bibfield  {journal} {\bibinfo  {journal} {Phys. Rev. Lett.}\ }\textbf
  {\bibinfo {volume} {61}},\ \bibinfo {pages} {2015} (\bibinfo {year}
  {1988})}\BibitemShut {NoStop}%
\bibitem [{\citenamefont {Jotzu}\ \emph {et~al.}(2014)\citenamefont {Jotzu},
  \citenamefont {Messer}, \citenamefont {Desbuquois}, \citenamefont {Lebrat},
  \citenamefont {Uehlinger}, \citenamefont {Greif},\ and\ \citenamefont
  {Esslinger}}]{Jotzu2014}%
  \BibitemOpen
  \bibfield  {author} {\bibinfo {author} {\bibfnamefont {G.}~\bibnamefont
  {Jotzu}}, \bibinfo {author} {\bibfnamefont {M.}~\bibnamefont {Messer}},
  \bibinfo {author} {\bibfnamefont {R.}~\bibnamefont {Desbuquois}}, \bibinfo
  {author} {\bibfnamefont {M.}~\bibnamefont {Lebrat}}, \bibinfo {author}
  {\bibfnamefont {T.}~\bibnamefont {Uehlinger}}, \bibinfo {author}
  {\bibfnamefont {D.}~\bibnamefont {Greif}},\ and\ \bibinfo {author}
  {\bibfnamefont {T.}~\bibnamefont {Esslinger}},\ }\bibfield  {title} {\bibinfo
  {title} {Experimental realization of the topological {Haldane} model with
  ultracold fermions},\ }\href {https://doi.org/10.1038/nature13915} {\bibfield
   {journal} {\bibinfo  {journal} {Nature (London)}\ }\textbf {\bibinfo
  {volume} {515}},\ \bibinfo {pages} {237} (\bibinfo {year}
  {2014})}\BibitemShut {NoStop}%
\end{thebibliography}

\begin{thebibliography}{8}%
\makeatletter
\providecommand \@ifxundefined [1]{%
 \@ifx{#1\undefined}
}%
\providecommand \@ifnum [1]{%
 \ifnum #1\expandafter \@firstoftwo
 \else \expandafter \@secondoftwo
 \fi
}%
\providecommand \@ifx [1]{%
 \ifx #1\expandafter \@firstoftwo
 \else \expandafter \@secondoftwo
 \fi
}%
\providecommand \natexlab [1]{#1}%
\providecommand \enquote  [1]{``#1''}%
\providecommand \bibnamefont  [1]{#1}%
\providecommand \bibfnamefont [1]{#1}%
\providecommand \citenamefont [1]{#1}%
\providecommand \href@noop [0]{\@secondoftwo}%
\providecommand \href [0]{\begingroup \@sanitize@url \@href}%
\providecommand \@href[1]{\@@startlink{#1}\@@href}%
\providecommand \@@href[1]{\endgroup#1\@@endlink}%
\providecommand \@sanitize@url [0]{\catcode `\\12\catcode `\$12\catcode
  `\&12\catcode `\#12\catcode `\^12\catcode `\_12\catcode `\%12\relax}%
\providecommand \@@startlink[1]{}%
\providecommand \@@endlink[0]{}%
\providecommand \url  [0]{\begingroup\@sanitize@url \@url }%
\providecommand \@url [1]{\endgroup\@href {#1}{\urlprefix }}%
\providecommand \urlprefix  [0]{URL }%
\providecommand \Eprint [0]{\href }%
\providecommand \doibase [0]{https://doi.org/}%
\providecommand \selectlanguage [0]{\@gobble}%
\providecommand \bibinfo  [0]{\@secondoftwo}%
\providecommand \bibfield  [0]{\@secondoftwo}%
\providecommand \translation [1]{[#1]}%
\providecommand \BibitemOpen [0]{}%
\providecommand \bibitemStop [0]{}%
\providecommand \bibitemNoStop [0]{.\EOS\space}%
\providecommand \EOS [0]{\spacefactor3000\relax}%
\providecommand \BibitemShut  [1]{\csname bibitem#1\endcsname}%
\let\auto@bib@innerbib\@empty
%</preamble>
\bibitem [{\citenamefont {Dunlap}\ and\ \citenamefont
  {Kenkre}(1986)}]{S-Dunlap1986}%
  \BibitemOpen
  \bibfield  {author} {\bibinfo {author} {\bibfnamefont {D.~H.}\ \bibnamefont
  {Dunlap}}\ and\ \bibinfo {author} {\bibfnamefont {V.~M.}\ \bibnamefont
  {Kenkre}},\ }\bibfield  {title} {\bibinfo {title} {Dynamic localization of a
  charged particle moving under the influence of an electric field},\ }\href
  {https://doi.org/10.1103/physrevb.34.3625} {\bibfield  {journal} {\bibinfo
  {journal} {Phys. Rev. B}\ }\textbf {\bibinfo {volume} {34}},\ \bibinfo
  {pages} {3625} (\bibinfo {year} {1986})}\BibitemShut {NoStop}%
\bibitem [{\citenamefont {Shirley}(1965)}]{S-Shirley1965}%
  \BibitemOpen
  \bibfield  {author} {\bibinfo {author} {\bibfnamefont {J.~H.}\ \bibnamefont
  {Shirley}},\ }\bibfield  {title} {\bibinfo {title} {Solution of the
  {Schr\"{o}dinger} equation with a {Hamiltonian} periodic in time},\ }\href
  {https://doi.org/10.1103/physrev.138.b979} {\bibfield  {journal} {\bibinfo
  {journal} {Phys. Rev.}\ }\textbf {\bibinfo {volume} {138}},\ \bibinfo {pages}
  {B979} (\bibinfo {year} {1965})}\BibitemShut {NoStop}%
\bibitem [{\citenamefont {Sambe}(1973)}]{S-Sambe1973}%
  \BibitemOpen
  \bibfield  {author} {\bibinfo {author} {\bibfnamefont {H.}~\bibnamefont
  {Sambe}},\ }\bibfield  {title} {\bibinfo {title} {Steady states and
  quasienergies of a quantum-mechanical system in an oscillating field},\
  }\href {https://doi.org/10.1103/physreva.7.2203} {\bibfield  {journal}
  {\bibinfo  {journal} {Phys. Rev. A}\ }\textbf {\bibinfo {volume} {7}},\
  \bibinfo {pages} {2203} (\bibinfo {year} {1973})}\BibitemShut {NoStop}%
\bibitem [{\citenamefont {Haldane}(1988)}]{S-Haldane1988}%
  \BibitemOpen
  \bibfield  {author} {\bibinfo {author} {\bibfnamefont {F.~D.~M.}\
  \bibnamefont {Haldane}},\ }\bibfield  {title} {\bibinfo {title} {Model for a
  quantum {Hall} effect without {Landau} levels: condensed-matter realization
  of the parity anomaly},\ }\href {https://doi.org/10.1103/physrevlett.61.2015}
  {\bibfield  {journal} {\bibinfo  {journal} {Phys. Rev. Lett.}\ }\textbf
  {\bibinfo {volume} {61}},\ \bibinfo {pages} {2015} (\bibinfo {year}
  {1988})}\BibitemShut {NoStop}%
\bibitem [{\citenamefont {Jotzu}\ \emph {et~al.}(2014)\citenamefont {Jotzu},
  \citenamefont {Messer}, \citenamefont {Desbuquois}, \citenamefont {Lebrat},
  \citenamefont {Uehlinger}, \citenamefont {Greif},\ and\ \citenamefont
  {Esslinger}}]{S-Jotzu2014}%
  \BibitemOpen
  \bibfield  {author} {\bibinfo {author} {\bibfnamefont {G.}~\bibnamefont
  {Jotzu}}, \bibinfo {author} {\bibfnamefont {M.}~\bibnamefont {Messer}},
  \bibinfo {author} {\bibfnamefont {R.}~\bibnamefont {Desbuquois}}, \bibinfo
  {author} {\bibfnamefont {M.}~\bibnamefont {Lebrat}}, \bibinfo {author}
  {\bibfnamefont {T.}~\bibnamefont {Uehlinger}}, \bibinfo {author}
  {\bibfnamefont {D.}~\bibnamefont {Greif}},\ and\ \bibinfo {author}
  {\bibfnamefont {T.}~\bibnamefont {Esslinger}},\ }\bibfield  {title} {\bibinfo
  {title} {Experimental realization of the topological {Haldane} model with
  ultracold fermions},\ }\href {https://doi.org/10.1038/nature13915} {\bibfield
   {journal} {\bibinfo  {journal} {Nature (London)}\ }\textbf {\bibinfo
  {volume} {515}},\ \bibinfo {pages} {237} (\bibinfo {year}
  {2014})}\BibitemShut {NoStop}%
\bibitem [{\citenamefont {Yuan}\ and\ \citenamefont {Fu}(2022)}]{S-YuanFu2022}%
  \BibitemOpen
  \bibfield  {author} {\bibinfo {author} {\bibfnamefont {N.~F.~Q.}\
  \bibnamefont {Yuan}}\ and\ \bibinfo {author} {\bibfnamefont {L.}~\bibnamefont
  {Fu}},\ }\bibfield  {title} {\bibinfo {title} {Supercurrent diode effect and
  finite-momentum superconductors},\ }\href
  {https://doi.org/10.1073/pnas.2119548119} {\bibfield  {journal} {\bibinfo
  {journal} {Proc. Natl. Acad. Sci. U.S.A.}\ }\textbf {\bibinfo {volume}
  {119}},\ \bibinfo {pages} {e2119548119} (\bibinfo {year} {2022})}\BibitemShut
  {NoStop}%
\bibitem [{\citenamefont {Legg}\ \emph {et~al.}(2022)\citenamefont {Legg},
  \citenamefont {Loss},\ and\ \citenamefont {Klinovaja}}]{S-Legg2022}%
  \BibitemOpen
  \bibfield  {author} {\bibinfo {author} {\bibfnamefont {H.~F.}\ \bibnamefont
  {Legg}}, \bibinfo {author} {\bibfnamefont {D.}~\bibnamefont {Loss}},\ and\
  \bibinfo {author} {\bibfnamefont {J.}~\bibnamefont {Klinovaja}},\ }\bibfield
  {title} {\bibinfo {title} {Superconducting diode effect due to magnetochiral
  anisotropy in topological insulator and {Rashba} nanowires},\ }\href
  {https://doi.org/10.1103/physrevb.106.104501} {\bibfield  {journal} {\bibinfo
   {journal} {Phys. Rev. B}\ }\textbf {\bibinfo {volume} {106}},\ \bibinfo
  {pages} {104501} (\bibinfo {year} {2022})}\BibitemShut {NoStop}%
\bibitem [{\citenamefont {Angehrn}\ \emph {et~al.}(2024)\citenamefont
  {Angehrn}, \citenamefont {Legg}, \citenamefont {Loss},\ and\ \citenamefont
  {Klinovaja}}]{S-Angehrn2024}%
  \BibitemOpen
  \bibfield  {author} {\bibinfo {author} {\bibfnamefont {L.}~\bibnamefont
  {Angehrn}}, \bibinfo {author} {\bibfnamefont {H.~F.}\ \bibnamefont {Legg}},
  \bibinfo {author} {\bibfnamefont {D.}~\bibnamefont {Loss}},\ and\ \bibinfo
  {author} {\bibfnamefont {J.}~\bibnamefont {Klinovaja}},\ }\bibfield  {title}
  {\bibinfo {title} {Relations between normal state nonreciprocal transport and
  the superconducting diode effect in the trivial and topological phases},\
  }\href {https://doi.org/10.1063/5.0216679} {\bibfield  {journal} {\bibinfo
  {journal} {Appl. Phys. Lett.}\ }\textbf {\bibinfo {volume} {125}},\ \bibinfo
  {pages} {023507} (\bibinfo {year} {2024})}\BibitemShut {NoStop}%
\end{thebibliography}
\end{document}